# AN INQUIRY INTO THE ECONOMIC LINKAGES BETWEEN THE SWEDISH AIR TRANSPORT SECTOR AND THE ECONOMY AS A WHOLE IN THE CONTEXT OF THE COVID-19 PANDEMIC

By

Rafael Andersson Lipcsey

A Thesis

Submitted to Università Commerciale Luigi Bocconi

In Partial Fulfillment of Requirements

For the Degree of Master of Science in Economics and Management of Government and International Organizations

April 2021

## Acknowledgements

Writing a thesis in the midst of a pandemic certainly is a unique experience. On the one hand you are locked at home, almost sensing the disappointed look of your thesis for every moment you don't spend writing it. On the other hand, you lack the human interaction that fuels your creativity. At times like this, the help and ideas you do receive are especially treasured.

I would like to start by expressing my sincere gratitude to my thesis supervisor, Professor Marco Percoco for his always rational and precise guidance throughout this journey.

I would also like to extend my sincere thanks to Professor Michael L. Lahr, Professor Manfred Lenzen and Professor Richard Wood for their valuable advice and suggestions related to the input-output framework used in this thesis.

I am also grateful to several of the passionate and patient people at the Swedish Statistical Office (Statistics Sweden). Without their datasets, parts of this thesis would simply not have been possible to complete. Thank you, Annika Damm, Andreas Poldahl, and Marcus Lundgren.

I also want to thank my close friends for keeping my spirits up at times where motivation was difficult to find, and I am as always eternally grateful for my parents' unconditional and limitless support.

Lastly, let's not forget about our house cat who brought joy and playfulness into the dull days of this past year.

# Abstract

This thesis aims to assess the importance of the air transport sector for Sweden's economy in the context of the COVID-19 pandemic. Two complementary research goals are formulated.
Firstly, investigating economic linkages of the Swedish air transport sector and secondly, estimating the effects of the pandemic on Swedish air transport, and the spin-off effects on the economy as a whole. Overview of literature in the field reveals that while a fair amount of research exists on the importance of air transport, unsurprisingly, pandemic effects have yet to be investigated. The methodological framework chosen is input-output analysis, thus, the backbone of data materials used is the Swedish input-output table, complemented by additional datasets. For measuring economic linkages, basic input-output analysis is applied. Meanwhile, to capture the pandemic's impacts, a combination of inoperability analysis and the partial hypothetical extraction model is implemented. It is found that while Swedish air transport plays and important role in turning the cogs of the Swedish economy, when compared to all other sectors, it ranks on the lower end. Furthermore, while the COVID-19 pandemic has a detrimental short-term impact on Swedish air transport, the spin-off effects for the economy as a whole are milder. It is concluded, that out of a value-added perspective, the Swedish government, and by extension policy makers globally need not support failing airlines and other air transport actors. Nonetheless, some aspects could not be captured through the methods used, such as the possible importance of airlines to national security at times of dodgy global cooperation. Lastly, to accurately capture not only short-term, but also long-term effects of the pandemic, future research using alternative, causality-based frameworks and dynamic input-output models will be highly beneficial.

# Table of Contents

# List of Tables

# List of Figures

# 1. Introduction

## 1.1 The COVID-19 Pandemic; A Transformative Force

January 2020. Dramatic scenes fill the reports of media outlets around the world. A "pneumonia of unknown cause" wreaks havoc in the city of Wuhan, China. The world watches as the city enters a complete lockdown, and those wishing to leave scramble to make it onto the last buses, trains and flights before all modes of transport are cut off. National governments also scramble to repatriate their citizens stranded in the city. While droves of aircraft, often provided by national carriers land and depart from Wuhan airport, an increasing number of cases of the "novel coronavirus" are confirmed globally. As the last aircraft with repatriated citizens return home, much of the world enters a lockdown of unprecedented magnitude. Airports fall quiet. Departure halls lay deserted, hundreds of aircraft are parked on the runways, and the world is hit by the first wave of the COVID-19 pandemic.

Airlines were strongly impacted already from the very beginning. By March 2020, global demand for passenger air services had plunged by 53% compared to the same period the year before (International Air Transport Association, 2020). Desperate calls for aid were heard all over, and governments now faced the dilemma of deciding how far to go in order to save their failing national airlines, a quandary remaining to this day, as the crisis continues.

## 1.2 The Swedish Air Transport Sector

Swedish air transport, despite the country's uniquely anemic restrictions was also heavily impacted with a 60% drop in passenger numbers in March, and 98% in April compared to the year before (The Swedish Transport Agency, 2019b). The Swedish air transport sector, chosen as the focus of this thesis serves an important role in moving passengers and goods. In this country of roughly 10 million inhabitants, there are 39 active airports, trafficked by 300 000 aircraft and 39 million passengers on a yearly basis. Out of these passengers, roughly 8 million travel domestically, and the remaining 31 million head towards international destinations. Furthermore, a touch over 140 000 metric tons of goods are freighted each year through air, out of which around 3 000 tons are domestic, and the remaining 137 000 are international cargo. (The Swedish Transport Agency, 2018; The Swedish Transport Agency, 2019; The Swedish Transport Agency, 2019a). Just strictly the air transport sector itself can be estimated to directly contribute with around SEK 7.4 billion and 6000 jobs to the Swedish economy (Statistics Sweden, 2020a), but when factoring in all the business activities supported by air transport, these contributions are calculated to be tenfold (Air Transport Action Group, 2018). Swedes are avid travelers, with air trips per capita standing at 2.25. Traveling to and from the country is rather easy, with an airport accessibility rating of 98.99% and rank 22 out of 189 countries in terms of air connectivity (Air Transport Action Group, 2018).

## 1.3 Questions and Objectives

It is by now clear that Swedish air transport is a vital sector for the economy, and that it was, and is, seriously impacted by the COVID-19 pandemic. But are basic statistics on direct contributions to GDP, and percentage drops in passenger numbers a salient enough base for policy makers to get the complete picture? How important is in reality the role played by air transport for the economy, in comparison with all the other sectors? And how great are the spin-off effects of the losses suffered by air transport for the entirety of the Swedish economy? What are the implications for Swedish, and international policy makers pondering on how much, if any, resources to pour into their airlines? The above problems can be compactly summarized in the following research question, chosen for this thesis:

*"How important is the Swedish air transport sector for Sweden's economy in the context of the COVID-19 pandemic?"*

An investigation into the above is logically divided up into two research goals. Firstly, finding the economic linkages of the Swedish air transport sector. How does air transport interact with other sectors of the Swedish economy, and how important is its role in ensuring added value, employment, income output and investment?

Secondly, what are the effects of the COVID-19 pandemic on the air transport sector, and how do the spin-off effects look like for the economy as a whole? The latter, of course depends on the economic linkages as inquired through the first research goal. And thus, these two research goals are necessary complements of each other to successfully answer the research question of this thesis.

## 1.4 Structure and Methods

This study contains the following structure. Firstly, a literature review, divisible into two main parts that align with the research goals. The first part investigates the economic linkages of air transport, and the second looks into the economic effects of the COVID-19 pandemic on the Swedish air transport sector, and the spin-off effects for the economy.

This is followed by the methods chapter, which contains an overview of the data materials, and theoretical frameworks used to complete the analysis. The Swedish Statistical office's, Statistics Sweden's various databases will serve as the pillars for data collection, complemented by resources from other institutions. The main methodological framework used will be input-output analysis, conducted on tables forged from Swedish national accounting statistics. Additional methods based on the input-output framework to assess the pandemic's impact will also be employed.

Up next is the findings chapter, containing results attained through analysis using the frameworks defined in the methods chapter. Then, in the discussion, these will be summarized, interpreted, and used to answer the research question and discuss the implications in a national and global context. Finally, a conclusion will aim to provide a condensed and clear summary of the key points of this thesis, and an appendix will also be added to hold complementary figures and tables.

# 2. Literature Review

## 2.1 Opening Reflections

Unsurprisingly, a large and varied body of literature exists on the air transport sector. It spans far and wide through fields of engineering, physics, political science, economics and more. For this reason, finding a focus on what is relevant for the purpose of this thesis is absolutely imperative. To start with, it can be expressed that this thesis is primarily one within the field of economics, although attributes importance to being of value to policy makers, and therefore somewhat stretches into the field of political science.

As already expressed, the research question of the thesis can be divided into two research goals. Firstly, the economic linkages between the Swedish air transport sector, and the economy as a whole. Secondly, the estimation of direct economic effects of a crisis such as COVID-19 on the air transport sector, and the following spin-off effects on the whole economy. Given this division, it is appropriate to attempt a similar division of the literature review, while acknowledging that there will always be overlaps. The literature review is divided into four main parts. The first one aims to primarily establish the body of non-academic literature available on the overall economic linkages between air transport end economies as a whole. The second one does the same for the available academic literature. The third part intends to capture the research available on the economic impacts of the COVID-19 pandemic and crises of similar scale on air transport, and economies as a whole. Lastly it is assessed in what way, and to what extent this thesis can prove to be a relevant contribution to literature in the field defined.

## 2.2 Economic Linkages of Air Transport: Non-Academic Literature

Non-academic literature on the economic linkages and contributions of air transport is largely comprised of reports drafted by industry organizations, government agencies and economic research consultancies. Generally, their aim is to provide an easy to understand, summarized overview of the air transport sector, and its benefits for society. The focus of these papers can be global, national, or company specific.

### 2.2.1 Global focus

Reports of global focus tend to follow a similar structure. They usually start with a historical overview of the air transport sector. Global Aviation Industry High-level Group (2019) and Air Transport Action Group (2018) both discuss for instance the exponential growth of air traffic in the past five decades and the significant reduction of air travel costs as a result of several key events, such as the entrance of passenger jets into the market, and US deregulation of the aviation sector.

Such overviews are often followed by a description of the air transport sector's contribution to global GDP and employment. Global Aviation High-level Group (2019) finds that air transport as of 2016 supported 65.5 million jobs worldwide, and contributed USD 2.7 trillion to global GDP, which can be compared to the size of the UK's GDP. Contributions of similar magnitude are also estimated in other reports (Air Transport Action Group, 2018). Such economic impacts are then divided up into four main categories: Direct, indirect, induced and catalytic. Direct impacts are described as those most closely linked to the air transport sector itself. These include contributions to employment by passenger and freight airlines, airports, air navigation services, and the manufacturing of aircraft and aircraft parts. Indirect contributions relate to sectors closely linked to air transport, such as suppliers of airplane fuel, warehousing services, and more. Induced effects meanwhile are the impacts spawned by the consumer spending of the employees directly or indirectly employed by the air transport sector. Since most sectors of an economy are highly dependent on consumer demand, the induced effects benefit an even wider array of actors, such as those in retail, or other service industries like telecommunication or banking. Finally, catalytic effects are the broadest and most difficult to define. They refer to all additional benefits not captured by direct, indirect and induced effects, such as for example increased efficiency of air transport, enabling private firms and public agencies to implement just in time delivery chains into their operations. Catalytic effects, in comparison with direct, indirect and induced are found to be by far the largest (Global Aviation Industry High-level Group, 2019).

Reports with a global focus also generally bring up social impacts and goals of air transport. Impacts include for instance the sector's key role in the provision of health and humanitarian aid in remote, otherwise inaccessible areas. Furthermore, the sector's ambition to align itself with the UN's sustainable development goals is also covered. This is to be done for example through carbon emissions offsetting (Air Transport Action Group, 2018; Global Aviation Industry High-level Group, 2019).

### 2.2.2 National focus

In terms of reports with a national focus, the most significant contribution is embodied in a very extensive report series created in the years 2009-2014 by economic forecasting and analysis company Oxford Economics. The series has a national focus and spans 41 countries of which Sweden is one (Chsherbakov & Gerasimov, 2019). Similarly to the likes of Air Transport Action Group (2018) and Global Aviation Industry High-level Group (2019), the Oxford series reports also contain estimations of the economic footprint of air transport. Note, these are now no longer global impacts, but rather impacts on national economies. The aviation sector is to start with split into three types of activity; airlines (cargo and passenger), ground based infrastructure and aerospace manufacturing. Then the direct, indirect and induced effects on the given country's economy are expressed as contribution to GDP, and employment. Catalytic effects are also measured and defined as spillover effects linked to the aviation sector. Examples given in the Oxford texts are the increased consumer demand due to tourism enabled by air transport, and the sector's contributions to increased efficiency in international trade flows (Oxford Economics, 2011b; Oxford Economics, 2014). Furthermore, the benefits of high air connectivity between the country of interest's cities and other major global hubs are also mentioned. These include increased foreign direct investment and a positive effect on long term GDP growth.

While most of the above in one way or another is also found in the global oriented reports of industry organizations, in addition to that, the Oxford papers also attempt to measure the total value enjoyed by consumers of air transport services. It is firstly established that finding the total value of consumer benefits is difficult. Passenger air fares and shipping charges only tell us so much. As the reports state, the total value must be defined as *"…the value placed on the service by the marginal passengers – those who would forgo the flight were prices to rise- and not the value that passengers as a whole place on air transport services."* (Oxford Economics, 2011, p. 7; Oxford Economics, 2014, p. 17) In other words the reports infer to the concept of diminishing marginal utility. For this reason, it is determined that a more ideal measure of value gained is the price elasticity of demand, which reflects consumer sensitivity to price changes, where a low elasticity, and therefore lower sensitivity to price changes would lead to a higher value to consumers (Oxford Economics, 2011; Oxford Economics, 2014). Once this is established the reports estimate the consumer benefits. In the case of Sweden, they are found to be SEK 114 billion for passenger air transport, and SEK 4.4 billion for cargo air transport. An interesting finding, considering how low the value is for the latter.

Furthermore, there are also more technically oriented reports in the field. They contain more discussion around methodology, and it can be theorized that they are possibly intended to a slightly different audience, more interested in technicalities. Malaysian Aviation Commission (2017) for instance aims at capturing the benefits of the air transport sector for the Malaysian economy. The impact, in synergy with Air Transport Action Group (2018), Aviation Industry High-level Group (2019), and the Oxford series is divided into direct, indirect and induced. (Note that in this case catalytic effects were not measured). Nonetheless, the report takes a more technical approach, in which it discusses the theory behind the calculation of output multipliers, which show how the spending on, or investment into a given sector spawns an effect of greater magnitude than the original amount spent/invested. The authors define output multipliers as *"…the total effects (direct, indirect and/or induced) divided by the direct effects…"* (Malaysian Aviation Commission, 2017, p. 3). The report finds the output multiplier for the Malaysian air transport sector to be 2.0, which implies that every unit of investment/spending on the air transport sector results in 2 units of output of the whole economy (Malaysian Aviation Commission, 2017).

### 2.2.3 Company focus

Lastly, at times reports don't have a global focus such as Air Transport Action Group (2018) and Global Aviation Industry High-level Group (2019), or a national focus such as the Oxford report series. Instead, they direct their attention to a single airline, such as a 2019 report written by Copenhagen based consultancy "Copenhagen Economics" on the economic benefits of Swedish, Danish and Norwegian national carrier Scandinavian Airlines. The goal of course, is to prove the airline's importance for these economies. When counting direct, indirect, induced and catalytic effects, SAS is estimated to contribute with 7800 jobs, and a SEK 16.6 billion GDP boost to Sweden's economy (Copenhagen Economics, 2019).

### 2.2.4 Non-academic literature: conclusions

Reports such as the ones covered above are many in number and variety. The stance they take, which is to demonstrate the benefits for society by air transport is unsurprisingly rather homogenous. Furthermore, as can be expected, methodology used, and a discussion around its potential downfalls, as well as a more in-depth reflection on the implications of findings is often limited. Still, one common denominator is that all reports to a smaller or larger extent make use of a method named input-output analysis, which investigates interdependencies between industry sectors, and can be used to measure the impacts of a given sector on the economy. This method, as will be seen is also a common theme to be found within the academic literature.

## 2.3 Economic Linkages of Air Transport: Academic Literature

Academic literature on economic linkages of air transport can be divided based upon four broad categories of methodology: literature reviews and summaries, methods of causality, the input-output framework, the computable general equilibrium approach and cost accounting methods.

### 2.3.1 Literature reviews and summaries

Predominantly, qualitative literature on the air transport sector seems to be mostly focused on emerging or developing economies, and undeveloped regions. Olariaga and Ávarez (2015) for example examine the development and evolution of Columbian airports, Bråthen and Halpern (2012) discuss the importance of air transport in remote regions for the purpose of fueling economic development, and Daley (2009) debates whether air transport development is done in a way where it enhances already existing economic inequalities. Put differently, the above papers address two key questions: how beneficial is air transport for emerging /developing nations, and what are the issues to look out for in order to leverage its benefits for economic growth in a fair and non-discriminating way?

Olariaga & Ávarez (2015) are more aimed at answering the former, in the context of Colombia. They consider the Colombian air transport sector in the period of the early 2000s up to 2014. The paper presents four main contributions: Firstly, a historical overview of the institutional framework of the Colombian air transport sector. Secondly, a timeline regarding public and private investment into Colombian air infrastructure. Thirdly, the historical evolution of passenger numbers and cargo volumes, as well as air connectivity, and fourthly, the Colombian air transport sector's contribution to GDP over the years. Overall, the authors present a fairly positive view of the sector. Institutional frameworks and investments are concluded to have somewhat solidified over the years, and cargo as well as passenger volumes are found to have increased as well, although less than expected. Furthermore, air transport's contribution to GDP is estimated to historically have stayed more or less constant, meaning it has mostly followed the development of the GDP changes over the years.

Overall, the paper provides a good overview of the state of the Colombian air transport sector, however it must be noted that it uses much the same structure as the Oxford paper series, and many of the figures are brought from a particular paper in the series on Colombia published in 2011 (Oxford Economics, 2011a). Perhaps, its greatest addition is a clear and illustrative explanation of the political context, as well as a detailed description of the main Colombian airports, and their evolution.

Daley (2009) and Bråthen & Halpern (2012), while still interested in economic linkages and economic contributions from air transport, also connect these to questions of economic inequality. It should also be noted that they take a global focus, in contrast to the national one in Olariaga & Ávarez (2015). To start with, Daley (2009) is aimed at assessing the claim that air transport drives economic development to an extent that outweighs environmental and social costs. The author theorizes that while air transport may bring economic benefits, these are unevenly distributed and could reinforce already existing economic inequalities. Daley (2009) goes through a list of studies, including the UK edition of the Oxford Series reports, and UNCTAD documents which argue that there is a disparity between air transport in poorer and richer regions. All in all, it is recognized that air transport is beneficial for economic development, but at the same time, Daley (2009) wants to challenge the view that air transport is uniformly beneficial, no matter the circumstances (Daley, 2009).

Meanwhile, perhaps as a further development of this issue Bråthen & Halpern (2012) examine the importance of air transport in remote regions for the purpose of creating economic development. The paper is centered around a literature review as well as experiences with air services in remote regions. It covers experiences and descriptions of a number of remote air service programs, such as the US centered Essential Air Service Program, and the EU based Public Service Obligation program. A number of studies in the field are also reviewed, which discuss for instance socio-economic effects of fare reductions in remote areas, aircraft and airport operating costs in remote regions, importance of marketing for airports, and the link between air transport and employment (Bråthen & Halpern, 2012). Overall, it can be argued that this paper is somewhat better researched, as it presents a literature review, and a number of other sources of good diversity and high quality in comparison to Daley (2009) who's conclusions are not as well supported. That being said, Bråthen & Halpern (2012) come to similar conclusions as Daley (2009). It is found that air transport is beneficial even for remote areas, but the development it spawns is unfortunately not completely uniform. The authors call for a clearer set of criteria to determine what classifies as "remote" in order to avoid currently on-going mis categorizations, where some remote areas are not given enough support, and some heavily trafficked areas are given too much of it. They also go on to suggest optimal models for calculating aircraft operating costs, subsidy levels, tender management mechanisms, and effective local marketing opportunities (Bråthen & Halpern, 2012). It can be concluded lastly that while the focus of this paper is not exactly what lies within the scope of this thesis, it contains an excellent collection of topics, that are of use also on a greater scale, not just for remote regions.

### 2.3.2 Causality studies

The next category of literature can be classified as one containing methods of causality to measure the economic linkages of air transport. More specifically, these papers, apart from their methodology, also seem to have two more aspects in common. Firstly, they use time series data. Secondly the key economic linkage they are interested in measuring is the possible connection between air transport and GDP growth, which if proven, establishes the sector's economic importance.

Brida et al. (2016) for example investigate the long-term causal relationship between air transport and economic growth in Italy during the period 1971-2012. The authors use methods such as the Johansen test[1] and the Granger Causality test[2] on time series GDP, and airport movements/air cargo volume. They find that on the long term, air transport has an important role in inducing GDP growth in Italy. In specific, a positive shock on air transport could increase GDP by as much as 10% after three years. The authors explain this effect by referring to the direct and indirect effects of air transportation, which was also covered in several non-academic reports (Air Transport Action Group, 2018; Global Aviation Industry High-level Group, 2019; Malaysian Aviation Commission, 2017; Oxford Economics, 2011; Oxford Economics, 2014). Overall, the paper uses a set of very relevant techniques to measure the long-term contribution of air transport to Italy's economy.

While not their overall focus, the authors also mention interesting findings in their literature review which are worth bringing up. It is expressed that external shocks such as the September 11 attacks caused on the short run large changes in passenger air traffic demand but had very little effect on cargo demand (Brida et al., 2016). This finding is not so relevant for research on the economic linkages of air transport. It can however prove useful for the second research goal, namely the effect the COVID-19 pandemic on Swedish air transport and in turn economy as a whole. While it is at this point still too early to draw any conclusions, it can be theorized that perhaps there are similarities to be found between the impact on air transport by the September 11 attacks, and the COVID-19 pandemic. Either way, it is a claim worth investigating.

Turning back to the main topic of this section however, it can be observed that the positive link between air transport and GDP growth, similarly to the findings of Brida et al. (2016) is also largely confirmed by others such as Tolcha et al. (2020) and Nasution et al. (2018) who focus on six Sub-Saharan economies, respectively the Indonesian economy. Both of these studies employ vector error correction and vector autoregression models[3], as well as the Granger Causality test used by Brida et al. (2016).

[1] A method for testing cointegration in time series data.
[2] A statistical hypothesis test to establish forecasting relationships between different sets of time series data.

To start with, Nasution et al. (2018) find clear evidence that both passenger air transport and cargo air transport bolster GDP growth in Indonesia. These results can be contrasted to Brida et al. (2016) in the sense that the latter did not find a significant relationship between air cargo demand and GDP growth, but rather only between air passenger demand and growth. Meanwhile, the findings of Tolcha et al. (2020) are more mixed. The authors express that the aim of their paper is to look into if the same causal relationship between air transport demand and economic growth that was established to exist for developed economies also exists also for economies with very immature aviation markets, such as those in the countries of Sub-Saharan Africa. These countries experience in general a lower volume of air transport traffic that is more highly regulated and also concentrated to a few major hubs. At the same time, air transport is key for many of these countries given that a large percentage of them are completely landlocked, with undeveloped land transport infrastructure. Therefore, as the authors state, air transport provides for them an important bridge to international markets. Tolcha et al. (2020) are able to establish that there exists a long run relationship between air transport demand and economic development, but also find that the strength of this linkage varies depending on the scale and accessibility of air transport in the given country. The effect is as such found to be stronger for countries with high degrees of air connectivity such as Ethiopia and South Africa, but lower for countries with low connectivity, in example Nigeria (Tolcha et al., 2020).

Overall, it can be stated that models of causality to evaluate the importance economic linkages of air transport prove to be of value for the research field. They are not however in majority. As will be seen below, input-output models are a more popular choice.

[3] The Vector Auto regression model is a version of classical OLS regressions but typically used when predicting multiple time series variables using a single model (Stock&Watson, 2001).

### 2.3.3 Input-output methodology-based studies

As noted, studies making use of input-output methodology within the specified field are many in number. They have much in common, as in that they account for direct, indirect, induced, and in some cases catalytic impacts. That being said, their focus varies considerably, and cat be placed into four main categories: general nation-wide studies, airport-specific research, papers on tourism impacts, and lastly a focus on climate effects.

Chsherbakov and Gerasimov (2019) belong to the first category, taking a general approach to measuring economy wide linkages of air transport. They account for the impact of air transport on the Russian economy between the years 2007 and 2016. In the paper, the aviation sector is divided into two main categories; air transport for cargo and passengers, and ground based infrastructure providing support activities for the former. Chsherbakov and Gerasimov (2019) find that overall, Russian airlines saw significant growth most years, with the exception of the years of the Russian financial (2013-14) crisis. Meanwhile Russia's GDP hovered more or less around the same average, hence, it is concluded that the importance, and contributions of the Russian air transport sector increased (Chsherbakov & Gerasimov, 2019). One of the main virtues of the paper, which the authors also point out themselves is that a longer period of time was assessed, which was traditionally rare for input-output analysis. Meanwhile however, the accuracy of their results is somewhat tainted by the lack of availability of data in Russia. For instance, the authors point out that many airlines and airports in Russia are not required to provide their financial reports publicly.

Other studies meanwhile focus on quantifying the impacts of a few specific airports. Bråthen et al. (2006) examine the direct, indirect, induced and catalytic contributions to employment from four Norwegian airports. Before continuing, it can be noted that Bråthen by now is a familiar name; after all, Bråthen and Halpern (2012)'s literature focused review of the link between the economic benefits of air transport to remote regions was already covered earlier on in this literature review. Steering back to the 2006 paper in question however, Bråthen et al. (2006) calculate the employment multipliers (direct, indirect and induced effects) for the airports, and find them to be smaller than those attained in many international studies. For instance, they attain an employment multiplier (a version of the output multiplier focusing on effects of employment) of 1.91 for Oslo airport, compared to Milan airport with a multiplier of 4.1. They also find an output multiplier of 1.99. The catalytic impacts are then estimated through an explorative, questionnaire-based case study linked to a smaller Norwegian airport. Catalytic impacts are by the authors defined to be location impacts, tourism and trade, productivity, and investment all brought upon the region by the air transport sector. The results show catalytic effects magnitudes larger than direct employment (around eight-fold). At the same time, these effects, as the authors also acknowledge are hard to measure. It is noted that the response rate for the survey was rather low, only 15%, which complicates the interpretation of results (Bråthen et al., 2006). Overall, the paper presents some intriguing findings, and can be considered a salient, early contribution to research in the field. It is noteworthy that multipliers for Norwegian air transport were found to be so much lower than elsewhere. One may theorize if this is equally the case for its Scandinavian neighbor, Sweden.

Another addition comes from Huderek-Glapska et al. (2016) who inquire into the impacts of three Polish airports in a similar fashion to Bråthen et al. (2006) with the only main difference being that this time catalytic effects were not estimated. Interestingly, employment multipliers for the three Polish airports chosen are found to range between .45 and 1.05, which are even lower than those found by Bråthen et al. (2006). It is difficult without additional research to theorize why this is the case, but it is possible for one that airports chosen as comparison by Bråthen et al. (2006) happen to have high employment multipliers in general, or that air transport multipliers have dropped in the years that passed between the two studies. Differences in methodology may also have led to significant discrepancies.

So far, two of the categories of input-output based research on air transport have been covered: studies with a general view as seen in Chsherbakov and Gerasimov (2019), and research focused on specific airports/regions as done by Bråthen et al. (2006) and Huderek-Glapska et al. (2016). A third category looks into links between air transport and tourism, and measures impacts on the economy. Dimitriou et al. (2017) investigate these impacts on tourism dependent regions in Greece. Using input-output methodology they find that the tourism sector is key for Greece's economic recovery, and air transport is key for the development of the tourist sector. They also measure catalytic effects by making use of data on international tourists' arrivals to the region, these tourists' average spending, and the percentage of air travelers to the region. It is interesting to contrast this method to Bråthen et al. (2006) who chose the survey alternative instead, and it further underlines existing ambiguity over what comprises catalytic effects.

Another interesting contribution is from Kronenberg et al. (2017). Their focus, similarly to Dimitriou et al. (2017) is tourism, however this time the air transport sector is not the only sector examined. The goal is instead to evaluate the importance of a range of different sectors for tourism in Sweden. While the paper perhaps steers a bit far from the strict air transport focus of this literature review, it is worth bringing up for two reasons. Firstly, the authors present a useful discussion around the evolution of input-output analysis, and its potential downsides. They express that what had been lacking in input-output research thus far had been the analysis of multipliers over several consecutive years, and a degree of regionalization. While the latter can perhaps be contested with examples such as the contribution from Dimitriou et al. (2017) there seems to be more evidence for the former. In fact, it can be observed that around this time, more and more efforts were put into multi-period analyses of the economic linkages of air transport. This is certainly a relevant aim, as actors like governments and firms alike are often interested in the long-term effects of for instance large investments into infrastructure. Secondly, Kronenberg et al. (2018) provide useful historical comparisons between the sectors with the highest output multipliers for tourism in Sweden. It is found that while in 2008 the multipliers stood at more or less similar levels between sectors such as air transport, accommodation and food, land transport and so on, in the years to come, air transport would experience a growth unlike any other, pushing its multiplier way above the average.

By this point, already a significant chunk of literature has been reviewed concerning the linkages between the air transport sector and economies as a whole and it is becoming fairly clear that air transport has several benefits for economic growth, tourism, and is an essential component of any economy, developed or developing. However, it is also important to recognize that despite all these benefits, the air transport sector is not all roses. Perhaps its most significant negative effect is its climate impact, and the sector is as of now showing limited progress in reinventing itself to alleviate this issue. This may partly be due to lack of willingness, incentives and regulations, but is also attributable to the engineering challenge of providing a method of propulsion powerful enough to be a viable replacement to kerosine (Verstraete, 2019). Challenges of engineering nature will of course not be covered, as they are far outside the scope of this thesis, but it can be concluded that the climate impact of air transport is possibly its greatest arch enemy at this day and age.

Sajid et al. (2019) investigate exactly this, using input-output methodology. They measure the transport sector carbon linkages of the EU's top emitters between 1995 and 2011, and also present emissions levels and intensities. While air transport is not the exclusive focus of the paper, it can be observed in the results that direct emissions of air transport when compared to inland and water transport are the highest and have also seen the most consistent rising trend amongst all countries covered. Furthermore, air transport is found to be the most carbon intensive sector. Sajid et al. (2019)'s contribution therefore proves to be a salient one, as it presents a good overview of the linkages between emissions and the air transport sector, over a period of time, in comparison with other transport sectors, and across several countries.

### 2.3.4 Computable General Equilibrium approach

Alternative methodologies, more specifically the Computable General Equilibrium (CGE) approach that uses the input-output framework as a base have also been used to assess the importance of air transport for growth and development (Ishikura & Tsuchiya, 2007; Njoya & Nikitas, 2020). CGE models are related to classical input-output models, but they view prices as more important. They are built on neo-classical assumptions such as that households receive income through a supply of labor and capital, and further differ from input-output analysis for instance on the basis that while in the input-output models fixed amounts of labor are required for the creation of a product, in the CGE model, wages can also affect labor demand (Dixon & Jorgenson, 2012) . Typically, GCE models are seen to be optimal to estimate the effects of specific measures on specific sectors (such as a tax) and in turn estimating the overall effect on the economy.

Ishikura and Tsuchiya (2007) apply the model to the Japanese economy and evaluate the impact of airport development. They find that airports have a significant economic potential for inducing growth, and underdeveloped airports can limit an economy's growth potential. Njoya and Nikitas (2020) in turn aim at capturing the role of air transport in the creation of jobs and growth in South Africa. The CGE model is in this case complemented by a SAM table (social account matrix), which is essentially a special input-output table that shows the relationships between the economy's sectors and the domestic and foreign institutions part of it. SAM models are beneficial for capturing policy impacts as they also provide pointers regarding the inefficiencies of transactions between sectors (Njoya & Nikitas, 2020). Meanwhile, the GCE model was used in a way as to simulate the impact of a 10% increase in air transport capital stock on the economy. The authors are able to identify rather narrow subsectors which are the most significantly impacted by this simulated change. They are also able to separate for instance which category of labor would see the greatest benefits. Limitations of the study include assumptions of full employment in South Africa, and a significant informal labor market not captured by the model, but overall, it is assessed to be a valuable and innovative contribution to the research in the field. It seems as if CGE models perhaps are starting to see a resurgence as they can be combined with other methodologies that complement it.

As can be concluded, input-output based models are commonly used to evaluate the long run economic linkages of, and the benefits gained from the air transport sector. Some papers decide to provide overviews of the impact of this sector on the economy as a whole, but we are also seeing a shift towards increased interest in regional effects, and the inclusion of the time variable into models.

**2.3.5 Cost accounting**

The cost accounting approach seems to be a popular choice for studies investigating the carbon linkages of air transport. In essence, it entails usage of accounting methods to effectively capture greenhouse gas (GHG) emissions of the air transport sector. The way this task is approached however depends on the paper at hand. Some are interested in capturing GHG emissions linked to a specific airport (Ivković et al., 2018) while others try to estimate total emissions coming from the civil aviation sector of a country (Larsson et al., 2018). The former choose Belgrade airport to focus on, and calculate emissions of various pollutants (CO2, HC and more) by aircraft type, based on landing and take-off statistics and emissions factors (Ivković, Čokorilo, & Kaplanović, 2018).

The latter also make use of airport statistics, but this time the number of non-domestic air trips made by Swedish residents, the length of these trips, and the greenhouse gas emissions generated per km travelled by each passenger were used for the analysis. Two key research goals can be identified. The first is to create an effective accounting method to quantify Sweden's greenhouse emissions linked to civil aviation. Larsson et al. (2018) find both an increase in international air trips made by Swedes, and that emissions per capita for these are seven times the global average. Given the yearly increase in passenger traffic, the results are found to be to be problematic in the context of increasingly strict emissions targets, and various solutions are suggested such as a frequent flyer tax that would imply a tax proportional to the number of trips taken by a person each year (Larsson et al., 2018). Their second research goal is to critique allocation mechanisms for GHG emissions options for international air travel. Already existing regulations are deemed ambiguous and not very clear. Larsson et al. (2018) recommend a system named "residency of the passenger", in which the allocation is made based on the country of residence of the passengers, a way to hinder countries with large transport hubs from receiving an unreasonably large quantity of options.

A link here may be made back to Bråthen and Halpern (2012) who as described investigate unequal political and financial support to airports in remote areas. As mentioned, they find that some very remote areas are not given enough support, while heavily trafficked areas are given too much of it. Although Larsson et al. (2018) do not investigate remote areas per se, but rather GHG emission option allocation systems, a parallel between these two papers may still be drawn. What emerges is that well trafficked airports enjoy unfair advantages in several areas in comparison to smaller airports. There is therefore a clear need for improved allocation mechanisms, whether that is emissions options for airports of developed nations, or financial support for airports of developing nations.

### 2.3.6 Final thoughts: academic literature on linkages

All in all, it can be concluded that academic literature on the linkages between air transport and whole, or parts of economies is fairly rich, and includes a number of perspectives and methodologies. The focus can lie on economies as a whole, but there has been a gradual shift in interest towards regional analyses, and some papers focus expressly on climate or tourism impacts. In addition, time series data has gained space in the literature. The most prominent framework used is input output analysis and the methods built around it.

## 2.4 Effects of the COVID-19 Pandemic on Air Transport

The second research goal of this thesis refers to the short-term effects of the pandemic on the Swedish air transport sector, and the spin-off effects on the economy as a whole. For this reason, this chapter is concerned with evaluating and exploring the body of literature available in the field. Primary focus will naturally be on the effect of the pandemic, but impacts of other historical crises will also be covered.

### 2.4.1 Effects of historical crises

Two relatively recent historical events with significant impacts on air transport are the September 11 attacks and the 2007-2008 financial crisis. Santos (2006) addresses the former with his research on the September 11 attacks' effect on US air travel. He takes a demand-oriented approach, aiming to measure the effects that reduced confidence in air travel had on the US air transport sector, and in turn the spin off effects on the economy as a whole. Santos (2006) considers a 33.2% reduction in passenger traffic and a 19.2% decrease on hotel occupancy, and uses an input-output based methodology, inoperability analysis to conduct his analysis. Most significantly affected are unsurprisingly the sectors on which he simulated the drop in demand, but other sectors are found to suffer large losses too, such as the oil and gas extraction sector, transportation and support activities, petroleum and coal products manufacturing, and administrative and support services.

Meanwhile Dobruszkes and Van Hamme (2011) take a more supply-oriented approach regarding the effects of the financial crisis and use regression analysis to find possible causal relationships between change in GDP and airline supply. Results indicate that there were clear geographical disparities in terms of how airline supply evolved in the years 2008-10, where countries of the traditionally Western world (Western Europe, North America) saw serious decline in supply whereas China and other countries of Southeast Asia saw their supply increasing. This is not very surprising, considering it is in line with how harshly each of these countries were affected by the crisis. Furthermore, Dobruszkes and Van Hamme (2011) are able to confirm a link of causality between economic growth and airline supply, but there are some interesting deviations, with the US seemingly never fully recovering from the September 11 attacks, and Middle Eastern countries seeing dampened damage because of hubbing strategies they employed. All in all, the results presented by the paper seemingly confirm that the proven positive link between air transport and growth can also be applied for times of crisis.

### 2.4.2 Effects of COVID-19: Introduction

Literature on the effects of the pandemic on air transport is unsurprisingly not all that developed yet, but is constantly expanding, and surely will continue to do so in the years to come. In the medical field, several studies already exist on the usage of aircraft for carrying infected people to hospitals, or the risks of air travel aiding the spread of the virus (Nakamura & Managi, 2020). Studies also exist on the impact on air traffic volumes (Sun, Wandelt, & Zhang, 2020). Taking into account the scope of this thesis however, it was decided that the focus will lie on papers that loosely speaking deal with economic impacts. Papers in the field will be categorized based on methodologies used.

### 2.4.3 Effects of COVID-19: SWOT

SWOT analysis was one of the methods used to evaluate the strength and resilience of air transport in the context of the pandemic. Li (2020) investigates how China's air cargo sector held up and finds that Chinese air cargo traffic experienced much less damage as a result of the pandemic than did air passenger traffic. This is due partly to the fact that air cargo traffic wasn't as strongly hindered by border closings and other restrictions as was air passenger traffic. The author notes that some cargo traffic is traditionally done through passenger aircraft belly holds, and since passenger air transport was heavily affected, this could also endanger air cargo flows. Some of these losses were compensated for however by early government investment into dedicated cargo aircraft. In general, Li (2020) predicts a bright future for the Chinese air cargo sector, in the wake of continuous government support, and rising demand for fast paced e-commerce deliveries which favor air cargo over other forms of transport. Some dangers faced by the sector do exist however and are identified to be the strained US-China trade relations and increasing pressures on profitability. Within this literature review, this is the only paper found that exclusively focuses on cargo air traffic impacts, and therefore provides valuable insights. A link can be made with Brida et al. (2016) who as expressed earlier in this review, found that the September 11 attacks, while strongly impacted passenger traffic, barely had any effect on cargo. One might therefore theorize that there are significant similarities between the effects on air transport caused by the September 11 attacks, and the COVID-19 pandemic.

#### 2.4.4 Effects of COVID-19: Interviews

Unsurprisingly, given the actuality of the pandemic, especially in the early stages, interview-based studies have proven to be fairly quick and insightful way to investigate the impacts on air transport. Suau-Sanchez et al. (2020) conducted interviews in March-April 2020 with experts within the air transport sector which show some interesting insights into how the situation at that point was perceived. On the demand side, interviewees expressed worry regarding the expected drop in in demand but seemed to believe in a rapid recovery of essential business travel, as this was deemed a necessary component to maintain relationships with clients and providers. At the same time, sentiments towards the recovery of non-essential travel such as conferences and other larger events were more pessimistic. The experts also expressed worries regarding a level playing field for airlines post-pandemic. It was believed that state aid, while necessary to rescue airlines, could also extend the life of carriers that would normally have disappeared as a result of market competition. Furthermore, it could provide unfair advantages to some airlines given that not every government wanted to commit equally to rescue packages (Suau-Sanchez et al., 2020).

It would be an interesting exercise to see if, with the third wave of the pandemic tightening its grip globally, the same executives would still view the path to recovery as optimistically as back in March. One cannot help but wonder if the transformative effect of the pandemic on the ways of working, and the ways of engaging with clients and partners which has stomped business travel to a bare minimum will allow for a full on rebound of business travel, or if its fate has been sealed in the long term. The interviews reflect very well the unexpected nature of the pandemic, and although as of December 2020, there seems to be a clearer path ahead, with vaccinations starting up in an increasing number of countries, much uncertainty still remains regarding the year and possibly years to come.

However even in such times, a silver lining can be discovered, and this is also elaborated on by the interviewees in Suau-Sanchez et al. (2020). It is the hope that the pandemic may uncover opportunities to bring the air transport industry more in line with sustainable development goals. Suau-Sanchez et al. (2020) bring up as an example the environmental considerations around the consumption of exotic fruits in developed countries, which require rapid air transport from the other side of the world. Other examples could include the sustainability conditions tied by governments to bailout packages for airlines, such as was the case for Air France (Flight Global, 2020).

### 2.4.6 Effects of COVID-19: Quantitative studies

There is understandably as of now only a very limited selection of studies with a quantitative nature available on the pandemic's effects on air transport. An example of the few studies to be found however is OECD (2020), which provides an overview of the impact on airlines in the short, medium and long term. The report starts by estimating revenue drops for passenger air transport as caused by the pandemic and describes the medium run uncertainties airlines face as a result. It also elaborates on short run costs which are primarily found to have to do with health-related measures such as requirements for social distancing, and other sanitary procedures.

Furthermore, policy interventions, more specifically government funded support packages for airlines are also discussed. Out of the countries covered, Sweden, with a projected support of USD 1 billion is fairly in the middle, with countries such as the USA (USD 80 billion), Germany (USD 11 billion) and France (USD 8 billion) in the front, and countries like Denmark (USD 200 million), Greece (USD 200 million) and India (USD 100 million) trailing behind. That being said, these figures remain uncertain, as they also contain proposed, but not yet realized governmental support. (In the case of Sweden only about USD 500 million is confirmed, and the rest is proposed) (OECD, 2020). Furthermore, it can be questioned if such support packages are backed by suitable research. Often times they are based on calculations made by the airlines themselves, which of course provide an accurate picture of their financial situation. However, before committing to large support packages from taxpayers' money, policy makers may also see the need to weigh in the costs of their national airlines failing for society, in order to make the most informed decisions. Understandably, this is difficult to do when only very limited time is given to react, however this underlines the great need for research on the importance of air transport for national economies, in the context of the pandemic.

Another study that goes in more depth is by Iacus et al. (2020) who take on the ambitious goal of estimating and projecting air passenger traffic during the pandemic, as well as calculating economic effects. The authors estimate changes in passenger numbers for 2020 using various scenarios, some of which are based on the level of measures implemented by countries in the times of the SARS and MERS epidemics of 2003, respectively 2015, in order to estimate actual passenger volumes. Next, in order to give a rough estimation of the global economic effects caused by the pandemic, the authors use a report from the International Air Transport Organization on the contributions of air transport, similar to the non-academic literature covered above, as it reports both on the direct, indirect, induced and catalytic contributions of air transport. The above defined scenarios were then applied to these figures in order to estimate the effects of the pandemic on air transport and its contribution to GDP and employment (Iacus et al., 2020). The study marks a very ambitious, early effort to capture the effects of the pandemic on air transport, and in turn the global economy. It is also clear, that the number of resources put into the project was quite significant. Perhaps one weakness of the paper not mentioned by the authors is the way the scenarios were applied to the aviation sector's direct, indirect and induced effects. The scenarios were seemingly applied in a uniform way on these figures, which may give cause to overestimation as the indirect and induced contributions may not see a drop in the same proportion as the direct contributions, or might on the other hand be underestimated, as the spin-off effects of a reduction in direct output are not taken into account.

#### 2.4.6 Concluding remarks

All in all, as can be observed, literature on the effects of the COVID-19 pandemic on air transport, and the spin-off effects for national economies is at this point in time sparce. At this stage, more common are qualitative studies on the immediate effects of the pandemic on air traffic volumes than research which also undertakes the estimation of economic effects. That being said, one can without much hesitation guess that a large body of literature in this field will emerge in the years to come. As can be observed, historically speaking, there has always been a great interest in evaluating effects of major events on air transport, such as was the case for the September 11 attacks, and various financial crises.

## 2.5 Contribution to Literature

Based on the literature review one can conclude that with the specified research question, this thesis can prove to be a valuable contribution to literature within this field. As detailed, this writing presents two main research goals: An inquiry into the overall economic linkages of Swedish air transport, and an assessment of the COVID-19 pandemic's economic effects on the Swedish air transport sector, and the spin-off effects caused for the whole economy. While there is clearly more literature available on the former, Sweden-specific studies are still few in number. As for latter, understandably, only now is literature on the economic effects of the pandemic starting to pop up, with the most significant contribution so far provided by Iacus et al. (2020). Here, the global effects are covered, which is undoubtedly an important undertaking, however regulators will surely have an interest in seeing the effects also on national economies, such as Sweden's.

Furthermore, this writing could also potentially contribute with some methodological advancements. As was made clear in the sections above, input-output methodology is by far the most popular choice when analyzing economic linkages of air transport. The input-output framework has also been proven to work for analysis of crisis effects, such as was done by Santos (2006) for the September 11 attacks. He used inoperability analysis, which is an advancement on top of the basic input-output method. Furthermore, Iacus et al. (2020) used figures attained by the International Air Transport Organization through input-output analysis, however they simply applied their scenarios to these, and did not use input-output methodology themselves. All in all, it is clear that the input-output framework carries potency for both research goals of this thesis, hence it is chosen to serve as its methodological pillar. As seen above, there may however be room for improvements, of only marginal, especially when it comes to its usage on crisis scenarios.

In the following chapter, all considerations regarding methodology will be presented, and these will further clarify what value this thesis can bring to the literature in this field. Before doing so however, it is also important to issue a foreword of reservation. The COVID-19 pandemic has proven to be more ambivalent, and harder to predict than initially expected. The situation is everchanging, as demonstrated by the almost time capsule worthy interviews of executives in Suau-Sanchez et al. (2020). Quantitative studies on this topic can therefore only be so accurate. It may be argued as such that they provide only limited value to policy makers, however another perspective is that the crisis is happening at this moment, and at the cost of delivering figures with somewhat lacking accuracy, the benefit to policy makers is gaining an overview of the current state, enabling them to react proactively, and on time.

# 3. Methodology

## 3.1 Input-Output Analysis: A Brief Introduction

In the following section a brief introduction to input-output analysis is provided. It was deemed beneficial to do so in order to set the stage for the methodological explanations to follow, which are all built on this foundation. This section is divided into two parts: a brief history of input-output analysis and its theoretical fundamentals.

### 3.1.1 History of Input-Output analysis

Input-output analysis traces its roots back to Harvard professor Wassily Leontief, who developed its analytical framework in the late 1930s. It can comfortably be argued that Professor Leontief was ahead of his time. With a method essentially based upon matrix algebra, he had very early on recognized the future importance of computers for large scale empirical studies (Polenske, 2004). Leontief, while he is today universally recognized as a significant contributor to economic theory, at his time was rather critical towards most economic theorists such as for instance advocates of neo-Keynesian frameworks. He believed, somewhere on their journey with a first-hand focus on theory, they had lost touch with empirics. At the same time, he also threw punches towards empirical analysists, whom he claimed tried to "*depict the operation of the entire economic system in terms of five, four, or even only three aggregative variables.*"(Leontief, 1966, pg. 40). For Leontief, a true and (natural) symbiosis between empirics and theory was the key to valuable research. In the coming decades, Leontief would make several additions and refinements to the model, his work eventually earning him the Nobel Prize in Economic Science in 1973 (Miller & Blair, 2009, p. 1).

### 3.1.2 Fundamentals of Input-Output analysis

While there exist these days a large selection of variations, and additions to the Leontief input-output model, the basics remain the same. For this thesis, the main piece of theoretical literature used will be Ronald E. Miller and Peter D. Blair 2009's "Input-Output Analysis; Foundations and Extensions", a book that provides a detailed, and rigorous overview of the basics, as well as more advanced methods based upon the input-output framework. This work will be complemented with additional theoretical papers, where deemed necessary.

The fundamental Leontief input-output model is based on collected economic data for a geographical area, which could be a country, region, etc. It can at this point already be established, that in the case of this thesis, the analysis will be based on the national Swedish input-output table. The input-output model captures the flows of products from each sector of the economy to other sectors, as well as consumers, the former called the "interindustry transactions table", and the latter denoted as "final demand" (Miller & Blair, 2009, p. 2). The below table is a highly simplified, condensed representation of the mechanics of an input-output table. The shaded area represents the flows between the industries of this simplified 3-sector economy. Its rows visualize how the output of each particular industry sector is allocated to be used by other industries as inputs to their production process. Its columns in turn illustrate the composition of inputs needed for each industry to create their product (their output). The additional columns under "Final Demand" show the flows from each industry to the final users: private consumers, governments, investments and finally, exports. The added rows labeled "Value Added" show contributions to production beyond those from industries, such as labor, taxes and imports.

***Figure 1**: Representation of the Input-Output Framework*

| | | PRODUCTS USED BY INDUSTRIES FOR PRODUCTION | | | FINAL DEMAND | | | |
|---|---|---|---|---|---|---|---|---|
| | | Manufacturing | Transportation | Agriculture | Private consumption | Investment | Government spending | Net Exports |
| PRODUCTS | Manufacturing | | | | | | | |
| | Transportation | | | | | | | |
| | Agriculture | | | | | | | |
| VALUE ADDED | Employees | Compensation of employees | | | Gross Domestic Product | | | |
| | Business Owners/Capital | Income from profits and capital consumption | | | | | | |
| | Government | Indirect business taxes | | | | | | |

Author's own creation, but inspired by (Miller & Blair, 2009, p. 3, figure 1.1)

Finally, as can be observed in the lower right corner, the total value of final demand adds up to the GDP of the economy, where:

$$\text{GDP} = \text{C} + \text{I} + \text{G} + (\text{X} - \text{M})$$ [4]

Let us now take the Swedish air transport sector as an example. It is firstly clear that it would provide its product, air transport services to several industries within the Swedish economy. It can for instance be theorized that manufacturing companies, travel agencies, and warehousing firms would need to rely extensively on-air transport services for their daily operations. Meanwhile, the air transport services sector would rely on inputs from for instance aircraft manufacturers, logistics support companies, as well as its own product, "air transport services" for its daily operations. In terms of final demand, consumers would use air transport services for private travel, and government agencies for business trips. Furthermore, globalized Swedish air transport service providers may also operate abroad, and therefore contribute also to exports.

All in all, input-output analysis, due to its nature of showing the complex interconnections between the main actors of an economy, is a framework that presents the user with a multitude of opportunities. In the next chapter, the data sources used for the analysis will be presented, and the chapter following it will cover in depth the methodologies used.

[4] Table was set up in this manner for representative purposes. More commonly, in Input-Output tables, while exports are indeed accounted for under final demand, imports are added under the value-added section.

## 3.2 Data Materials

### 3.2.1 The Swedish Input-Output table

The 2017, domestic, industry by industry, Swedish input-output table is the main dataset used for the analysis. It is the most up-to-date input-output table available on the Swedish economy. The table is symmetric, as in that it spans 65 by 65 sectors. The sectors are defined based on ESA 2010 (European System of Accounts), which also match the Swedish SNI 2007 classification. When compared to other advanced economies, this level of granularity is above average, and the age of the table is considered low. In accordance with the example table provided in the previous section, the Swedish input-output table also includes interindustry flows, final demand, and value added. The units of the table are millions (M) of SEK (Swedish Kronor), and the amounts are given in basic prices. One sector, "Activities of extra territorial organizations and bodies" was omitted from the analysis as it does not contribute to output, nor does it provide any value added to the economy.

### 3.2.2 The Swedish Air Transport Sector

The Swedish air transport sector is represented in the table under code "H51" and is labeled "Air transport". No data on subcategories is provided, as is the norm with most input-output tables. That being said, using resources outside of the input-output table, it is possible to break down which subsectors H51 actually consists of. The below able illustrates precisely these, and their production value.

***Table 1**: Breakdown of the Swedish Air Transport Sector*

| | H51: Air transport | | | |
|---|---|---|---|---|
| | 51.101: Scheduled passenger air transport | 51.102: Non-scheduled passenger air transport | 51.211: Scheduled freight transport | 51.212+51.220: Non-scheduled freight air transport and space transport[5] |
| **Production value** (2017, SEK, M) | 22 805 | 4 183 | 1 988 | 108 |

Source: Statistics Sweden, 2018

[5] Note that non-scheduled freight and space transport are combined due to national security concerns, in order not to reveal too much about the latter.

It is clear from the table that passenger air transport takes a lion's share of the Swedish air transport sector, with cargo trailing behind. More discussion about these figures' implications for methodology will be covered at a later stage.

### 3.2.3 Additional use data

An additional breakdown of intermediate transactions and final demand for the air transport sector was requested and received from (Statistics Sweden, 2020d). The dataset allows us to obtain detailed information on the usage of air transport services by each and every industry of the economy. Additionally, more detailed data is also provided on final demand for air transport services. The main value of this dataset is that it contains information to enable to calculation of the ratio between cargo air transport and passenger air transport both for intermediate and final demand. It can here be noted that while data in the use table is provided not in basic prices but rather in purchasers' prices, in this particular scenario it is the ratio and not value of passenger air transport services to cargo air transport services that is to be used. In other words, the dataset is simply used to calculate what percentage of the use of air transport services by each sector can be attributed to passenger transport, and what percentage to cargo transport.

### 3.2.4 Employment data

Employment data was also collected from a database operated by Statistics Sweden, as this data matches the sectors defined in the input-output table (Statistics Sweden, 2020f). On the whole, no significant hurdles were encountered for the majority of sectors, however, a few times only aggregate data was retrieved, which combined one or more sectors into one bloc.[6] Aggregated figures such as these were broken down into estimates for each sector using the 2017 Swedish use table, which is a building block for the input-output table and contains the number of hours worked in each sector (Statistics Sweden, 2020b). For sectors that had combined employment numbers, the ratio between them in terms of number of hours worked was calculated using the use table, and this ratio was then assumed to be the same for number of employees. In such a way, it was possible to estimate number of employees for these (few) sectors as well. Finally, employment data on the public sector was collected from a separate database, also provided by Statistics Sweden (Statistics Sweden, 2020e).

[6] It may be worth noting that this is a very common issue faced by economists wishing to perform input-output analysis.

### 3.2.5 Demand projections for air transport services

A projection by the Swedish Transport Agency of a 74% decrease in volume of passenger air traffic for the year of 2020 was used as a proxy to attain the simulated a drop in both final and intermediate demand for passenger air transport services (Swedish Transport Agency (Transportstyrelsen), 2020). The forecast is based upon a combination of known relationships between GDP and passenger air traffic, as well as a number of assumptions. It is firstly assumed that under 2021, mass vaccinations on a global scale will start. Furthermore, it is assumed that there will not be another powerful wave of the virus sweeping over Europe, but rather more local outbreaks. It is also assumed lastly that as a result of the pandemic competition will decrease, and that alternative ways of meeting will replace in-person trips. Note that the report was released in October of 2020. As of December 2020, it can be evaluated that these assumptions still largely hold, perhaps with the exception of an assumption of new “wave” of the pandemic. Given this, it may be expected that the drop in passenger demand for air transport services for the year of 2020 is slightly underestimated, however since disruptive outbreaks were still factored into the equation, the figure provided is still expected to be fairly accurate.

Additionally, data on cargo traffic had to be collected (The Swedish Transport Agency (2018)[7], as the Swedish Transport Office only provides a forecast for passenger air traffic, but not cargo traffic. While this is in itself may seem problematic, the circumstances lead to the assessment that a passenger traffic only forecast is still suitable for the simulation. Based on the monthly data provided by the Swedish Transport Office on air cargo transport, it is found that cargo volumes remain more or less unchanged compared to previous years. In the first 1-2 months of the pandemic, volumes did see a limited dip but over time they had returned to normal levels, and in some cases even surpassed volumes of the previous year. This isn't a surprising finding. It is echoed for instance by Brida et al. (2016) who found that air cargo transport was not impacted by the September 11 attacks. Li (2020) also came to similar conclusions within the context of the pandemic.

[7] Please see Annex A for complete air cargo statistics.

Hence, the drop in demand for air transport as a whole is exclusively driven by the drop in demand for passenger services. Furthermore, since detailed use data concerning the usage of air passenger versus air cargo services was obtained as detailed in section 3.2.2, it is possible to accurately simulate the drop in demand for Swedish air transport services overall. A further point to strengthen this argument, and to explain why this domestic forecast was favored over non-domestic ones (even if these might be more exhaustive) is the advantage, that the classification of the air transport sector as a whole, and its subsectors relating to passenger air, and cargo air respectively exactly matches the classification in the input-output table. Meanwhile international classifications may vary slightly from country to country, and that in turn may have a significant effect on the findings. That being said, once a figure of the overall drop in demand is obtained, it will be compared to a EUROCONTROL, and found not be radically different.

Finally, worth noting is also the size of the Swedish air cargo transport. As seen in table 1 of section 3.2.2, it makes up only a small part of the total production value of the Swedish air transport sector, which computes to be around 7%. It is clear, that Swedish air cargo is not very big, in fact, much of the air cargo market in Sweden is dominated by foreign companies.

### 3.2.6 The composition of final demand for Swedish air transport services

In this section, the composition of final demand for Swedish air transport services is presented. It is useful and necessary to do so, as final demand will play an important role in several of the methods used during analysis. Final demand for Swedish air transport services for the year of 2017 totaled approximately SEK 18 billion. The below figure illustrates its various components, and their size.

***Figure 2:*** *Composition of Final Demand for Swedish Air Transport Services*

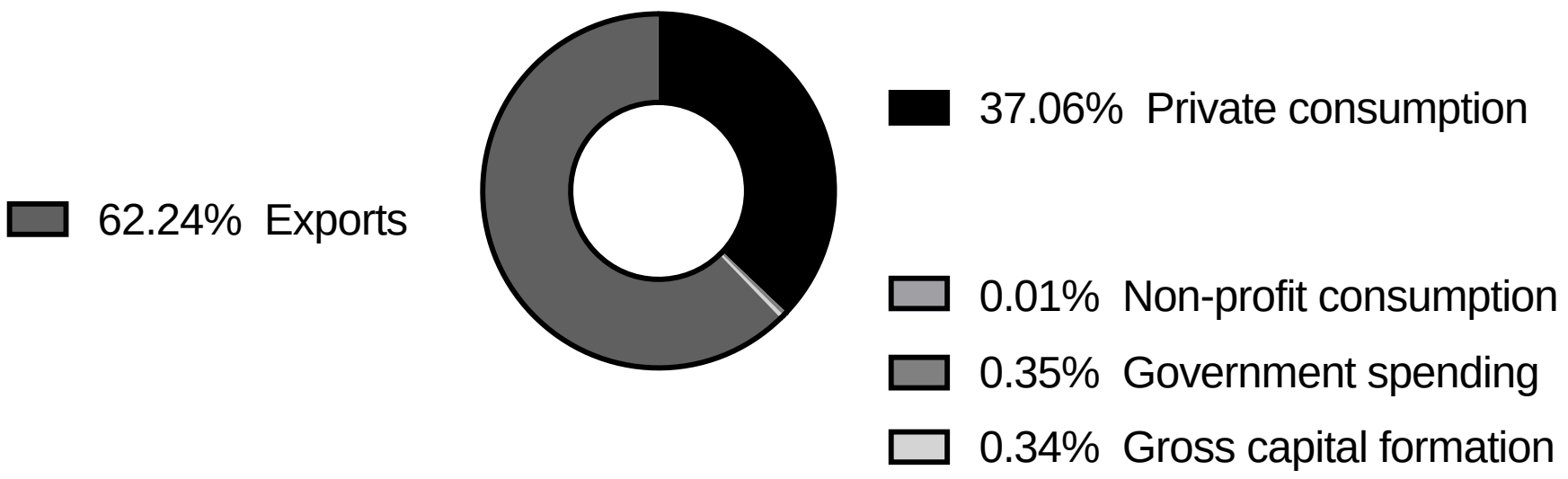

Source: Statistics Sweden, 2020a

As can be observed, exports and final consumption expenditure by households combined make up the great majority, 99.3% of final demand for air transport. Furthermore, based on use data acquired from Statistics Sweden[8], it is also known that 100% of household consumption and 93% of total exports can be attributed to passenger air transport, versus cargo air transport. Since no such information was obtained for non-profit organizations and government demand, these were assumed to solely demand passenger transport services.

[8] See section 3.2.3

## 3.3 Classical Input-Output Analysis

The following chapter discusses the methodological framework used for performing classical, Leontief input-output analysis of the data collected and described in the previous sections. It should at this point also be noted that the chief software of choice for this thesis will be MATLAB, complemented by Microsoft Excel.

### 3.3.1 Model construction with regards to consumer demand

Before diving into the methodological explanations however, it is essential to make a distinction in regard to the way the input-output table was approached. The classical input-output framework treats the components of final demand as exogenous. However, these days an alternative approach also exists which is to treat parts of final demand as endogenous. For instance, most commonly, private consumption is treated as so. This is done in order to take into account the changes in household income, and therefore consumer spending that a final demand change can cause. It is a way to include a bit of Keynesian theory into the analysis (EUROSTAT, 2008), which certainly has its pros in terms of perhaps correcting for underestimations caused by the omission of this perspective. That being said this time, the classical model was chosen with consumer demand as exogenous as the time resources needed to ensure compatibility with the methods used was deemed to be unrealistic. The implications of this choice will be discussed at a later stage in this thesis.

### 3.3.2 The input-output layout

Now that the choice of model has been completely outlined, the structure of the full input-output table, on which all the methods used will be based can be established, and at the same time, useful notation can also be introduced.

***Table 2:*** *The Input-Output Layout*

| INDUSTRY SECTORS | 1 | 2 | FINAL DEMAND (f): Private consumption | Investments | Government spending | Exports | **Total output (x)** |
|---|---|---|---|---|---|---|---|
| 1 | $z_{11}$ | $z_{12}$ | $c_1$ | $i_1$ | $g_1$ | $p_1$ | $x_1$ |
| 2 | $z_{21}$ | $z_{22}$ | $c_2$ | $i_2$ | $g_2$ | $p_2$ | $x_2$ |
| Imports | $m_1$ | $m_2$ | $m_C$ | $m_I$ | $m_G$ | $m_P$ | M |
| Value Added | $va_1$ | $va_2$ | $va_C$ | $va_I$ | $va_G$ | $va_P$ | VA |
| **Total Uses (x')** [9] | $x_1$ | $x_2$ | C | I | G | E | **X** |
| | | | | | | | |
| Income | $s_1$ | $s_2$ | | | | | |
| Employment | $e_1$ | $e_2$ | | | | | |
| Gross fixed capital formation | $cf_1$ | $cf_2$ | | | | | |

Source: Author's own creation, but inspired by Miller & Blair, 2009, p. 14, Table 2.2)

As can be observed, this time, a simple two sector economy was used. Imports are denoted with **M**, and value added with **VA**. Final demand (**f**) is divided into the same four categories as in Figure 1, namely private consumption (**C**), investments (**I**), government spending (**G**) and finally exports (**P**).

Next, as can be seen, interindustry flows between sectors 1 and 2 are shown with $\mathbf{z}_{nn}$. Furthermore, "Total Uses" show what goes into the production for sectors 1 and 2 ($\mathbf{x}_1$ and $\mathbf{x}_2$ respectively), as well as how the different parts of final demand are composed. For instance, private consumption demand is composed of purchases from sectors 1 and 2 ($\mathbf{c}_1$ and $\mathbf{c}_2$ respectively), imports ($\mathbf{m}_C$) and the various components of value added ($\mathbf{va}_C$), and can therefore be expressed as follows:

$$C = c_1 + c_2 + m_C + va_C$$

[9] Note that horizontal vectors will always be denoted with the symbol $'$.

"Total outputs" in turn account for the total gross output of each component. For instance, sector 1 outputs its product to meet interindustry demand ($z_{11}$ and $z_{12}$), private consumption, investments, government spending and exports. Therefore:

$$x_1 = z_{11} + z_{12} + c_1 + i_1 + g_1 + p_1$$

All in all, however, total uses and total outputs are just different ways of summing up the elements in this symmetric table, and therefore:

$$X = x_1 + x_2 + C + I + G + E = x_1 + x_2 + M + VA$$

### 3.3.3 Additional indicators

Lastly, it can also be observed that three satellite (additional) accounts have been included into the table. These are income, employment and gross fixed capital formation. Satellite accounts are accounts that are added into the table in order to provide further insights when the analysis is performed. The term satellite account is perhaps not necessarily the most accurate one, as satellite accounts more commonly refer to external data brought into the table. That being said, it is not clear if this always holds true, therefore the author decided to use this definition both for strictly external and also internal data, as it is argued that the term provides a salient illustration of the purpose of these accounts. Notwithstanding this however, below, a clarification is made in regard to the origin of each account.

The income satellite account is taken simply from the input-output table. It is present as a part of value-added under the name "Compensation of employees". The employment satellite account, as detailed in in section 3.2.4 is collected externally. Gross fixed capital formation is a part of final demand, and more specifically investments. Finally, note that value added will also be used in the analysis in a similar way as the above-mentioned ones in order to provide additional insights, but it was decided not to include it as a satellite account into the table, as it already is part of it, under the same name.

### 3.3.4 The interindustry flows matrix

Input-output analysis uses matrix algebra as its core. The first matrix that was taken from the table is the interindustry flows (intermediate purchases) matrix, **Z**, which can be visualized as follows:

$$
\begin{array}{cc}
 & \text{Sectors} \\
\text{Sectors} &
\begin{array}{c|ccccc}
 & 1 & \cdots & j & \cdots & n \\
1 & z_{11} & \cdots & z_{1j} & \cdots & z_{1n} \\
\vdots & \vdots & & \vdots & & \\
i & z_{i1} & \cdots & z_{ij} & \cdots & z_{in} \\
\vdots & \vdots & & \vdots & & \\
n & z_{n1} & \cdots & z_{nj} & \cdots & z_{nn}
\end{array}
\end{array}
$$

As can be seen, "j" represents the columns, and "i" the rows both adding up to "n" since the table is symmetric, in other words, contains the same number of sectors both in its columns and rows. From this follows that the Z matrix can be described in the following way:

$$\mathbf{Z} = \begin{bmatrix} z_{11} & \cdots & z_{1n} \\ \vdots & \ddots & \vdots \\ z_{n1} & \cdots & z_{nn} \end{bmatrix}$$

### 3.3.5 The satellite accounts

Now that the interindustry flows matrix has been obtained, the relevant satellite accounts for value added, income, employment and capital formation can be extracted. They are denoted by the following expressions[10]:

$$\mathbf{va}' = [va_1 \quad \cdots \quad va_n]$$

$$\mathbf{s}' = [s_1 \quad \cdots \quad s_n]$$

$$\mathbf{e}' = [s_1 \quad \cdots \quad s_n]$$

$$\mathbf{cf}' = [cf_1 \quad \cdots \quad cf_n]$$

[10] Note that $'$ symbolizes a reference to a row vector.

### 3.3.7 The technical coefficients matrix

One of the most important assumptions of the input-output model is the relationship between the interindustry flows and the output of the given sector. The relationship is given by the following expression:

$$a_{ij} = \frac{z_{ij}}{x_j}$$

where $a_{ij}$ is a technical coefficient, $z_{ij}$ is the interindustry flow between two sectors, and $x_j$ is the output of the sector receiving the input. What this tells us is essentially that the ratio, $a_{ij}$, between the interindustry flows relating to a specific sector, and the same sector's output, is fixed. For instance, $z_{ij}$ may be the value of fertilizer bought by crop producers, in which case $x_j$ would be the value of crop production over the same period of time. The technical coefficient $a_{ij}$ therefore represents the SEK worth of input from a given sector per the SEK worth of output of the sector receiving this input. In input-output analysis, this relationship is assumed to be completely static. Hence, economies of scale are not taken into account. The relationship between a sector's output and its input does not change depending on the value of inputs it receives and the value of outputs it produces. This assumption, necessary for the functioning of the model, is also seen as a weakness of it, which will be discussed at a later stage in this thesis.

In matrix notation, when considering all the elements of the interindustry flows matrix, the same relationship can be expressed as below, and is called the technical coefficients matrix.

$$\mathbf{A} = \mathbf{Z}\hat{\mathbf{x}}^{-1}$$

where $\hat{\mathbf{x}}$ is a diagonal matrix with the elements of the vector along the diagonal of the matrix, all other elements being 0 as seen below:

$$\hat{\mathbf{x}} = \begin{bmatrix} x_1 & \cdots & 0 \\ \vdots & \ddots & \vdots \\ 0 & \cdots & x_n \end{bmatrix}$$

From this follows that $\hat{\mathbf{x}}^{-1}$ can be defined as:

$$\hat{\mathbf{x}}^{-1} = \begin{bmatrix} \frac{1}{x_1} & \cdots & 0 \\ \vdots & \ddots & \vdots \\ 0 & \cdots & \frac{1}{x_n} \end{bmatrix}$$

The technical coefficients matrix **A** is a significant component of input-output analysis. Its columns tell us the input-recipes for each sector. In other words, they tell us how much of the products of all the sectors are used by a specific sector to produce its product. In addition to the technical coefficients matrix, technical coefficients for all satellite accounts were also created. For instance, taking value added as an example, its technical coefficients (row) vector can be defined as:

$$\mathbf{va'}_{\mathrm{c}} = \mathbf{va'}\,\hat{\mathbf{x}}^{-1}$$

where the subscript "c" stands for "coefficient".

### 3.3.8 The final demand vector

Next, a final demand vector, summing up all its components (C, I, G, E) was created as such:

$$\mathbf{f} = \begin{bmatrix} f_1 \\ \vdots \\ f_n \end{bmatrix}$$

where "$f_n$ " represents the total final demand for the product of every industry sector.

### 3.3.9 The output vector

Furthermore, the output vector was taken from the table, which contains the outputs of all the industry sectors, and can be described as follows:

$$\mathbf{x} = \begin{bmatrix} x_1 \\ \vdots \\ x_n \end{bmatrix}$$

### 3.3.10 The Leontief inverse

The connection between interindustry flows, final demand and output is fairly straightforward if keeping in mind Table 2. Here, we can see that the output of each sector is the sum of the interindustry flows and the final demand. Since we also know the relationship between the interindustry flows and the output ($z_{ij} = a_{ij}x_j$), it is now possible to fully show the relationship in matrix form:

$$\mathbf{x} = \mathbf{A}\mathbf{x} + \mathbf{f}$$

If we utilize an $n \cdot n$ identity matrix $\mathbf{I}$ with ones along the diagonal and zeroes elsewhere as given by $\mathbf{I} = \begin{bmatrix} 1 & \cdots & 0 \\ \vdots & \ddots & \vdots \\ 0 & \cdots & 1 \end{bmatrix}$, we are able to rewrite the above expression in the following way:

$$(\mathbf{I} - \mathbf{A})\mathbf{x} = \mathbf{f}$$

Then, using the definition of an inverse for a matrix, it is possible to arrive to the following form:

$$\mathbf{x} = (\mathbf{I} - \mathbf{A})^{-1}\mathbf{f} = \mathbf{L}\mathbf{f}$$

where $(\mathbf{I} - \mathbf{A})^{-1} = \mathbf{L}$ is dubbed the "Leontief inverse". It is clear from this expression that the gross output of each sector is dependent on the final demand for its product.

### 3.3.11 Direct, indirect and induced effects

The above expression also captures one of the fundamental characteristics of input-output analysis, which is that it can be successfully used to measure different "levels" of the impact of a change in final demand. For instance, if the final demand for cars goes up by a given number of units, to meet this extra demand, the car manufacturing sector will need to increase its production by at least the same number of units. This is called the direct effect. However, as the car manufacturing sector increases its production, it also requires more inputs to its production. These inputs, as we know are partly inputs from other industries (in this case for example car parts manufacturers), and partly the sector's own product as defined by the technical coefficients matrix. This therefore causes not only the car manufacturing industry to increase its production, but also other industries which's inputs it uses, which in turn require more inputs from other industries to scale up their production.

The result can amply be described as a chain reaction, and therefore the added production on top of the direct effect is called the indirect effect. In this thesis, direct and indirect effects will be captured by the methods used, and as detailed in section 3.3.1, induced effects will not be measured.

### 3.3.12 Multipliers

The column sums in the Leontief inverse signify the output multiplier for each industry. The multiplier shows essentially the total units of output gained from a one unit increase in final demand for that industry's product. Multipliers are a valuable tool of basic input-output analysis because they give us a hint of the importance of an industry sector for the economy. A sector with a high multiplier may be seen as one worthy of for instance government spending, as it gives the highest "bang for the buck" in terms of the achieved output. In matrix form, the output multiplier for a given sector $j$ can be given by the following expression:

$$m_j = \sum_{i=1}^{n} l_{ij}$$

where $l_{ij}$ represents the column vector for the sector of interest.

The multipliers for all the other factors, namely, value added, income, employment and capital formation can also be easily found once the Leontief inverse has already been obtained by multiplying the row vector of technical coefficients for the satellite account of choice by the Leontief inverse. Firstly, as we already know the technical coefficients for satellite account $h$[11] is found in the following way:

$$h'_c = h'\hat{x}^{-1}$$

Then the multipliers can be found in the following way:

$$\mathbf{m}(\mathrm{h}) = \mathbf{h'}_c\mathbf{L}$$

Note that these multipliers essentially are just a conversion of the already known output multipliers into different forms to reflect income, value added, employment and capital formation.

[11] Note that *h* is simply a general term, it can be replaced with *va*, *s*, *e* and *cf*.

## 3.4 Final Demand Scenarios

### 3.4.1 Why use two demand scenarios?

Up until now, when dealing with final demand, a simple percentage drop in final demand for Swedish air passenger services was assumed. However, if we view final demand changes in the context of an economic shock such as the COVID-19 pandemic, it is worth considering how closely this might align with reality. It is after all possible that while consumer demand for air transport services is sharply reduced, to some extent, it could be re-routed towards other sectors instead as a result of the drastic change in our lifestyles. For this reason, two demand scenarios will be considered throughout the analysis.

### 3.4.2 Final demand: Scenario 1

This is the simple scenario. It is assumed that in these times of uncertainty, households that no longer demand Swedish air transport services choose to put 100% of their excess disposable income that was freed up into savings. The below table illustrates the changes made to the components of final demand for air transport services. Recall, that the figure of 74% relates to a drop in passenger air demand only, and not cargo demand. For some components, we are in the knowledge of the ratio between passenger air traffic demand, and freight air traffic demand. In other cases however we are not, and therefore we must make assumptions.

***Table 3:*** *Scenario 1: Changes in Final Demand for Swedish Air Transport*

| Component | Ratio: Passenger to Freight | Assumptions | Change |
|---|---|---|---|
| Private consumption | 100/0 | - | -74% |
| Investments | Not known | No change | 0.0% |
| Nonprofit org. demand | Not known | Ratio: 100/0 | -74% |
| Government demand | Not known | Ratio: 100/0 | -74% |
| Exports | 93/7 | - | 93%*-74%=-69% |

As seen in the table above, when it comes to private consumption, the ratio between passenger air demand and freight air demand was known. Hence, it is reduced simply by 74%. In terms of investments, no data was available on the ratio, and no change was assumed either. As for government demand and demand from nonprofit organizations, no data was available on the ratio, but it was assumed to be the same as for private consumers, 100/0. This is arguably the weakest assumption, as it is possible that there is some government demand for cargo traffic. However, seeing how in general, demand for passenger air takes a lion's share, it is argued that it is not an unreasonable assumption to make. Lastly, the ratio for exports was known, 93 to 7 in favor of demand for passenger air services, which gives a total drop of 69% in exports of air transport services

### 3.4.3: Final demand: Scenario 2

At the same time, another possibility is that consumers decide to spend some of the excess disposable income they now possess after reducing their demand for air transport services. As a result of the lifestyle changes forced upon us by the pandemic, certain trends can already be observed in consumer spending patterns. Real winners of the pandemic, such as telecommunications company Zoom, demonstrate that such services are in high demand. Additionally, it is also clear that the pandemic has induced an increase in health consciousness, and that demand for pharma products has gone up. It may also be theorized that local travel (especially in a country like Sweden where throughout the pandemic, restrictions limiting free movement have been lax) has increased in popularity. Lastly, it can also be argued that as Swedes are largely unable to travel abroad, they may choose to rely more heavily on postal services, to keep in touch with friends and family. Of course, it is absolutely necessary to be aware that these assumptions are only assumptions, and nothing more. The author was unable to get hold of any data that may be of help to evaluate exactly how consumers have chosen to spend what they would normally had spent on air travel. It is clear therefore that this method has its limitations, that will be discussed more in depth at a later point in this thesis. That being said, it is deemed to still be of value as it can provide another, alternative view than is provided by scenario 1.

The below table illustrates the assumptions made. Note that the reallocation only concerns final consumption, which to the largest part is private consumption, and to a very small extent also contains demand from non-profit organizations and governments. Gross capital formation is not expected to see a reallocation so quickly. Lastly, the drop in demand in exports is not reallocated either. This is because it is assumed that consumers of foreign countries, as they have reduced their demand for Swedish air transport services (such as Swedish organized tourist trips), spend their excess disposable income domestically.[12] In practice, for final consumption, it was assumed that 50% of the excess disposable income still goes to savings, in these times of uncertainty. This demand therefore "disappears" from the final demand vector. The other 50% however, is expected to be reallocated as detailed by the table below. Note, that the changes in Table 3 are still applied, Table 4 only describes the additional modifications to final demand for sectors, on top of those already simulated for Swedish air transport services.

***Table 4:*** *Reallocation of Excess Consumer Income*

| **Sector** | **Share of excess disposable income gained** |
|---|---|
| Warehousing and support activities for transportation | 20% |
| Telecommunications | 20% |
| Manufacture of basic pharmaceutical products and pharmaceutical preparations | 10% |
| Retail trade, except of motor vehicles and motorcycles | 10% |
| Land transport and transport via pipelines | 10% |
| Postal and courier activities | 10% |
| Human health activities | 10% |
| Manufacture of food products, beverages and tobacco products | 5% |
| Water transport | 5% |

As can be observed firstly, warehousing and support activities for transportation receive a share of the excess disposable income as people spend more on online purchases rather than travelling; this would also explain the increase in demand for retail trade. Furthermore, telecommunications also receive some "re-routed" demand as people need a way to keep in touch with family, friends and colleagues rather than travelling and meeting them in person. This also explains a boost to postal and courier activities. As health consciousness increases, human health activities and pharma also receive some of the excess income. Local travel (land transport and water transport) also receives a slice of excess demand. Lastly, as consumers stay home more, instead of spending money on food experiences abroad, they may spend a bit more at home, hence the increase in demand for food products, beverages and tobacco products.

[12] It may be possible to argue that some foreign buyers, instead of spending on Swedish air transport services, spend on other Swedish products instead. However, this effect is seen as minimal, as Sweden isn't a large exporter of any products in the sectors of table 4, and therefore it is much more likely that foreign buyers spend excess disposable income locally, or on the products of more significant exporters.

## 3.5 Inoperability Analysis

The above detailed frameworks, linked to fundamental, Leontief type input-output analysis can all be thought of as serving primarily to answer the first research goal; finding the economic linkages of the Swedish air transport sector. They will be used to establish the role it plays, and the importance it has for other sectors, and the economy as a whole. Meanwhile, inoperability analysis, and the hypothetical extraction method which will be detailed in section 3.6, can be thought of as methods to fulfil the second research goal; estimating the effects of the pandemic on air transport, and the spin-off effects for the economy.

### 3.5.1 The fundamentals of inoperability analysis

Inoperability analysis, a method built on the input-output framework has been developed to assess vulnerabilities of an economy in the context of a disaster (Percoco, 2011). In other words, it is essentially used in a way as to simulate the effect of a crisis on demand, and due to the nature of input-output analysis, also allows the user to capture which other sectors, apart from the one(s) seeing the direct drop in demand are the most affected by the event. The method can be traced back to Haimes and Jiang (2001) however Santos and Haimes (2004) and Santos (2006) expanded on the model to be useful for investigating the economic losses caused by crises, such as the September 11 attacks on the US air transport sector. While naturally different in many ways, it is clear that both the COVID-19 pandemic and the September 11 attacks are crises with a detrimental impact air traffic on a global scale. Hence, it was deemed appropriate to use inoperability analysis for this thesis.

Inoperability analysis, as its name also implies, is centered around measuring inoperability. Inoperability can be defined as "*the level of the system's dysfunction, expressed as a percentage of the system's intended production level.*" (Santos, 2006) In order to capture the direct and indirect impacts of a crisis, the following model structure is used:

$$\mathbf{q} = \mathbf{A}^{*}\mathbf{q} + \mathbf{c}^{*}$$

In this model, firstly, $\mathbf{c}^{*}$is seen as a demand-side perturbation vector. While Santos and Haimes (2004) consistently express demand side disruptions in this way, for the purpose of staying consistent with earlier notations of final demand, it was deemed appropriate to use **f**. Therefore, the above model can be expressed in the following alternative way:

$$\mathbf{q} = \mathbf{A}^{*}\mathbf{q} + \mathbf{f}^{*}$$

where $\mathbf{f}^{*}$ is given by the following expression:

$$\mathbf{f}^{*} = \hat{\mathbf{x}}_{\mathrm{p}}^{-1}(\mathbf{f}^{\mathrm{p}} - \mathbf{f}^{\mathrm{r}})$$

As can be seen it essentially entails the multiplication of the inverse of the output vector along the diagonal of a matrix by the planned final demand minus the reduced final demand, caused by the given crisis. The *i*th element of $\mathbf{f}^{*}$then represents the disruption in final demand for sector *i*. This element can be expressed in the following way:

$$f_i{}^{*} = \frac{f_i{}^{p} - f_i{}^{r}}{x_i{}^{p}}$$

Meanwhile, $\mathbf{A}^{*}$ is the "interdependency matrix", which demonstrates the degree to which each industry sector is connected to all others. It is related to Leontief technical coefficient matrix **A** and is defined as follows:

$$\mathbf{A}^{*} = \left[\left(\hat{\mathbf{x}}^{-1}\right)(\mathbf{A})\right]\left(\hat{\mathbf{x}}\right)$$

As can be observed, it is essentially the inverse of the output vector along the diagonal of a matrix multiplied by the technical coefficient matrix, and in turn multiplied by the output vector along the diagonal of a matrix. An individual element of $\mathbf{A}^{*}$ is given by the following expression:

$$a_{ij}{}^{*} = a_{ij}\left(\frac{x_j}{x_i}\right)$$

where $a_{ij}$ is a technical coefficient, $x_j$ is the planned production for industry *j* and $x_i$ is the planned production for industry *i*.

Furthermore, $\mathbf{q}$ is the inoperability vector, the percentage loss in output caused by the disruption in final demand. It is given by the following expression:

$$\mathbf{q} = \left(\hat{\mathbf{x}}^{-1}\right)(\mathbf{x} - \mathbf{x}^{\mathrm{r}})$$

In other words, it is the difference between the planned output and the reduced output normalized to the level of output. An individual element of vector $\mathbf{q}$ can therefore be found in the following way:

$$q_i = \frac{x_i - {x_i}^r}{x_i}$$

Finally, the total economic loss of the economy is then given by the inoperability suffered by all sectors times their original, or "planned" output.

$$\boldsymbol{x}_{loss} = \left(\hat{\boldsymbol{q}}\right)(\boldsymbol{x})$$

### 3.5.2 Critique of the inoperability model

At the same time, it has been argued by Dietzenbacher and Miller (2015) that inoperability analysis does not need to use different notation and frameworks than classical input-output analysis. It can be argued that $c^* = f^*$ is just $\triangle f$ normalized to the original output level. Firstly, we take the basic equation from the Leontief input-output model:

$$\triangle \mathbf{x} = \mathbf{L}(\triangle \mathbf{f})$$

It is clear that $\triangle x = (x - x^r)$, and therefore:

$$\mathbf{q} = \hat{\mathbf{x}}^{-1}(\mathbf{x} - \mathbf{x}^{\mathrm{r}}) = \hat{\mathbf{x}}^{-1}(\triangle x)$$

If we then substitute in the Leontief expression for change in output, we arrive to the following:

$$\mathbf{q} = \hat{\mathbf{x}}^{-1}(\mathbf{x} - \mathbf{x}^{\mathrm{r}}) = \hat{\mathbf{x}}^{-1}\mathbf{L}(\triangle \mathbf{f})$$

As can be observed, normalized output loss, or inoperability can be found also using the simple Leontief model, through normalizing the standard equation to the output level.

Lastly, in the knowledge of this, the total output loss, which is by Santos and Haimes (2004) given as $x_{loss} = \left(\hat{q}\right)(x)$ is also simply just equal to $\triangle x$, therefore we can express the relationship in the following way:

$$x_{loss} = \left(\hat{q}\right)(x) = \triangle x = \mathbf{L}(\triangle \mathbf{f})$$

### 3.5.3 Further applications of the model

It is also possible to translate the change in output to changes in value added, income, employment and capital formation using the input coefficient vectors that were already attained in the classical input-output analysis. We know that the input coefficients for the satellite accounts are given by the following expression:

$$\mathbf{h}'_c = \mathbf{h}'\hat{\mathbf{x}}^{-1}$$

In words, the product of the row vector containing the satellite account and the diagonal matrix of the inverse of the output vector. Then the change in one of the satellite accounts is given by the product of the vector of input coefficients for the given satellite account along the diagonal of a matrix, and the change in output as expressed below:

$$\triangle \mathbf{h} = \left(\hat{\mathbf{h}}_c\right)(\triangle \mathbf{x})$$

## 3.6 Partial Hypothetical Extraction

### 3.6.1 The classical hypothetical extraction model

The second, input-output based model that was chosen for the purpose of this thesis is the hypothetical extraction model. As its name also implies, it is fundamentally based around the removal of certain defined parts from an input-output table in order to understand their importance for the economy. The method has its roots in the late 60s, first written about by the likes of Miller (1969), Paelinck et al. (1965) and Strassert (1968). It too uses the classical input-output framework as its base:

$$\mathbf{x} = \mathbf{Zs} + \mathbf{f}$$

where $\mathbf{x}$ is the output vector, $\mathbf{Z}$ is the interindustry flows matrix, $\mathbf{f}$ is the final demand vector, and $\mathbf{s}$ is the summation vector consisting of ones. Again, as before, the input coefficients matrix (also called technical coefficients matrix) is given by the expression below:

$$\mathbf{A} = \mathbf{Z}\hat{\mathbf{x}}^{-1}$$

Furthermore, as before, the Leontief inverse is given by:

$$\mathbf{L} = (\mathbf{I} - \mathbf{A})^{-1}$$

And therefore, output is given by:

$$\mathbf{x} = \mathbf{Lf}$$

Hypothetical extraction essentially entails the removal of the row and column for a specific industry from the direct requirements matrix **A**. Once that is done, an alternative direct requirements matrix $\overline{\mathbf{A}}$ is obtained[13]. Furthermore, in most cases, the final demand for this industry's product is also removed from the final demand vector, giving us the new final demand vector $\overline{\mathbf{f}}$. For instance, if the classical hypothetical extraction method was to be used to investigate the dependance of the Swedish economy on its air transport sector, its column and row in the technical coefficients matrix, and its final demand would be removed.

[13] Note that throughout, the overbar will refer to a matrix or vector, post-extraction.

After the extraction is complete, the new output vector is then given by the following expression:

$$\bar{\mathbf{x}} = \left(\mathbf{I} - \bar{\mathbf{A}}\right)^{-1}\bar{\mathbf{f}}$$

The drop in output, or put differently, the difference between the original output vector and the now obtained output vector can as such be obtained as detailed below:

$$\mathbf{s}'(\bar{\mathbf{x}} - \mathbf{x})$$

where $\mathbf{s}'$ is the summation vector of ones.

The fairly apparent advantage of hypothetical extraction compared to other methods is that it allows us to also account for intermediate demand. For instance, one can theorize that in the wake of a crisis such as the COVID-19 pandemic, not only does demand from consumers and governments decrease, but so does also demand from industries normally relying on air transport services.

### 3.6.2 The partial hypothetical extraction method

In the case of the Swedish air transport, it can for instance by hypothesized that some part of the intermediate demand for its services can be attributed to demand for business travel, which is similarly impacted by the pandemic as private consumer travel. Therefore, a simple final demand cut does not capture the full effect of the COVID-19 pandemic on the air transport sector, and therefore underestimates the economic impact. That being said, since the classical hypothetical extraction model only allows for a complete elimination of an industry, doing so with the Swedish air transport sector would only be an exercise of showing the overall importance of the sector for the economy, and would not be an estimate the actual effects of the pandemic. However, Dietzenbacher and Lahr (2013) suggest an expansion to the hypothetical extraction model which allows for a partial extraction. This gives researchers the freedom to remove the desired effect (such as a change in intermediate and final demand) from the table, recalculate and attain the change in output. Firstly, we assume that intermediate deliveries by an industry $\mathbf{k}$ (in this case, the air transport sector) are decreased by α. These deliveries or no longer demanded, or instead met by imports. Therefore, the percentage change in demand is given by the following expression:

$$\alpha \cdot 100\%$$

Since the industry's output is now reduced, so does its demand for inputs. The reduction in its intermediate demand for inputs is proportional to the decrease in output; its input recipe does not change. Therefore, the *k*th column of the technical coefficient matrix **A** remains unchanged as detailed below. For all $i = 1, \dots, n$ we obtain:

$$\overline{a}_{ik} = \frac{\overline{z}_{ik}}{\overline{x}_k} = \frac{(1-\alpha)z_{ik}}{(1-\alpha)x_k} = a_{ik}$$

What does change however, is the *k*th row of **A**. After all, this row contains how much of industry *k*'s product is demanded by all the other industries. Therefore, all its elements, with the exception of the diagonal element of the *k*th row (because this is part of industry *k*'s input recipe) decrease by $\alpha \cdot 100\%$. We can as such obtain the following expression for all $j = 1, \dots, n$:

$$\overline{a}_{kj} = \frac{\overline{z}_{kj}}{\overline{x}_j} = \frac{(1-\alpha)z_{kj}}{(1-\alpha)x_j} = (1-\alpha)a_{kj}$$

In matrix notation, the same can be expressed in the following way:

$$\overline{\mathbf{A}} = \mathbf{A} - \boldsymbol{\alpha}\mathbf{e}_{\mathrm{k}}\mathbf{b}'_{\mathrm{k}}$$

where $\mathbf{e}_k$ stands for the *k*th unit vector with a one in element *k* and zeroes everywhere else and

$$\mathbf{b}'_{\mathbf{k}} = \left(a_{k1}, \dots a_{k,k-1}, 0, a_{k,k+1} \dots, a_{kn}\right)$$

Finally, final demand can be changed freely, depending on the assumptions made. It can be left unchanged, or it can be cut. Just as with the classical hypothetical extraction model, and with the inoperability model, the new output level can now be obtained:

$$\overline{\mathbf{x}} = \left(\mathbf{I} - \overline{\mathbf{A}}\right)^{-1}\overline{\mathbf{f}}$$

And again, the drop in output can be denoted by:

$$\mathbf{s}'(\overline{\mathbf{x}} - \mathbf{x})$$

In addition, as also discussed in regards to the inoperability model, the changes in output can be translated into changes in value added, income, employment and capital formation. We know that the input coefficients for the satellite accounts are given by the following expression:

$$\mathbf{h'}_{\mathrm{c}} = \mathbf{h'}\hat{\mathbf{x}}^{-1}$$

Then the change in one of the satellite accounts is given by the product of the vector of input coefficients for the given satellite account along the diagonal of a matrix, and the change in output.

$$\triangle\, \mathbf{h} = \left(\hat{\mathbf{h}}_{\mathrm{c}}\right)(\triangle\, \mathbf{x})$$

### 3.6.3 Extraction of demand for the Swedish air transport sector

Before progressing, it must be noted that while the above theoretical process used a uniform $\alpha$, Dietzenbacher and Lahr (2013) express that it does not have to be that way if we are in the possession of data regarding the change in intermediate demand for every sector. In this case, as stated in section 3.2.3, statistics on the usage of air transport services by each sector of the Swedish economy is available. $\alpha$, for each sector will depend on the ratio between passenger and cargo air transport within each sector's demand for air transport services. For instance, for the sector H52, "Warehousing and support activities for transportation", 7% of inputs from the air transport sector are attributable to passenger transport, and the remaining 93% to freight transport. We know, furthermore that $\alpha = .74$ which is the reduction in demand for passenger air transport. This builds upon the assumption that each industry sector's demand for Swedish passenger air services is affected in a similar way by the pandemic as was consumer demand. One could for instance hypothesize that most of interindustry demand for passenger air services accounts for business travel, which was also heavily reduced as a result of the pandemic.

Continuing on with the above example, we know that 7% of the given sector's demand for air transport services is attributable to passenger demand. Let this be denoted by *r*. The new input coefficient is therefore given by the following expression:

$$\overline{a}_{kj} = a_{kj} - (a_{kj})(r)(\alpha)$$

After this process is repeated for every industry, an accurate reduction in input coefficients for air transport can be attained. Now that $\overline{\mathbf{A}}$ has been found, what remains is the change in final demand. The change in final demand is treated in the same way as for the inoperability analysis, where two different final demand vectors were created based on the two demand scenarios. Worth mentioning at this point is also that the losses in intermediate demand are not reallocated in a similar way as was done with the two final demand scenarios, because if attempted, assumptions for the reallocations mechanisms of every single sector would have to be made, which is deemed to cause much more harm than benefit. Lastly, as detailed before, the new, reduced output can be found by multiplying the new Leontief inverse with the final demand for Scenario 1 and for Scenario 2 as seen below:

$$\overline{\mathbf{x}} = \left(\mathbf{I} - \overline{\mathbf{A}}\right)^{-1}\overline{\mathbf{f}}$$

## 3.7 The Issue of Aged Input-Output Tables

### 3.7.1 Final demand blowup

The last methodological theme concerns a well-known downside of input-output analysis, the age of input-output tables. As known by this point, the latest Swedish input-output table describes the 2017 Swedish economy. This isn't unusual in itself. Input-output tables take time to set up, in fact, many countries have even larger time lags when it comes to the creation and publishing of input-output tables. That being said, it is undoubtedly the case that a 2019 table would have been the most ideal for the completing of this analysis, after all, it is obvious that between the years 2017 and 2019 the Swedish input-output table has changed somewhat. There are various suggested approaches to solving this issue. Many of them are rather complex and require the use of large-scale surveys or complex calculations and are deemed to be beyond the capabilities of this master's thesis. However, simpler methods are also available. Miller and Blair (2009) write the following on the issue: "*It is (also) apparent that for aggregate kinds of measures, such as total economy-wide output associated with a specific vector of final demand, the error introduced by using an "old" table may not be large. On the other hand, there are other much simpler methods for forecasting total output that are not much worse.*" (Miller & Blair, 2009, p. 310)

One such method is the "final-demand blowup" approach presented by Conway Jr (1975) that essentially entails blowing up the change in output by the same percentage as we expect the final demand to change by. If total output of 2017 is given by $s'x(2017)$[14] and final demands for 2017 and 2019 are given by $s'f(2017)$ and $s'f(2019)$ respectively, then the estimate of output is given by the expression:

$$s'x(2019) = \left(s'x(2017)\right)\left(\frac{s'f(2019)}{s'f(2017)}\right)$$

[14] As always s' indicates the summation vector.

In this particular case, it is clear that we do not have an exact vector for 2019 final demand, however it is possible to estimate it. This can be done by investigating input-output tables from previous years, and also factoring in GDP changes to make the prediction more accurate. For this estimation, firstly, the change in final demand between years 2015-2016, and 2016-2017 is obtained (Statistics Sweden, 2020a). Then these figures are compared to the GDP changes for the same years (Statistics Sweden, 2020c). The ratio between $\triangle\, final\ demand$ and $\triangle\, GDP$ for years 2015-2016 and 2016-2017 is found, and an average ratio is calculated. This ratio, in the knowledge of the GDP changes for years 2017-2018 and 2018-2019 can then be used to estimate final demand changes for the same two periods. The total final demand increase between the years 2017-2019 is as such found to be 9.2%, which is the estimate that will be used for the analysis. Furthermore, it is assumed that all input coefficients remain unchanged. This assumption is supported by Carter (1970) who compared the output vector needed to satisfy a set of finals demands for years 1939/47/58 and 61. She concluded that the coefficients barely changed.

Given the above, the resulting output changes attained by inoperability analysis and partial hypothetical extraction are inflated by 9.2%. This is possible because we assume unchanged input coefficients, therefore the percentage change in output will reflect the percentage change in final demand. It becomes clear that once we have obtained the percentage change in demand, and calculated the nominal change in output, this output change may be inflated in the way described below. Let *b* denote the increase in final demand. Let $\triangle\, x$ denote the output change as a result of the pandemic. Next, let $x$ denote the original output vector, and let $\overline{x}$ denote the reduced output. Let also $L$ denote the original Leontief inverse, $\overline{L}$ the post-extraction inverse, $f$ the original final demand vector and $\overline{f}$ the post extraction/perturbation final demand vector.

We know that the following holds true:

$$\mathbf{x} = \mathbf{Lf}$$

$$\bar{\mathbf{x}} = \overline{\mathbf{Lf}}$$

$$\triangle \mathbf{x} = \mathbf{x} - \bar{\mathbf{x}} = \mathbf{Lf} - \overline{\mathbf{Lf}}$$

Therefore, the following must also hold true:

$$\mathrm{b} \triangle \mathbf{x} = \mathrm{b}\mathbf{x} - \mathrm{b}\bar{\mathbf{x}} = \mathrm{b}(\mathbf{x} - \bar{\mathbf{x}}) = \mathbf{L}\mathrm{b}\mathbf{f} - \bar{\mathbf{L}}\mathrm{b}\bar{\mathbf{f}} = \mathrm{b}(\mathbf{Lf}) - \mathrm{b}(\overline{\mathbf{Lf}})$$

Furthermore, if we take the equation for calculating changes in satellite accounts $\triangle h = \left(\hat{h_c}\right)(\triangle x)$, we can arrive at the following expression:

$$\mathrm{b} \triangle \mathbf{h} = \left(\hat{\mathbf{h}_{\mathrm{c}}}\right)(\mathrm{b})(\triangle \mathbf{x})$$

Given all the above, it should be briefly discussed what kind of results we are left with. The normalized output loss (also called inoperability) will be the primary result both for inoperability analysis and hypothetical extraction. This is an exact measurement because input coefficients are the same, and no matter what initial output we start with is, the percentage change, or change in operability will always be the same. Furthermore, estimates of the nominal changes in output, value added, income, and capital formation will be obtained. These are estimates because the nominal changes are based on the nominal assumed change in the original output vector. That being said, since the changes in value added and more are linked to output, the percentage change in each of them will be the same as for output, in other words, the sectors most affected percentually will always be the same.

## 3.8 Motivation of Methodological Choices and Expectations

### 3.8.1 Motivation of choice of methodology

In the literature review chapter of this thesis, it was expressed that the most suitable framework for the purpose of tackling the issues presented by the research question was the input-output framework. This framework, as was concluded is the most common one used in the field. However now, that all the theoretical concepts of methodologies chosen have been presented, it is possible to establish exactly why this was done, and what kind of value these choices are expected to have in answering the research question.

To start, classical input-output analysis, which in this case includes the examination of the technical coefficients matrix, the Leontief inverse and the multipliers is primarily intended to provide responses to the first research goal; evaluating the economic linkages of the Swedish air transport sector. The technical coefficients matrix, as demonstrated can provide valuable information on the input recipes of all the sectors of the economy. It can therefore be used for one to find the most important sectors providing inputs for Swedish air transport, and also the sectors for which Swedish air transport's inputs are the most salient. The Leontief inverse, containing the output multipliers for each sector, is also a beneficial way to establish a sector's importance for the economy, as it shows how much total output the economy gains for a given increase in final demand for the chosen sector's product. Furthermore, additional multipliers concerning other metrics such as value added, employment and more can enrich the picture of the role a sector plays in a national economy. Next, the usage of two final demand scenarios is motivated by a strive towards the provision of two different perspectives in results. An important aspect not always considered in this field is the possible reallocation of demand from one sector to other sectors, caused by serious changes in lifestyle, such as the one we have experienced during the pandemic. Once the consumer gains excess disposable income as a result of reduced spending on air transport services, this sum can be allocated to investments or consumption, and there are various factors affecting this decision as also theorized by Sjöström et. al (2016).

Meanwhile, inoperability analysis and hypothetical extraction are the backbone to the second research goal; measuring the effects of the COVID-19 pandemic on Swedish air transport, and the spin-off effects for the whole economy. There are two main reasons for why both methods were chosen to be included into the analysis. The first reason has to do with the strive for an, admittedly modest, but nevertheless valuable contribution to input-output theory, more specifically, harmonization of inoperability and hypothetical extraction techniques. The linking of inoperability analysis to classical input-output analysis was already very well presented in Dietzenbacher and Miller (2015). Furthermore, Dietzenbacher and Lahr (2013) have shown the links between hypothetical extraction and classical input-output analysis. This thesis, meanwhile, attempts to advance a framework, and common terminology for tying together inoperability analysis and partial hypothetical extraction that enable the user to take a compare and contrast approach when using them. It can be argued that the two methods can both be used to measure normalized output losses (traditionally called "inoperability" in the case of inoperability analysis), and normalized output gains as results of crises or demand increases for certain products. While the terminology initially seems rather different when comparing the two, it can be harmonized into a common framework, as was demonstrated in the methodology section.

Secondly, it can be argued that the usage of both methods simultaneously can in some cases provide additional value for the most accurate conclusions to be made. These are for instance situations with high degrees of uncertainty, cases in which data quality may for various reasons be lacking, and therefore a fairly substantial number of assumptions need to be made to carry on with the analysis. One such example is the assumption that within the interindustry flows table, sectors in need of air transport services would reduce their demand for air passenger services with the same amount as passengers did. This assumption is applied when using the hypothetical extraction method and allows us to also factor in the effect of a decrease in demand from industry sectors, which inoperability analysis does not allow for. However once again it must be noted that it is an assumption and may lead to an overestimation of effects if found that interindustry demand for air transport services is more rigid, as it is perhaps to a larger extent contract based. For this reason, it is useful to provide a range within which results are expected to fall, and this is enabled by the dual usage of inoperability analysis and hypothetical extraction presented in this thesis, a possibility induced by the harmonization of their methodological frameworks.

It can as such be concluded that the combination of basic input-output methodology, and the dual usage of inoperability analysis and the partial hypothetical extraction model is deemed to provide the ideal methodological toolkit to answer the research question of this thesis.

### 3.8.2 Expectations (Hypotheses)

Input-output analysis is much different from methods looking at causality between various variables, and therefore does possibly not lend itself to hypotheses in the traditional sense. However, it was, after the detailed discussion of methodology to be used for the analysis, still deemed a contribution of value to discuss in short expectations for the upcoming results section. As for the technical coefficients table, it is firstly theorized that important sectors providing inputs for Swedish air transport will include sectors that can be thought of as providing support functions necessary for its operations. Such sectors may for instance include the logistics/warehousing sector and sectors providing repair services for machinery, such as aircraft. Other important inputs to air transport are expected to come from sectors providing various forms of real estate, and buildings for air transport operations. Lastly, manufacture of aircraft and aircraft equipment is also theorized to prove to be an important input. Meanwhile, sectors for which air transport services are important are expected to be logistics firms that once again heavily rely on air transport services. Furthermore, travel operators, and sectors which require a lot of business travel, such as scientific activities, create activities and real estate services are also expected to come out on top. In terms of the output multiplier, based on results found by Bråthen et al. (2006) and Malaysian Aviation Commission (2017 the Swedish air transport sector is expected to rank in the lower end.

As for inoperability analysis, and partial hypothetical extraction, firstly, for both methods, the effects on air transport itself are expected to be detrimental. Its output is expected to drop significantly, and it is expected to have to let go of the majority of its labor force. Spin-off effects are expected to be large as well, with sectors proven to be closely linked with air transport to be affected the most. Overall, total effects are expected to be larger when hypothetical extraction is conducted, since it also contains the drop in intermediate demand. Finally, for both methodologies, impacts are expected to be slightly dampened in scenario 2, which "recirculates" some of the lost demand into the economy.

# 4. Findings/Results

## 4.1 Fundamental Input-Output Analysis

### 4.1.1 Inputs to the air transport sector

As detailed in the methodology chapter, the input coefficients matrix **A** can give us some early hints about the Swedish air transport sector's importance. The sector's column vector gives us an insight into the inputs that go into creating one unit of output of air transport services, and also demonstrates which industries are the most essential for the creation of this product. The below table illustrates the ten most important sectors for the input recipe of the air transport services sector, in descending order.

**Table 5:** *Swedish Air Transports' Input Recipe*

| Industry sector | Importance |
|---|---|
| Warehousing and support activities for transportation | 0.06544 |
| Education | 0.03353 |
| Real estate activities excluding imputed rents | 0.02484 |
| Wholesale trade, except of motor vehicles and motorcycles | 0.02160 |
| Manufacture of other transport equipment | 0.01309 |
| Rental and leasing activities | 0.01124 |
| Computer programming, consultancy and related activities; information service activities | 0.00919 |
| Air transport | 0.00661 |
| Security and investigation activities | 0.00496 |
| Manufacture of food products, beverages and tobacco products | 0.00485 |

It can be observed firstly that "Warehousing and support activities for transportation" is the most significant industry in the Swedish air transport's input recipe, with a value of 0.06544. In other words, for example for the creation of SEK 1 000 worth of air transport services, the sector uses SEK 65 worth of services from warehousing and support. Furthermore, as can be observed, another important input is education, with a value of 0.03353. Therefore, for the creation of SEK 1 000 worth of air transport services, SEK 33 worth of education services are needed. Further sectors of importance are also wholesale trade and manufacture of other transport equipment, with values in the .02-.025 range. Rental and leasing activities are also a fairly significant input, as well as computer programming services, security and investigation activities and manufacture of food products, beverages and tobacco products. It can also be observed that air transport also uses its own service as an input.

It is also important to note just how important imports are for the air transport sector's production recipe. They make up a whole 49%, which is very significant. In other words, for SEK 1 000 worth of air transport services, SEK 490 worth of products are imported. The domestic input-output table used for this thesis does not contain a breakdown of imports. However, the Statistics Sweden also provides a separate table for imports. From this it is possible to get an idea of which and how much of imported products the air transport sector requires for its functioning. It is found that the most important product comes from the sector "Manufacture of coke and refined petroleum products". Other significant inputs also include "Manufacture of other transport equipment", and "Warehousing and support activities for transportation".

### 4.1.2 Air transport as an input to other sectors

Meanwhile, the row for the air transport sector shows how important inputs coming from the air transport sector are for the other sectors of the economy. The below table illustrates the top ten sectors for which the air transport sector's product is a significant input. Note that the row vector was transposed to enhance the viewing experience.

***Table 6:*** *Air Transport Services as an Input to Other Sectors*

| Sector | Importance |
|---|---|
| Travel agency. tour operator reservation service and related activities | 0.07373 |
| Warehousing and support activities for transportation | 0.00821 |
| Air transport | 0.00661 |
| Legal and accounting activities; activities of head offices; management consultancy activities | 0.00632 |
| Wholesale trade. except of motor vehicles and motorcycles | 0.00283 |
| Other professional. scientific and technical activities; veterinary activities | 0.00166 |
| Creative. arts and entertainment activities | 0.00157 |
| Activities of membership organisations | 0.00137 |
| Architectural and engineering activities; technical testing and analysis | 0.00137 |
| Scientific research and development | 0.00130 |

Out of all the industry sectors, air transport carries the most importance for the travel sector. For the creation of SEK 1 000 worth of travel services, SEK 74 worth of air transport services are needed. Next, it can be observed that in terms of importance, it is the warehousing and transport support sector that uses the second most of air transport services as an input. It is followed by air transport itself, which was already discussed above. Additional sectors for which air transport is a fairly important input are wholesale trade, other professional scientific and veterinary activities, art and entertainment activities, activities of membership organizations, architect firms and scientific research and development.

## 4.2 Multipliers

### 4.2.1 Output multipliers

Once the Leontief inverse is obtained, its column sums can be interpreted as the multipliers for production. The multipliers contain both direct and indirect effects. The column sum for the air transport sector is 1.428, which means that a one unit increase in final demand for air transport services would lead to an output of 1.428 units. For instance, a SEK 1 billion increase in government spending on air transport would lead to an economy-wide increase in output of SEK 1.428 billion. The below table contains an extract from all the multipliers, with the two lowest, and two highest multipliers, as well as the air transport sector's multiplier.[15] Note that overall, the air transport sector's multiplier ranks 14th lowest out of the 64 sectors in the economy.

***Table 7:*** *Output Multipliers of Swedish Industry Sectors*

| Sector | Output multipliers[16] |
|---|---|
| Manufacture of coke and refined petroleum products | 1.237 |
| Social work activities | 1.279 |
| … | … |
| Air transport | 1.428 |
| … | … |
| Forestry and logging | 2.088 |
| Manufacture of wood and of products of wood and cork | 2.197 |

The industry with the lowest multiplier of 1.237 is the manufacture of coke and refined petroleum products. For every SEK 1 invested into the sector, or alternatively, for every SEK 1 increase in final demand for its product, the economy gains in total SEK 1.237 worth of output. Social work activities also have a low multiplier of only 1.279. Meanwhile, the two sectors with the highest multipliers are forestry and wood manufacturing. If investing SEK 1 billion into the latter, a total of SEK 2.197 billion of production is gained, which is SEK 767 million more as if the same amount was invested into air transport.

[15] Sector "activities of households as employers" with a multiplier of 1 was omitted as it is not very illustrative in terms of this comparison.

[16] Please see Annex B for full table.

### 4.2.2 Further multipliers

Based on the output multiplier, we can also create other multipliers that allow us to measure the effects of changes in final demand for air transport on factors such as value added, income, employment and capital formation. The below table details the various multipliers for the Swedish air transport sector.

***Table 8:*** *Additional Multipliers of Swedish Air Transport*

| **Multiplier Type** | **Multiplier value** |
|---|---:|
| Value Added | 0.450 |
| Income | 0.225 |
| Employment | 0.394 |
| Capital Formation | 0.031 |

As can be observed, firstly, the value-added multipliers comes out to be .45. Since the units of the input-output table are millions of SEK, it is interpreted as follows. A SEK 1 million increase in final demand for air transport would lead to an increase of SEK 450 million in value added for the whole economy. The multiplier for income is interpreted in the same way, for every SEK 1 million increase in final demand for air transport, income of all employed in the economy goes up by SEK 225 000. The interpretation of the employment multiplier is different. This is because the units of employment are not given in millions, but rather just shows the actual number of employees. Therefore, a SEK 1 million increase in final demand for air transport would lead to the creation of .4 jobs. Or a SEK 10 million increase in final demand for air transport would lead to the creation of 40 jobs. Finally, a SEK 1 million increase in final demand for air transport would lead to a SEK 31 000 increase in gross fixed capital formation (investments).

## 4.3 Final Demand Scenarios

In the following chapter, it is detailed how final demand is affected once the two scenarios are applied to it.

### 4.3.1 Scenario 1

As for Scenario 1, the following changes occur:

***Table 9:*** *Scenario 1: Changes in Final Demand for Air Transport*

| Final demand component | Nominal change (M SEK) | Percentage change |
|---|---|---|
| Private consumption | -4 955 | -74% |
| Nonprofit org. demand | -1.5 | -74% |
| Government demand | - 48 | -74% |
| Exports | -7 748 | -69% |
| **Total** | **-12 752** | **-71%** |

As seen in the table, the drop in total final demand for air transport is 71%, or SEK 12 752 million. This is primarily driven by the drop in exports and private consumption, which take the lion's share for air transport final demand, and therefore have the biggest impact.

### 4.3.2: Scenario 2

Scenario 2, as known, contains the same drops in demand, with the only difference being that half of the drop in final consumption (private consumption, nonprofit and government), SEK 2 500 million is reallocated according to the assumptions made. As a result, on top of the changes to final demand for air transport, the below changes to final demand for products of other sectors can be observed:

***Table 10:*** *Scenario 2: Additional Changes to Final Demand*

| Industry | Nominal increase (M SEK) |
|---|---|
| Manufacture of food products, beverages and tobacco products | 125 |
| Manufacture of basic pharmaceutical products and pharmaceutical preparations | 250 |
| Retail trade, except of motor vehicles and motorcycles | 250 |
| Land transport and transport via pipelines | 250 |
| Warehousing and support activities for transportation | 500 |
| Postal and courier activities | 250 |
| Telecommunications | 500 |
| Human health activities | 250 |

## 4.4 Inoperability Analysis

In this section, the sectors most and least affected in terms of operability will be presented. For ease of viewing, the industry sectors' full names will be shortened. The below table presents the abbreviations used. The same table can also be found in Annex E.

***Table 11:*** *Sector Abbreviations*

| Abbreviation | Full description |
|---|---|
| ACM | Accommodation and food service activities |
| AMR | Advertising and market research |
| AT | Air transport |
| CP | Computer programming, consultancy and related activities; information service activities |
| CST | Construction |
| CRO | Crop and animal production, hunting and related service activities |
| ED | Education |
| EMP | Employment activities |
| HH | Human health activities |
| LTP | Land transport and transport via pipelines |
| MP | Manufacture of basic pharmaceutical products and pharmaceutical preparations |
| MF | Manufacture of food products, beverages and tobacco products |
| MMA | Manufacture of machinery and equipment n,e,c, |
| MM | Manufacture of motor vehicles, trailers and semi-trailers |
| MOT | Manufacture of other transport equipment |
| POS | Postal and courier activities |
| RAI | Repair and installation of machinery and equipment |
| RE | Real estate activities excluding imputed rents |
| RL | Rental and leasing activities |
| RT | Retail trade, except of motor vehicles and motorcycles |
| SEC | Security/services to buildings/, landscape, office administration and support |
| WHS | Warehousing and support activities for transportation |
| TEL | Telecommunications |
| WHT | Wholesale trade, except of motor vehicles and motorcycles |
| WRT | Wholesale and retail trade and repair of motor vehicles and motorcycles |

### 4.4.1 Scenario 1: Top losers

Before presenting the results, for easier interpretation a note regarding the demand change is necessary. As known, in the case of inoperability analysis, final demand drops by 71%, but intermediate demand does not change. This results in an overall drop of 42.3% in demand for Swedish air transport services. The below figure, in turn, demonstrates the ten sectors that in scenario 1 suffered the largest loss in operability, or in other words, normalized output loss. [17]/[18]

***Figure 3:*** *Inoperability Analysis, Scenario 1 – Largest Losses in Operability*

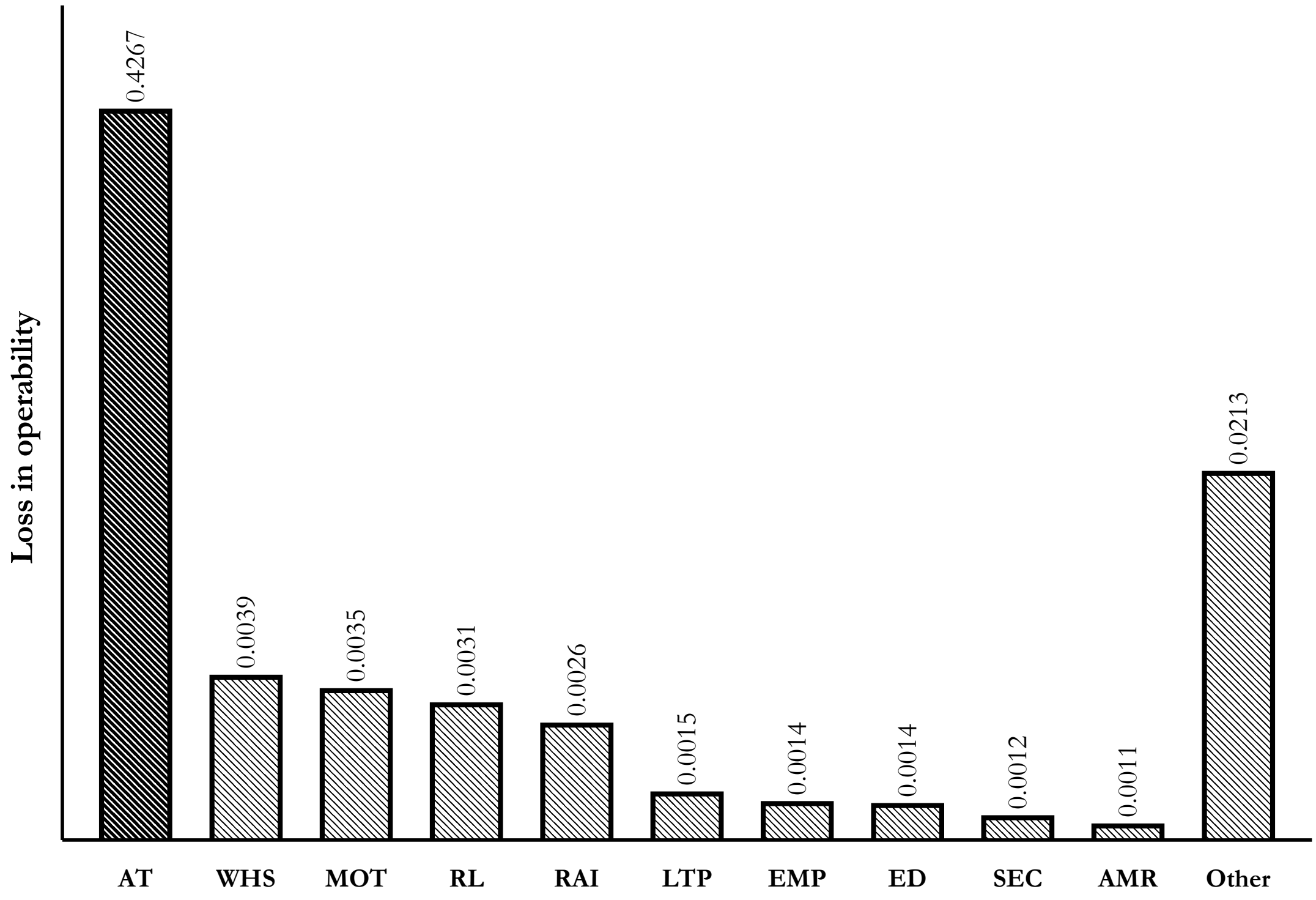

As can be observed, by far, the air transport sector is the most affected by the final demand loss. Air transport as seen, experiences an almost 43% loss in operability, or, in other words, sees its output drop by 43%. Next in line is warehousing and support activities for transportation which sees a .39% loss in operability. Manufacture of other transport equipment also sees a significant loss in operability of .35%. Rental activities are affected, as well as repair and installation of machinery and equipment. These are now followed by sectors less impacted, which see a drop in operability between .11% and .15%, and include land transport, education, security services and advertising and market research. Lastly, all the remaining 54 sectors, combined see a drop in operability of 2%.

[17] Note that the scale was logged for easier readability.
[18] Please see Annex C for complete list.

As detailed in the methods chapter, it is also possible, once we know the drop in operability, to calculate the nominal loss in output, as well as effects on value added, employment, income and capital formation. Below, these will be presented. Note that the results have been inflated in order to take into account the age of the input-output table as detailed also in the methods chapter.

***Table 12**: Inoperability Analysis, Scenario 1's Greatest Impacts – Loss in Income, Value Added, Employment and Gross Capital Formation*

| Sector code | **Income loss (M SEK)** | Sector code | **Value added loss (M SEK)** |
|---|---|---|---|
| AT | 1 711 | AT | 3 441 |
| ED | 264 | WHS | 353 |
| WHS | 137 | ED | 342 |
| WHT | 114 | RE | 292 |
| LTP | 91 | WHT | 211 |
| SEC | 73 | LTP | 165 |
| CP | 56 | MOT | 100 |
| EMP | 56 | CP | 96 |
| CST | 52 | RL | 96 |
| RE | 42 | SEC | 91 |
| Sector code | **Loss in employment (#emp)** | Sector code | **Capital formation loss (M SEK)** |
| AT | 2 560 | CST | 123 |
| WHS | 228 | CP | 77 |
| LTP | 205 | WHT | 44 |
| SEC | 198 | MOT | 42 |
| WHT | 189 | AT | 31 |
| WRT | 176 | WHS | 17 |
| EMP | 137 | RE | 16 |
| ED | 113 | MM | 8 |
| CST | 110 | MMA | 6 |
| ACM | 105 | RAI | 6 |

Firstly, as can be observed, on the whole, many of the sectors that experienced the largest drop in operability also top the list of most affected sectors for the above metrics. Of course, there will always be differences however given that all these sectors have different outputs to begin with, through which we calculate income, value added and more by using he input coefficients for satellite accounts. In terms of loss in income it can be observed that air transport tops the list with a loss of SEK 1.7 billion. That is followed by education with a drop of around SEK 260 million. Strongly affected is also warehousing and support activities for transport. These, as may be recalled all appeared amongst the top ten most affected sectors in terms of loss in operability. There are however some sectors that see very significant nominal drops but did not lose the most operability. These are for instance, wholesale, computer programming and construction. Lastly, real estate also makes the list, even though it is not amongst the top ten for inoperability.

As for loss in value added, the composition of sectors affected is rather similar as well. Air transport experiences a SEK 3.4 billion drop in value added, warehousing sees a drop of SEK 350 million and education a decrease of SEK 340 million.

When it comes to employment, air transport once again tops the list with a decrease in employment of 2 560. This is nearly half of its labor force of 6 000. Warehousing and land transport both experience a drop in the 200 range, followed by security services, wholesale, employment activities, education and construction which experience drops in the 100-200 range. Last on the list is accommodation and food service activities with a drop of 105.

Lastly, when it comes to drop in capital formation, the list of sectors affected is very different. This is due to the fact that air transport, and the sectors in its vicinity are not necessary the most capital intensive. That being said, these results too can provide some interesting insights. Construction sees the largest drop of SEK 123 million, followed by computer programming, wholesale, manufacture of other transport equipment, and only at 5$^{th}$ place, air transport. Real estate, manufacture of motor vehicles, and lastly repair and installation of machinery and equipment are all sectors that can traditionally be thought of as capital heavy.

### 4.4.3 Scenario 2: Most affected sectors

Scenario 2 in turn, as we know introduces some changes to the final demand vector, and therefore we may also expect to see differences in how operability is affected. Below graphed are the sectors that see the largest loss in operability for scenario 2.[19/20]

***Figure 4:*** *Inoperability Analysis, Scenario 2 – Largest Losses in Operability*

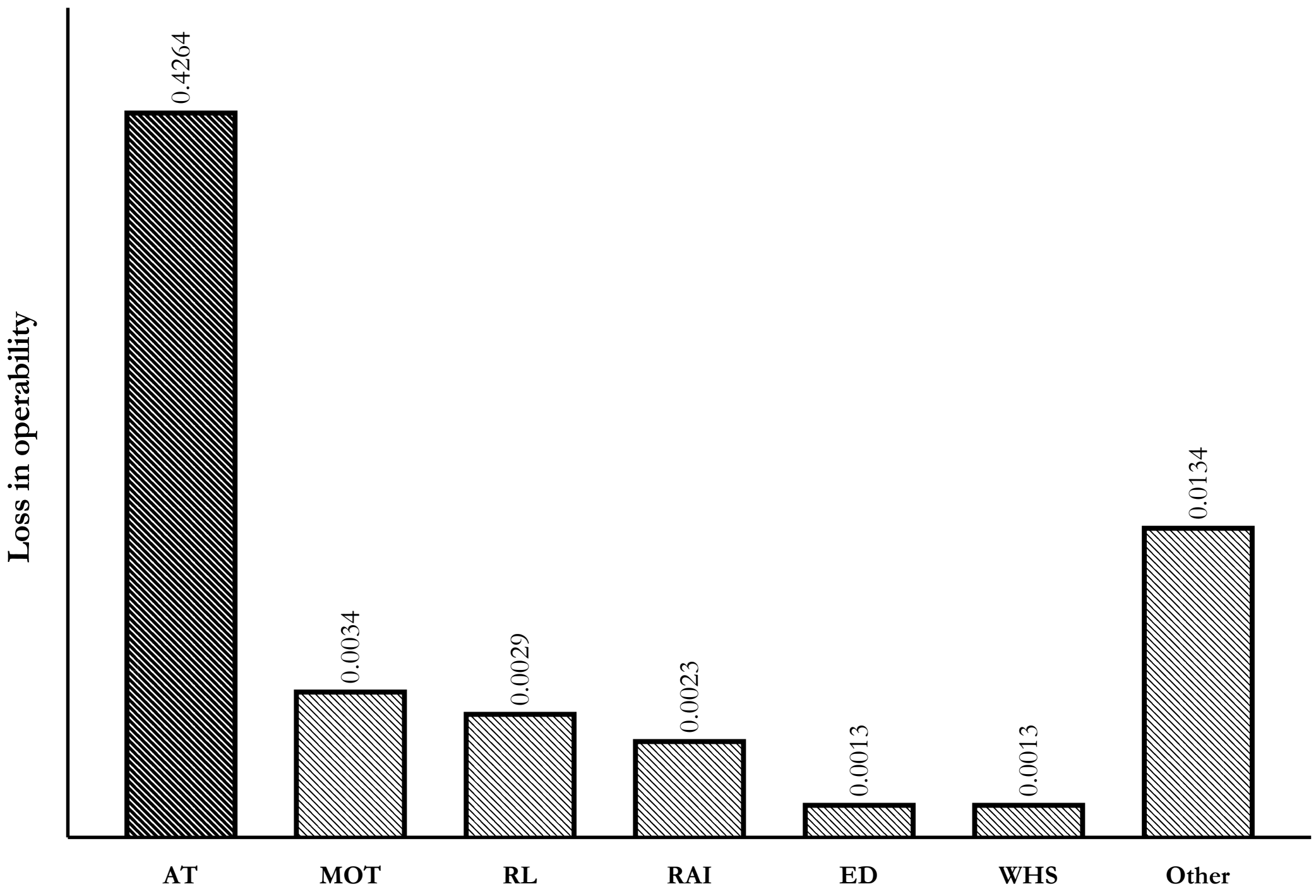

As can be observed, less sectors were included. This is because the effects on individual sectors beyond warehousing were negligible. Again, air transport takes the lead with a drop in operability of 43%. Next up is the manufacture of other transport equipment with a drop of .34%, followed by rental and leasing activities, repair and installation of machinery and equipment, education and warehousing and support activities for transportation. The remaining sectors experience a drop in operability in total of 1.3%.

[19] Note that the scale was logged for easier readability.
[20] Please see Annex C for full list.

As for scenario 2 however, there were also some sectors that instead experience a gain in operability. These are graphed below:

***Figure 5:*** *Inoperability Analysis, Scenario 2 – Largest Gains in Operability*

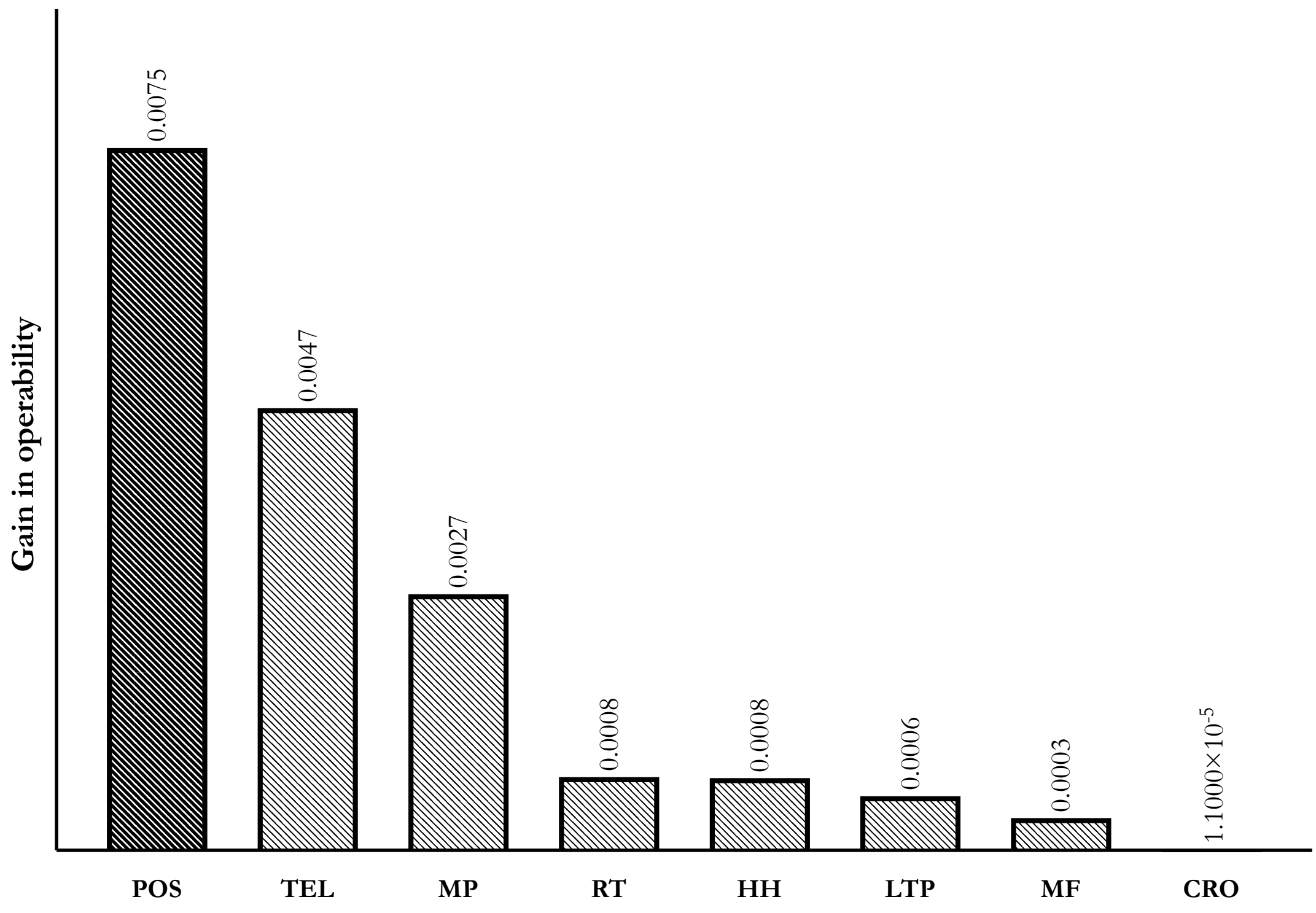

The sector that sees the largest increase in operability is the postal and courier activities sector, with an increase in operability of .75%, followed by the telecommunications sector with a gain of .45% and the manufacture of pharma products with a gain of .27%. Retail trade, human health activities, land transport and manufacture of food products also see positive, but very marginal gains. These sectors all received a part of reallocation of final demand for scenario 2.

Lastly, crop and animal production was also included. It sees a very small gain, but this is the only sector that did not receive a direct reallocation of demand. This may be due to its close interrelationship with the sector "manufacture of food products" (MF), and not very tight links with air transport. Looking at the input recipe of crop and animal production, we can notice that manufacture of food products is its second most important input, while air transport is ranked as low down as 48th. It can therefore be theorized that the positive impact from the increase in demand for manufacture of food products outweighed the negative impact of the drop in demand for air transport, and for this reason operability increased for crop and animal production.

### 4.4.4 Scenario 1 versus Scenario 2

It is by this point clear that in the two scenarios we can observe differences in the degree to which various sectors were affected. However, it may be worthwhile comparing exactly how different the picture looks. For this purpose, the below figure depicts the operability drop in the most affected sectors for scenario 1, and the change in operability for the same sectors in scenario 2. Note that air transport was not included for easier readability of the figure. Air transport, with a drop in operability of .4267 for scenario 1, and .4265 for scenario 2 did not see a radical change between the two scenarios.

***Figure 6:*** *Change in Operability - Comparison Between the Two Scenarios*

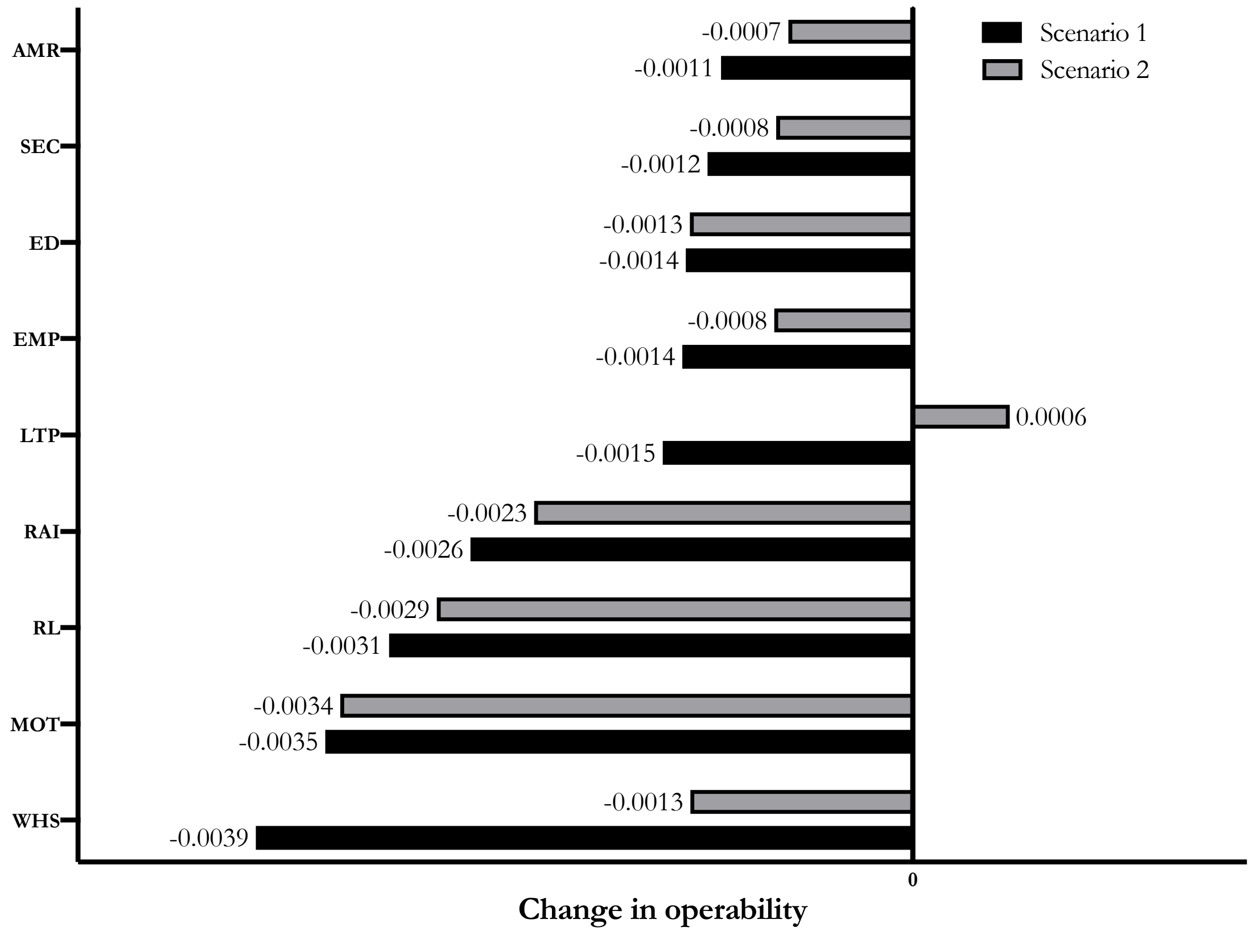

As can be observed above, all sectors perform better under scenario 2. The first group of sectors worth pointing out are WHS (warehousing and support activities for transportation) and LTP (land transport and transport via pipelines). Both of these received a slice of the demand reallocation for scenario two.

The second group of sectors worth observing are those that did not receive a demand reallocation, and at the same time did not see a large difference between the two scenarios. These are MOT (manufacture of other transport equipment), and ED (education). Lastly, the remaining sectors, RL (rental and leasing activities), RAI (repair and installation of machinery and equipment), EMP (employment activities), SEC (security/services and more) and AMR (advertising and market research) all saw moderate to significant relief in scenario 2.

#### 4.4.5 Inoperability analysis: summary

Below presented are the summarized impacts (direct and indirect) for the Swedish economy of the drop in final demand for air transport services, as found through inoperability analysis.

***Table 13:*** *Inoperability Analysis - Aggregate effects of the COVID-19 Induced Drop in Demand for Swedish Air Transport Services*

| Effects | Scenario 1 | Scenario 2 |
|---|---|---|
| Change in output (M SEK) | -19 886 | -15 548 |
| Change in output (%) | - 0.24 | - 0.19 |
| Change in value added (M SEK) | - 6 259 | - 4 264 |
| Change in income (M SEK) | - 3 129 | - 2 156 |
| Change in employment (number of employees) | - 5 483 | - 3 507 |
| Change in capital formation (M SEK) | - 437 | - 269 |

The effects are presented for both final demand scenarios and are measured as nominal and percentage change in output, and change in value added, income, employment and lastly capital formation. Note that the results presented also take into account the age of the input-output table and have therefore been inflated as described in the methods chapter. As can be seen above, the drop in final demand for the air transport sector causes for scenario 1 a SEK 19.9 billion drop in total output, a drop of 0.24%. The same figures for scenario 2 are SEK 15.5 billion and 0.19%. Value added is reduced by SEK 6.3 billion and SEK 4.3 billion respectively, which can also be interpreted as a GDP drop. SEK 3.1 billion, respectively SEK 2.2 billion of income is taken from citizens pockets. 5500, respectively 3500 jobs are lost. Lastly capital formation is reduced by SEK 437 million, respectively SEK 269 million.

## 4.5 Partial Hypothetical Extraction

### 4.5.1 Partial extraction results

The partial hypothetical extraction, as detailed in the methods chapter, results in a SEK 7 255 million drop in intermediate demand for air transport services, a 60% fall. The same final demand drop as for inoperability analysis is also applied here, a drop of SEK 12 752 million, a 71% fall. The combined drop in final use of air transport therefore becomes SEK 20 007 million, a 66.4% drop (which can be compared with the overall drop of 42.3% for inoperability analysis). This figure of 66.4% can be compared to a data projection requested and received from EUROCONTROL (The European Organisation for the Safety of Air Navigation (EUROCONTROL, 2020). This data was not used for reasons detailed in the methods section, that being said it can prove to be useful as a way of checking the feasibility of the demand changes found by this thesis. The EUROCONTROL forecast projected a 58% decrease in demand for Swedish air transport services. As can be observed, this figure very much falls between the finding of inoperability analysis and hypothetical extraction. Nevertheless, it shows that these two figures are not out of the ordinary.

### 4.5.2 Scenario 1: Top losers

In similar fashion to inoperability analysis, sectors most affected in terms of normalized output loss (in the case of inoperability analysis this was called loss in operability) were obtained. The below figure illustrates exactly these. [21/22]

***Figure 7:*** *Hypothetical Extraction, Scenario 1 – Largest Losses in Normalized Output*

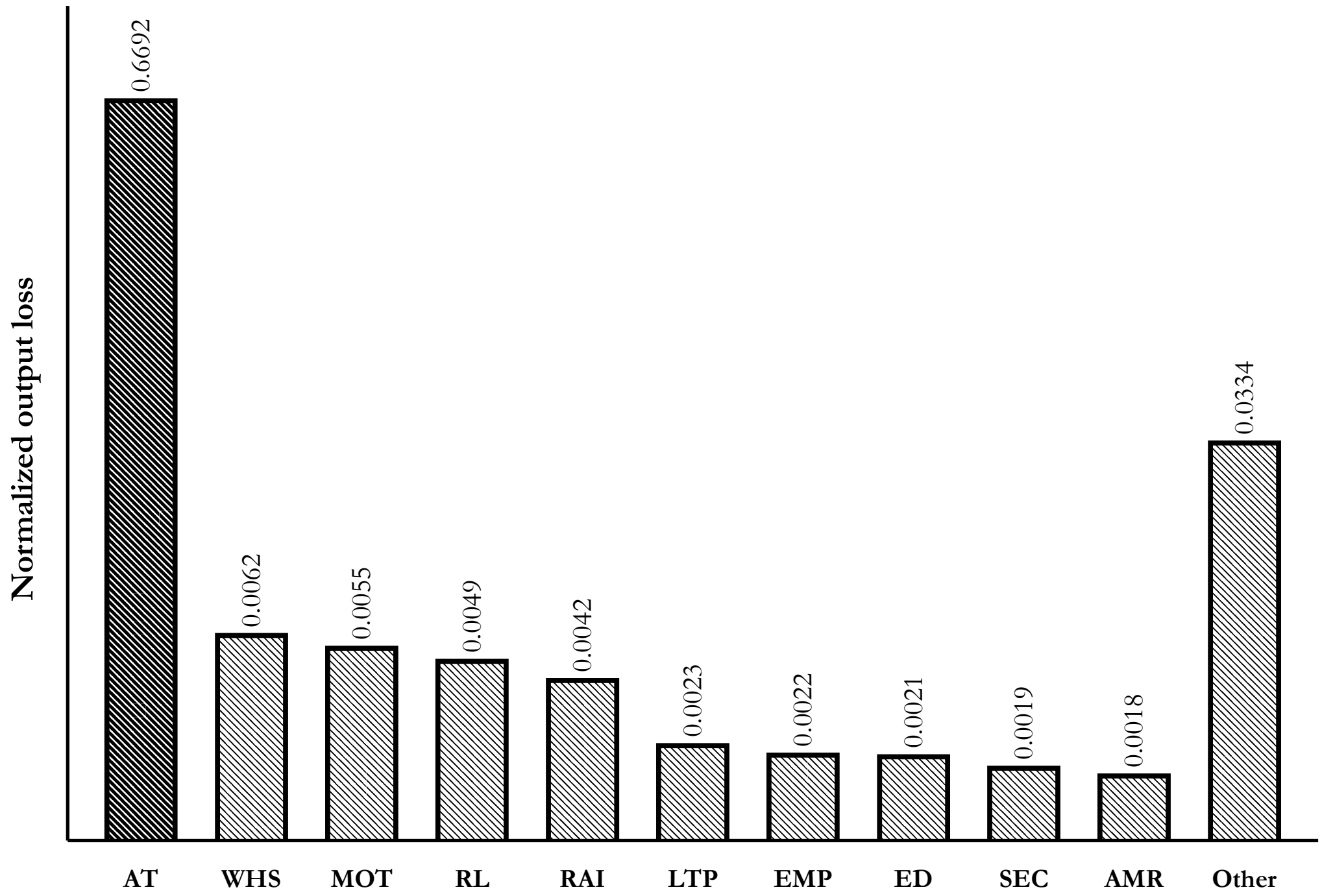

As can be observed, air transport sees a normalized output loss of almost 67%, a significant disruption. Warehousing and support activities for transportation is up next with a normalized output loss of .62%, followed by manufacture of other transport equipment, rental and leasing activities, repair and installation of machinery and equipment, land transport, employment activities, education, security, advertising and market research. The remaining 54 sectors combined saw a loss of 3.34%.

[21] Scale logged for easier viewing experience

[22] Please see Annex D for full list.

The below table in turn represents the most impacted sectors in terms of the other metrics used throughout this analysis.[23]

***Table 14:*** *Hypothetical Extraction, Scenario 1's Greatest Impacts – Loss in Income, Value Added, Employment and Gross Capital Formation*

| Sector code | **Income loss (M SEK)** | Sector code | **Value added loss (M SEK)** |
|---|---|---|---|
| AT | 2 684 | AT | 5 397 |
| ED | 413 | WHS | 553 |
| WHS | 214 | ED | 537 |
| WHT | 179 | RE | 458 |
| LTP | 142 | WHT | 330 |
| SEC | 115 | LTP | 259 |
| CP | 88 | MOT | 157 |
| EMP | 87 | CP | 150 |
| CST | 82 | RL | 150 |
| RE | 66 | SEC | 143 |

| Sector code | **Loss in employment (#emp)** | Sector code | **Capital formation loss (M SEK)** |
|---|---|---|---|
| AT | 4 385 | CST | 193 |
| WHS | 390 | CP | 121 |
| LTP | 350 | WHT | 69 |
| SEC | 339 | MOT | 66 |
| WHT | 324 | AT | 48 |
| WRT | 301 | WHS | 27 |
| EMP | 234 | RE | 25 |
| ED | 194 | MM | 13 |
| CST | 188 | MMA | 9 |
| ACM | 180 | RAI | 9 |

Starting with income loss, employees in the air transport sector experience the biggest drop in income, a total of almost SEK 2.7 billion. Education sees a large drop of SEK 413 million, as well as warehousing and support for transport activities with a drop of SEK 214 million. Wholesale trade is next on the list, although if we recall figure 6, it did not appear in the top ten in terms of normalized output loss. Up next are then land transport and security services. Once again, computer programming, consultancy and related services, and real estate also make the list, but did not in terms of normalized output loss.

[23] Note again that the values have been inflated to take into account input-output table age.

As for value added, air transport experiences a decrease of nearly SEK 5.4 billion, followed by warehousing, education, real estate wholesale trade and land transport. Manufacture of other transport equipment, computer programming, rental and leasing activities, and lastly security services see a smaller drop relatively speaking.

It can also be seen that air transport loses almost 4 400 employees, which is almost three fourths of all employees in the sector. Warehousing and support activities for transportation, land transport, security services, wholesale and retail trade and employment services also see significant impacts.

Lastly, as before, slightly different sectors show up in terms of loss in capital formation. This is because a rather different set of sectors are capital heavy than many of those in the vicinity of air transport. Topping the list is the construction sector with a drop in capital formation of SEK 193 million.

### 4.5.3 Scenario 2: Most affected sectors

Below presented are the most affected sectors in terms of normalized output loss for scenario 2.[24/25] It can be observed that once again air transport experiences a significant drop of 66.9%, followed by manufacture of other transport equipment with a drop of .54%. Rental and leasing activities, repair and installation of machinery and equipment, and warehousing and support activities for transportation are also significantly impacted. Education, employment activities, security, real estate activities and advertising are also affected but to a smaller extent. The remaining sectors combined experience a normalized output loss of 2%.

***Figure 8:*** *Hypothetical Extraction, Scenario 2 – Largest Losses in Normalized Output*

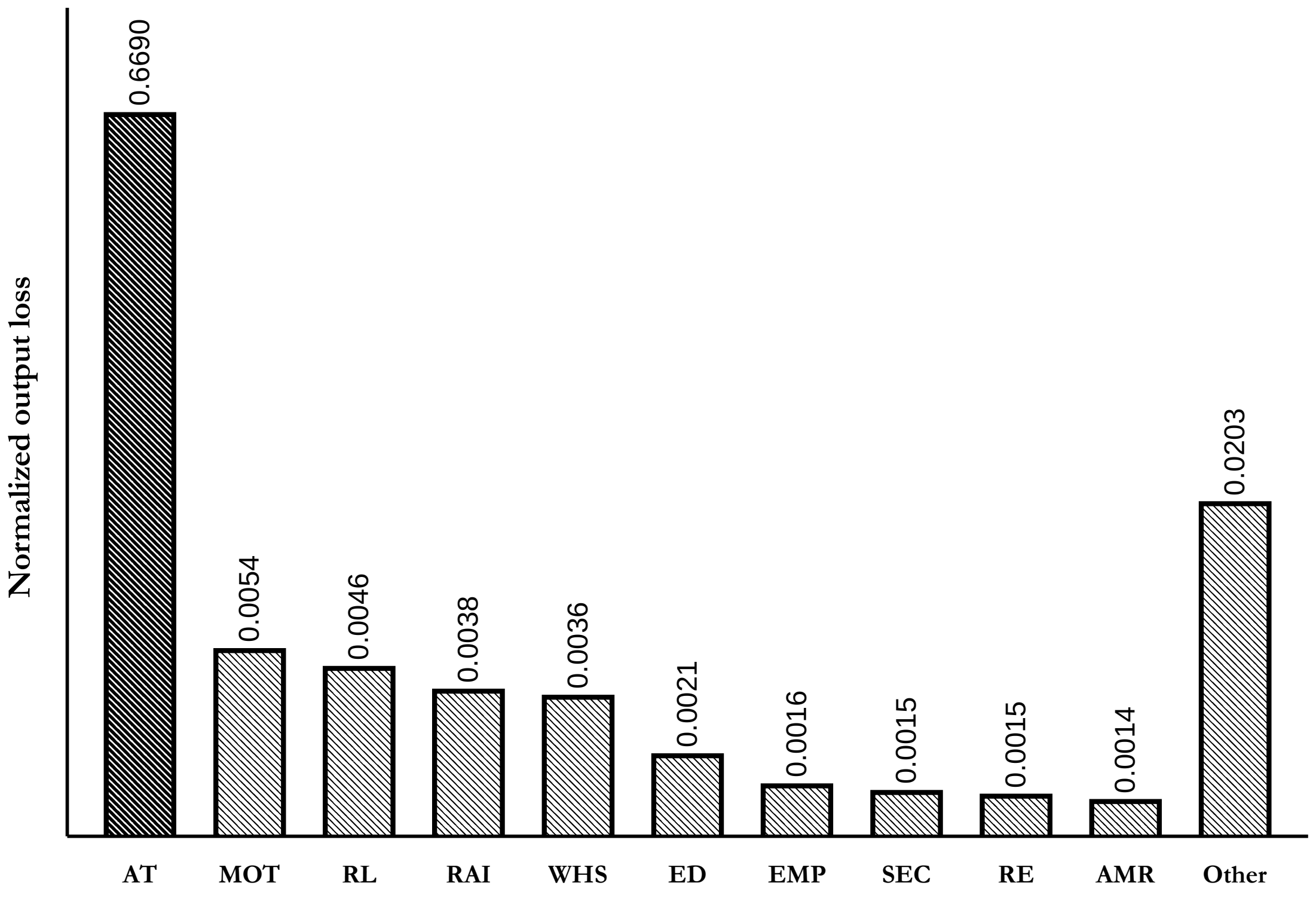

[24] Logged for ease of viewing
[25] Please see Annex D for full list.

Furthermore, below are presented the sectors that saw the most significant gains in scenario 2.

***Figure 9:*** *Hypothetical Extraction, Scenario 2 – Largest Gains in Normalized Output*

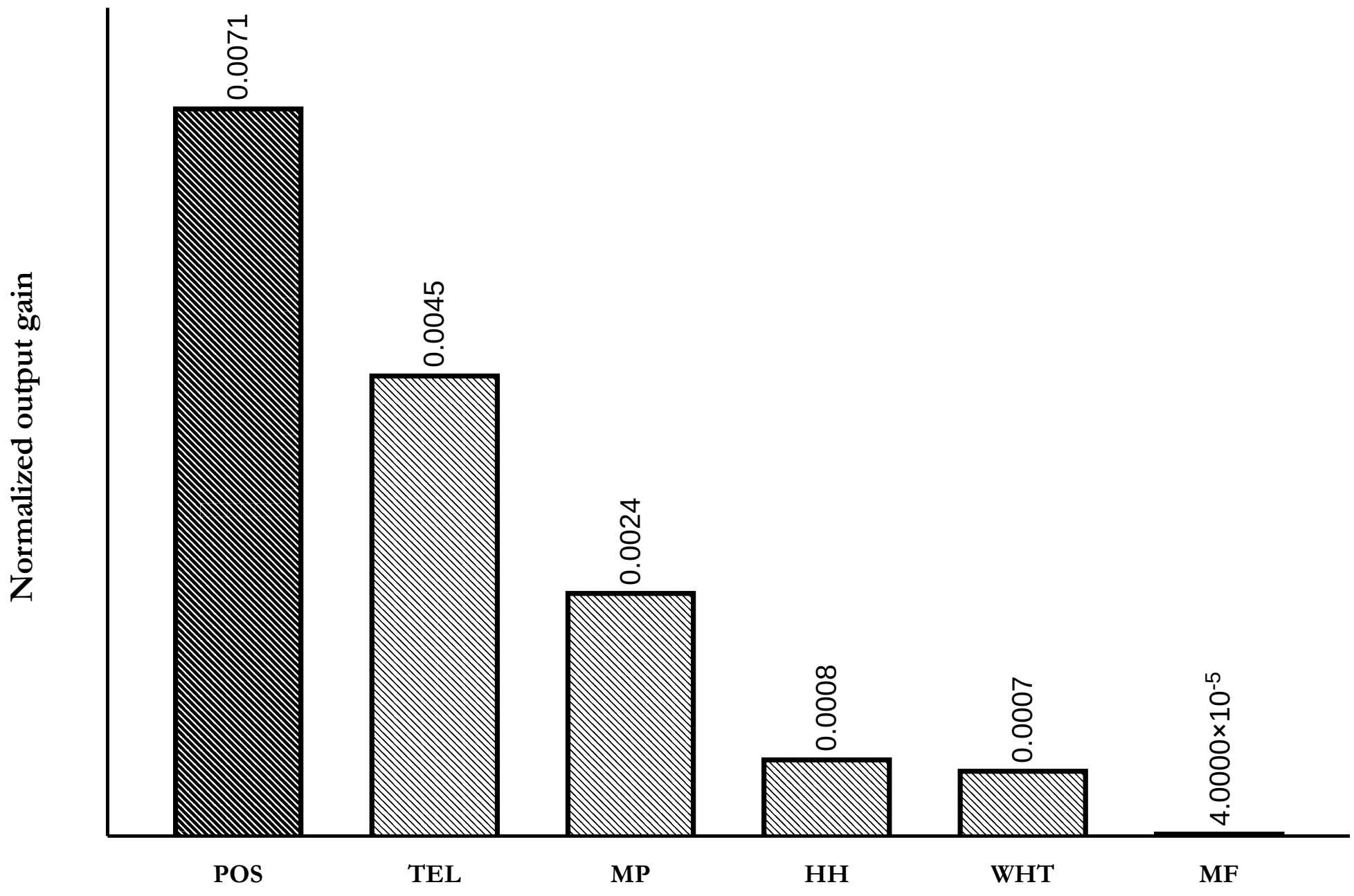

As can be observed above, a similar set of sectors are those positively affected as when it comes to inoperability analysis (scenario 2). Postal and courier activities see a gain in output of .71%, followed by telecommunications with .45%, manufacture of pharma products with .24%, and negligible effects on human health, wholesale trade and manufacture of food products.

### 4.5.5 Scenario 1 versus Scenario 2

Once again, comparisons between two scenarios can be made, as shown below.

***Figure 10:*** *Change in Normalized Output - Comparison Between the Two Scenarios*

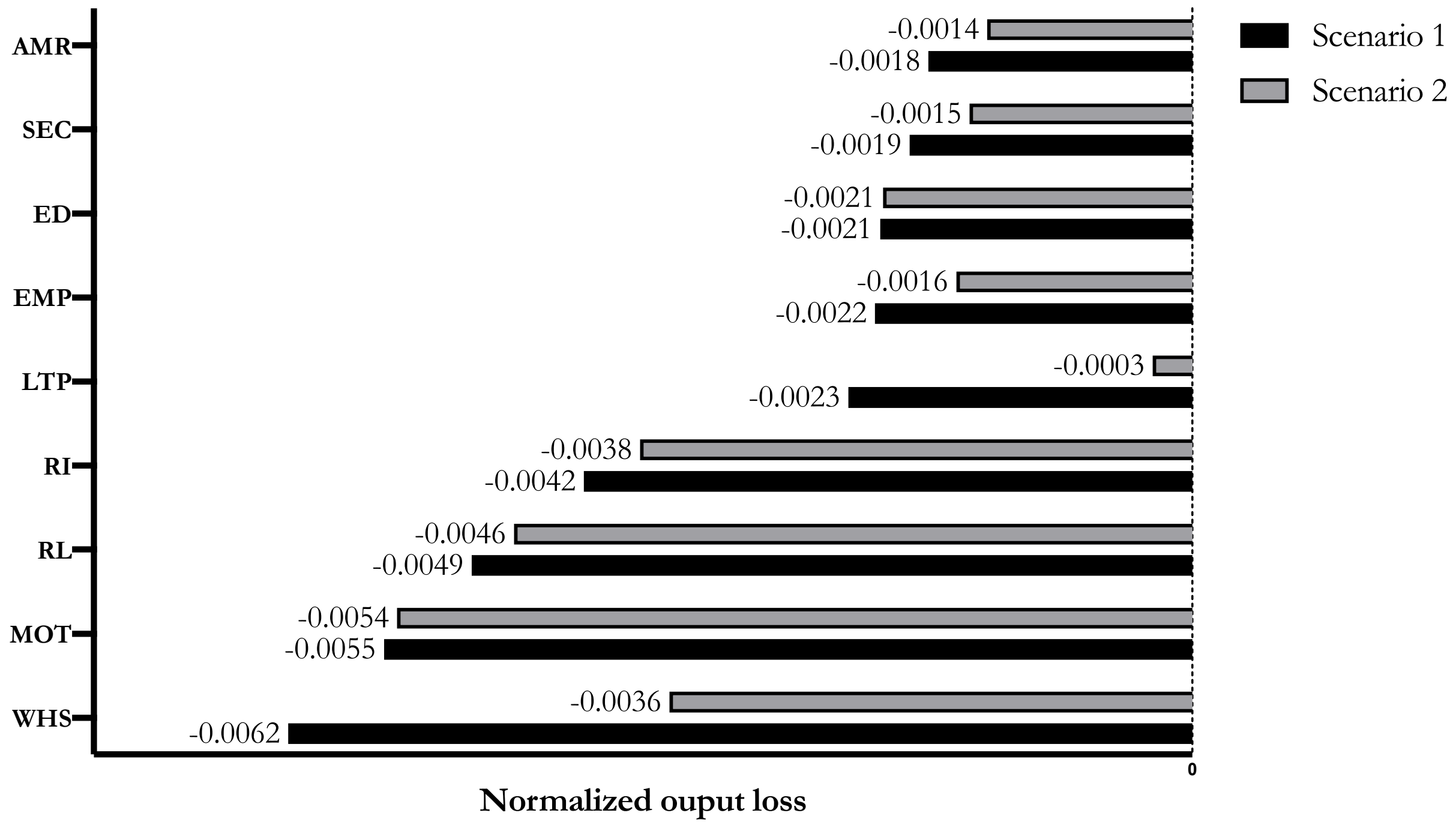

It can be observed that all sectors are better off in scenario 2. While in scenario 1, warehousing experiences a normalized output loss of .62%, the figure is only .36% in scenario two, and the figures are .23% and .003% for land transport. Advertising and market research, security services, and employment activities also experience significant improvements, whereas the remaining sectors only see minor changes.

### 4.5.6 Hypothetical extraction: summary

Apart from the individual sectors affected, the aggregate effects on the economy are also of interest. The below table summarizes the total effects of the extraction in terms of output, value added, income, employment and capital formation.

***Table 15:*** *Hypothetical Extraction - Aggregate effects of the COVID-19 Induced Drop in Demand for Swedish Air Transport Services*

| Effects | Scenario 1 | Scenario 2 |
|---|---:|---:|
| Change in output (M SEK) | - 31 189 | - 26 854 |
| Change in output (%) | - 0.34 | - 0.29 |
| Change in value added (M SEK) | - 9 817 | - 7 823 |
| Change in income (M SEK) | - 4 908 | - 3 935 |
| Change in employment (number of employees) | - 8 600 | - 6 624 |
| Change in capital formation (M SEK) | - 686 | - 517 |

As can be seen, the drop in final and intermediate demand for air transport services causes a total drop of almost SEK 31.2 billion in scenario 1, and almost SEK 27 billion in scenario 2. Percentwise, this is a .34% respectively .29% drop. Value added sees a decrease of SEK 9.8 billion, respectively SEK 7.8 billion. SEK 4.9 billion respectively SEK 3.9 billion is taken from employees' pockets, and 8 600 respectively 6 624 jobs are cut. Lastly, capital formation is reduced by SEK 690 million, respectively SEK 517 million.

## 4.6 Cross Methods Comparison and Summary

### 4.6.1 Individual sector effects

As final results for both methods have been obtained, these can now also be compared.

The below table provides an overview of the most affected sectors for scenario 1, for both the methods used.[26]

***Figure 11:*** *Scenario 1's Most Affected Sectors: Cross-Method Comparison*

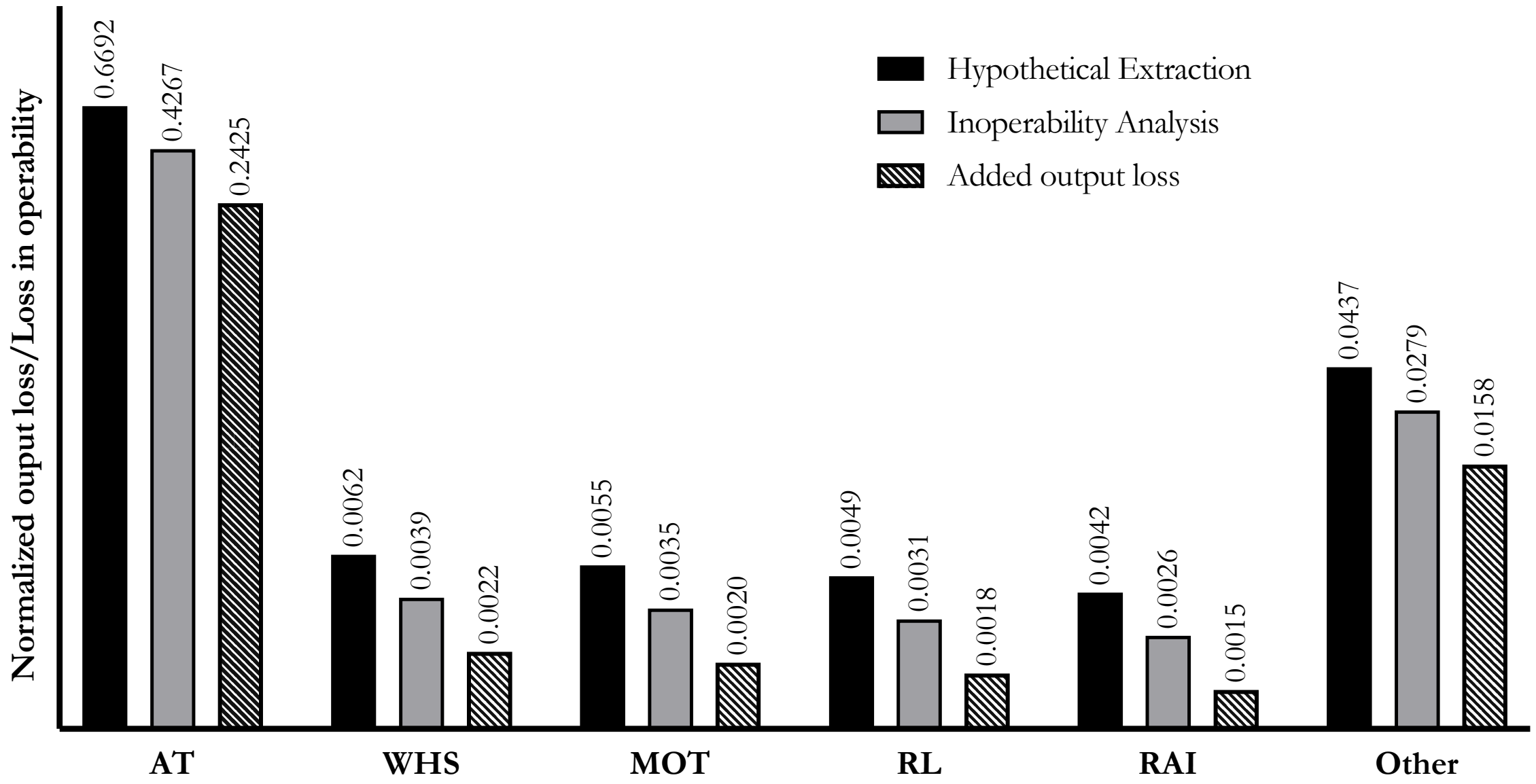

It can be observed that the added loss in intermediate demand implemented through the hypothetical extraction model is significant for all the sectors. The added output loss can also be interpreted as the normalized output loss caused by the drop in intermediate demand for air transport services. The air transport sector for instance experiences and added normalized output loss of around 24% using the hypothetical extraction model versus the hypothetical extraction model. For warehousing the figure is .22% added loss, and as can be observed, overall, for most sectors the added output loss lies between .15% and .2%.

[26] Logged scale for easier viewing.

### 4.6.2 Aggregate effects

Finally, the aggregate effects between the two methods can also be compared. The table below compares the effects for scenario 1.

***Table 16:*** *Scenario 1: Cross Method Comparison of Aggregate Effects of the COVID-19 Induced Drop in Demand for Swedish Air Transport Services*

| Effects | Hypothetical extraction | Inoperability | Difference |
|---|---|---|---|
| Change in output (M SEK) | -31 189 | -19 886 | 11 303 |
| Change in output (%) | - 0.34 | - 0.24 | 0.10 |
| Change in value added (M SEK) | - 9 817 | - 6 259 | 3 558 |
| Change in income (M SEK) | - 4 908 | - 3 129 | 1 779 |
| Change in employment (number of employees) | - 8 600 | - 5 483 | 3 117 |
| Change in capital formation (M SEK) | - 686 | - 437 | 249 |

On the aggregate level too, the differences are clear. In terms of output, hypothetical extraction generates exactly a .1% larger loss in economy-wide output. It also causes a SEK 3.56 billion larger loss in value added, and nearly SEK 1.8 billion more lost in terms of income. The impact on employment is also significantly different, with a difference of more than 3 000 jobs. Capital formation too is impacted very differently. Note, that a comparison for scenario 2 is not meaningful as the difference between the two methods remains the same for all metrics. Therefore, results for scenario 2 are simply presented below for the two methods, with the "difference" column omitted:

***Table 17:*** *Scenario 2: Cross Method Comparison of Aggregate Effects of the COVID-19 Induced Drop in Demand for Swedish Air Transport Services*

| Effects | Hypothetical extraction | Inoperability |
|---|---|---|
| Change in output (M SEK) | -26 854 | -15 548 |
| Change in output (%) | - 0.29 | - 0.19 |
| Change in value added (M SEK) | - 7 823 | - 4 264 |
| Change in income (M SEK) | - 3 935 | - 2 156 |
| Change in employment (number of employees) | - 6 624 | - 3 507 |
| Change in capital formation (M SEK) | - 517 | - 269 |

# 5. Discussion

In the penultimate chapter of this writing, firstly, reflections relating to the methodologies used in this thesis will be discussed. Then, findings will be summarized and interpreted in the context of the research question, and previous research in the field. Furthermore, implications, limitations and suggestions for further research will also be discussed.

## 5.1 Theoretical Evaluations

In section 3.8.1: "Motivation of choice of methodology" of this writing, it was in detail discussed exactly why the given methodologies were chosen to be used for the analysis. Apart helping produce novel results which will be evaluated in the sections to come, a hope was also expressed that this thesis can be of modest value to the methodological advancements in the field. As expressed, in the knowledge of the link between traditional input output analysis and inoperability analysis/partial hypothetical extraction, the notation of these two methodologies was harmonized in a way as to be able to use both as complements to each other. After the completion of the analysis and the obtaining of the results, it is concluded by the author that this slight advancement in methodology can claim to be of some value for the field. Firstly, the prospect of the dual usage of these methods was established by the author, and secondly the links between them were further demonstrated through the harmonization of their notations. Their dual usage, which was made easier through these marginal methodological advancements enables a more complex view of a situation with high degrees of uncertainty.

## 5.2 Economic Linkages of Air Transport

In the below two sections, the findings concerning both research goals will be summarized and interpreted. As detailed, the first objective is to establish and evaluate the overall economic linkages between the Swedish air transport sector and the economy, while the second goal is to estimate the effects of the pandemic on Swedish air transport, and the spin-off effects for the economy. The first task is appropriately approached through basic input output analysis, more specifically, analysis of the air transport sector's input recipe, the importance of air transport as an input to other sectors, and finally air transport multipliers.

### 5.2.1 Swedish air transport's input recipe

In terms of the most important inputs going into air transport, some results don't raise any eyebrows, while others need additional thought. For instance, the most important input can be traced back to the "warehousing and support activities for transportation" sector, which already in name indicates its close connection to air transport. When digging for more details, one can also discover that one of the subsectors of warehousing and transport support is called "52.22: Service activities incidental to air transportation" (Statistics Sweden, 2007). Such services may for instance include ground logistics operations, storage, fueling, and more. Other easily explainable sectors include real estate, rental/leasing activities, wholesale trade, and manufacture of other transport equipment, which may in practice be for instance office buildings, hangars, logistics centers, buildings for staff training, lease of aircraft, purchases of food and beverages, information and communications equipment, and aircraft machinery.

Perhaps slightly harder explained are sectors such as education (the second most important input), computer programming/consultancy activities and security and investigation activities. Regarding education, one might theorize that much of it is attributable to staff training. Airlines require highly trained staff, and often, sets of skills not taught at universities are needed. Cabin crew training programs, and even pilot trainings are often sponsored by airlines, incurring significant costs. As for the other two sectors they are rather broad, and it is therefore not easy to pinpoint exactly what types of services may be of most importance. It is uncontestable however that airlines, both cargo, and passenger, heavily rely on IT-services, and security/investigation activities are also demanded, both for crew and passenger safety, and investigations into operational safety. The results indicate that air transport also uses its own services as an input, which isn't of much surprise given the necessity to transport airline staff, and executives to various locations where they are required.

Finally, worth noting is also air transport's great reliance on imports. At 49%, imports nearly stand for half of inputs, a significant portion. As discussed in the findings chapter, these are dominated by three main sectors; "manufacture of coke and refined petroleum products", "manufacture of other transport equipment" and "warehousing and support activities for transportation." The first can without much doubt be linked to airline fuel. This "liquid gold" is by far one of the most vital ingredients to the air transport recipe, and since Sweden produces no crude oil (Trading Economics, 2020), airlines exclusively have to rely on imports. Neither is Sweden a significant manufacturer of aircraft and aircraft parts, and so these too need to be found elsewhere. Lastly, warehousing and support activities are seemingly demanded both domestically and acquired from foreign service providers.

### 5.2.2 Swedish air transport services as an input to other sectors

Meanwhile, air transport is likewise unsurprisingly used as an input by other sectors for their operations. Travel agencies/tour operators stand out, with air transport constituting slightly over 7% of the input recipe. Next up is warehousing and support activities, but with a figure of only .8%. Other sectors include legal and accounting activities, wholesale, professional and scientific activities, creative activities, membership organizations, architectural and engineering activities and scientific research. A common theme that can be observed here is that much of these can be linked to passenger travel, more so than to cargo services. It is noteworthy that various manufacturing sectors which one would hypothesize are in great need of air cargo services don't show up amongst the top ten. However, this can at least in part be explained by the miniscule size of domestic Swedish air cargo services. As known, they only account for 7% of production value in the sector, and so these types of services are usually imported by Swedish industry sectors.

### 5.2.3 Multipliers

The findings on output multipliers also provide insight into Swedish air transport's economic linkages. As mentioned, air transport ranks 14$^{th}$ lowest out of the 64 sectors used in the analysis with a multiplier of 1.4. This is lower than the 2.0 found in Malaysian Aviation Commission (2017) and the 1.99 for Oslo airport found by Bråthen et al. (2006). Here it must be noted that some of the difference surely is attributable to the choice of model. These studies use models with household consumption endogenous to the interindustry matrix, therefore their multipliers capture not only direct and indirect, but also induced effects, in other words, effects of additional demand caused by increased income for employees. At the same time, amongst the 30 sectors included in Malaysian Aviation Commission (2017), aviation ranks the third lowest. These findings are fairly consistent with those of this thesis, where air transport is ranked in the lower range in terms of its output multiplier.

Additional multipliers for value added, employment, income and capital formation also provide some insight. These findings could be compared to the employment and income multipliers found by Huderek-Glapska et al. (2016). Although they calculate multipliers for three specific airports, and their findings show a very significant disparity in results depending on the airport, overall, the income multiplier and employment multiplier found in this writing is somewhat lower in comparison. Again however, this is most likely due to the model choice, where induced effects were not measured.

## 5.3 Effects of the COVID-19 Pandemic on Air Transport

In the following section the findings concerning the impact of the pandemic on the Swedish air transport will be discussed.

### 5.3.1 Direct effects

The extent to which demand (final and intermediate) for air transport services is impacted gives us a good first impression of the gravity of the crisis, and it is clear from the results that the impact is significant. Final demand for air transport services is calculated to drop a whole 71% overall, which accounts for a total disruption of 42.3%, the figure used for inoperability analysis. In addition, intermediate demand is expected to drop by 60%, which in combination with the final demand shrinkage leads to a total disruption of 66.4%, used for hypothetical extraction.

### 5.3.2 Inoperability analysis and hypothetical extraction

The findings concerning the total effect on Swedish air transport measured by both inoperability analysis and hypothetical extraction end up confirming the grave picture already projected through the direct effects. Inoperability analysis indicates a drop in output between 42.6-42.7% for air transport services (depending on the scenario), and hypothetical extraction, which also includes the shrinkage in intermediate demand finds a drop of 66.9% (for both scenarios). As can be noted, the extent to which air transport is affected barely varies between scenario 1 and 2 within each method. This finding is somewhat surprising as a significant part of the final demand loss was reallocated in scenario 2, and one could have expected to see a larger alleviation of pressure on the air transport sector. However, it can be theorized that because the reallocation primarily benefited sectors not very strongly linked to air transport, the positive indirect effects spawned by added demand for their products were not enough to seriously improve the situation for the air transport sector.

The above results may be compared to those found by Santos (2006) who simulated a 33.2% drop in demand for air transport, and a 19.2% decrease in demand for accommodation services as a result of the September 11 attacks. It must as such be noted that the focus is slightly different, as Santos (2006) simulated demand decreases for two sectors rather than just one. That being said, it still serves as a useful comparison as Santos (2006) also used inoperability analysis to obtain his results. Through his analysis, he found a 33.49% drop in operability for the US air transport sector. Overall, it is clear that the direct effects of the fall in demand for air transport services in the case of Santos (2006), 33.2%, and in the case of this thesis, 42.3%, respectively 66.4%, account for the majority of the total effect (direct and indirect). The indirect effects, for Santos (2006), 0.29%, and for this writing, 0.4%, and 0.5% respectively are rather small.

Furthermore, the impact on air transport was also expressed with a range of other indicators. Here as well, the effects are significant. Employees in the sector are expected to lose between SEK 1.7 billion and SEK 2.7 billion of income depending on the method used, and between ½ and ¾ of the sector' labor force is expected to be cut.

## 5.4 Spin-Off Effects

In the previous section, the effects of the COVID-19 pandemic on the Swedish air transport were summarized and interpreted. In this section, the same will be done for the spin-off effects. Firstly, sector-specific effects will be summarized and interpreted, followed by an analysis of results regarding aggregate, economy-wide effect.

### 5.4.1 Sector specific spin-off effects

As seen from the findings, the pandemic's impact on Swedish air transport also induces negative spin-off effects for other sectors of the economy. In general, sectors with close economic linkages to air transport are impacted the most. Warehousing and support activities for transport, a significant input to air transport, and a sector for which air transport services are key, manufacture of other transport equipment, rental and leasing activities, repair and installation of machinery and equipment, and education, all experience output losses in the .1%-.6% range. Additional sectors negatively impacted also include those of security services, employment services, advertising and market research. Again, these results may be compared to the results of Santos (2006). In his case, the most affected sectors apart from those on which he simulated the direct demand drop experienced losses in operability in the 1-3% range. It is clear therefore that the results of this thesis indicate smaller sector-specific spin-off effects than those found by Santos (2006). Again, it is important to underline however that Santos (2006) simulated demand drops in two sectors, which are sure to produce larger spinoff effects.

According to the results, some sectors also see a boost in operability as a result of the demand reallocation in scenario 2. The sector with the biggest gain is the postal services sector, with an increase in output of .71-.75%, followed by the likes of telecommunications and manufacture of pharma products. These sectors seemingly are not very closely linked to air transport, which is why the demand boost they receive gives the biggest positive percentual effect on their output. In the context of the pandemic, these results are not very surprising. The Swedish postal service for instance, at the end of Q4 2020 reported its best quarter yet (PostNord, 2021). It is clear that while demand for certain services, such as air transport has been severely reduced, Swedes' spending have shifted towards other alternatives, or substitutes.

### 5.4.2 Economy-wide spin-off effects

The total effects on the economy are significant. The total loss in output ranges between 0.19% and 0.34% depending on the method and scenario used. Changes in value added, income, employment and capital formation are also large in magnitude, with calculated value-added losses between SEK 4.3 billion and SEK 9.8 billion, potential layoffs between 3500 and 8600 employees, a decrease in capital formation between SEK 269 million and SEK 686 million, and a drop in employee income between SEK 2.2 billion and SEK 4.9 billion. As can be observed, these ranges given are fairly wide. They are the result of the dual usage of two methodologies, and two demand scenarios. To recap, hypothetical extraction includes an additional layer of effects as caused by a simulated demand drop from other sectors, apart from the reduced demand from final consumers. Furthermore, scenario 2 mildens the effects somewhat as it assumes that some of the lost demand for Swedish air transport is reallocated to other sectors. It is therefore clear that the lower bound for these results is the combination of inoperability analysis and scenario 2, whereas the higher bound are the findings from hypothetical extraction in scenario 1. It may of course still be argued that this research could have been of additional value if it managed to better pinpoint the effects more exactly. However, as motivated earlier on, this fairly complex choice of methods and scenarios was done to provide a more dynamic, more detailed picture of the pandemic's effects at a stage where the crisis is still ongoing, data is in some cases lacking, and we are not yet completely sure of the final outcome. As could be observed in the literature review at this stage, scenario building isn't uncommon. In fact, it was done by Iacus et al. (2020), considered to be a valuable and significant contribution in assessing the global effects of the pandemic on air transport, and spin-off effects for the global economy.

## 5.5 Global Implications

It is at this point possible to link together the two research goals of this thesis, which are to investigate the economic linkages of the Swedish air transport sector, and the effects of the COVID-19 pandemic on air transport and spin-off effects for the whole economy. Overall, the results seem to indicate that other sectors of the economy are more important for Swedish air transport than Swedish air transport is for them. This idea is supported both by fundamental input-output analysis, and more advanced methods using the same framework (inoperability analysis and hypothetical extraction.). This finding also is in line with for instance those of Malaysian Aviation Commission (2017) where air transport is ranked low down in terms of its output multiplier. The effects of the pandemic are very grave for the air transport sector, but the spin-off effects, while most certainly significant, are somewhat less dire for the economy as a whole.

Does this indicate that governments should refrain from support packages to national airlines, and let them fall into depths of failure? Not necessarily. Firstly, in a global context, based on the literature review, and the findings of this thesis, (while also acknowledging same differences in methodology), it seems as if air transport sector output multipliers for Sweden are on the lower side. The higher the multiplier, the stronger the impact on the economy, whether positive or negative. Therefore, this would imply that the same demand drop for air transport that took place in Sweden would have a larger impact in many other countries.

There are also non-economic arguments on the table for policy makers, some rational, others less so. These include for instance concerns for national security. The COVID-19 pandemic itself has been a testimony to the importance of national carriers in crisis situations. At the early stages when governments were presented with the excruciatingly complicated logistical challenge of efficiently expatriating their citizens from affected areas, national carriers were heavily relied on. These stages of the pandemic also tested the on paper solid global and regional cooperation between nations at times of need, but unfortunately presented a somewhat different picture of reality where many countries behaved in an egocentric way at the cost of others. (Think for instance hindering of mask transports, and more). Since crises will take place in the future as well, one might ask if the pandemic will cause a shift towards increased global cooperation, or will instead cement the need for national resources such as airlines in times of crisis?

Lastly, there is a further aspect to consider pointing towards that many governments are, and will be hesitant to let national carriers fail, and that is the notion of national pride. National carriers undoubtedly still represent a salient part of national identity, comparable to achievements in space exploration, sports competitions and more. Even though low-cost carriers operating from several countries often provide more economically efficient ways of air transportation than do national carriers, and it is clear that we are experiencing a slow shift in which such carriers are gaining more and more dominance on national and regional routes, the author believes that national flagship carriers will remain important to the national identities of several countries for a considerable time to come.

Based on the above, the final takeaway may therefore be expressed as follows: Policy makers worldwide must not necessarily completely refrain from crisis support to their airlines. However, they are advised not to take for granted that they should always be saved, and at any cost. Instead, they should carefully consider both economic and non-economic arguments in order to make more informed, and less hasty decisions than they have done in the past year.

## 5.6 Limitations and Suggestions for Further Research

In this sections, limitations of this thesis, as well as suggestions for further research will be covered. Limitations are divided up into two parts, general limitations typical to input-output analysis, and additional limitations specific to this thesis.

### 5.6.1 General theoretical limitations of input-output analysis

Input-output theory allows for unique perspectives on economies but isn't without its weaknesses (Kronenberg et al., (2018). These weaknesses are well known and brought up by several researchers in the field (Huderek et al., (2016), (Kronenberg et al., 2018), (Miller & Blair, 2009). One such limitation is the assumption of linear input structures. As known, the technical coefficients for each industry are always fixed, and do not change depending on output levels. In other words, no rules of diminishing returns are present here. In addition, prices do not play a role in the model. The costs for labor and capital are fixed, and not affected by output levels. Therefore, the model allows for the provision of unlimited amounts of labor and capital at fixed prices.

Furthermore, another assumption (industry technology assumption) is that each industry only produces one type of product. This is done out of necessity, in order to simplify the collection, organization and categorization of massive amount of data from a whole range of actors within a national economy. This process of course presents great challenges in order to ensure consistency and accuracy of data.

Another significant downside of input-output analysis has to do with the age of input-output tables. Such tables take significant resources to create, as they require gathering and organizing very large amounts of data. Hence the most up to date tables are always a few years old, as is the case for Sweden as well, with a table that is two years older than ideal. In this case, the final demand blowup method was employed in order to correct for an underestimation of effects that would have resulted if it wasn't. That being said, additional assumptions were required during the process, which are never positive for the accuracy of results. Still, as also mentioned in Miller and Blair (2009), one is still generally better of using such methods, than not using them at all.

### 5.6.2 Additional limitations specific to this writing, and suggestions for further research

Additional possible weaknesses of this thesis also include the definition of the input-output model. As described in detail, a model with consumer demand exogenous was chosen as the time resources to ensure the compatibility with methods used of a model with it endogenous were deemed to be out of reach. This undoubtedly still brings with it the downside that induced effects, in other words effects that reach beyond direct, and indirect impacts could not be shown. It is clear that were it used, the impacts would have been found to be higher, and therefore it is possible to theorize that the total effects on output, jobs, income and capital formation found in this thesis provide a milder picture than what is reality.  Further research in this field may therefore provide value if this alternative model is used.

Another point worth discussing is the assumption that interindustry demand of Swedish passenger air services behaves in the same way as does consumer demand. Had additional data on the possible differences in behavior between these two groups been available, then accuracy of results could have been improved. Still, the way interindustry demand was handled also has some positives that aren't necessarily to be taken for granted. The fact that use data on the ratio between air passenger and air cargo transport was obtained for every single industry, is deemed to have meant a very significant improvement to accuracy, as further assumptions were not needed to make.

A further possible weakness are the assumptions made regarding demand scenario 2. It is clear, that the way some demand was reallocated to various sectors could have been done on a more objective basis. Information regarding demand flows to and from Swedish air transport were requested from the Swedish National Institute of Economic Research precisely to tackle this weakness, but the author was denied access. Since Sweden has a fairly small space transport sector, and data regarding Swedish air transport and Swedish space transport is often lumped together, the institute expressed worry that giving out such data would be of concern to Swedish national security. (It should be noted that this policy isn't unique to this specific institute, but in fact all government bodies and instructions consistently refrain from providing statistics on the Swedish space sector). Needless to say, such data could have improved the accuracy of the assumptions for scenario 2. At the same time however, it is deemed that the benefits brought by the additional view provided by its inclusion are valuable. Especially in terms of total effects, it provides a more realistic picture of the way final demand works.

Lastly, it is suggested for further research to make use of models which include a time variable, to fully capture the long-term effects of crises such as the one examined in this writing. As has been made clear earlier, the aim of this thesis is to estimate the short-term effects of the pandemic on air transport and, the spin-off effects for the whole economy. However, in order to expand the research horizon further up in time, it is suggested that models such as the dynamic input-output model as discussed in Santos (2006) could be used to provide a better picture of the long-term effects.

# 6. Conclusion

SARS-CoV-2, a virus initially shrouded in mystery, and perhaps underestimated, has during the year of 2020 grown to become an unfortunate reality that almost every human on Earth has to deal with on a daily basis. The pandemic has had a transformative effect on our daily lives to an extent that will guarantee it a prominent place in history books of decades, or even centuries to come. Many aspects of a post-Cold War, globalized, and interconnected society, up until now taken for granted, have had to be discarded for the time being. One such aspect is unrestricted, international air travel, which had been seen as a hallmark of this era. Airlines have experienced unprecedented plunges in demand, and several desperate cries for help have been directed towards policy makers, who face the dilemma of evaluating how much (if any) of their taxpayers' money to allocate to these failing national carriers. It is as such certainly a relevant time to take a close look at how important the air transport sector actually is for the well-being of national economies. In the case of this particular research, focus is directed to the Swedish air transport sector, however, implications can be extended to a global scale.

The research question chosen to be investigated was how important the Swedish air transport sector is for Sweden's economy in the context of the COVID-19 pandemic. Additionally, two research goals were defined to complement each other in the analysis. Firstly, finding the economic linkages of the Swedish air transport sector to outline its role in the Swedish economy. Secondly, estimating the effects of the COVID-19 pandemic on air transport, and the spin-off effects for the economy as a whole.

The literature review conducted exposes important gaps in current research on the topic. Firstly, while literature on the economic linkages of air transport is abundant when viewing the international playing field, it is rather limited when it comes to Sweden. Secondly, and unsurprisingly, not many papers of quantitative nature exist on the economic effects of the pandemic on air transport, and spin-off effects for national economies.

When it comes to contribution to methodological advancements however, the role of this thesis is more modest. The framework used, input-output analysis, is already well structured and explained by excellent researchers in the field. However, it can be argued that this thesis' dual application of two input-output based methodologies, inoperability analysis and partial hypothetical extraction is unique in this field, and the author's harmonization of these methodologies have further illustrated links that exist between them, and how they can potentially be used in a complementary way at times of crisis, when data is not always in abundance.

The findings of this thesis lead to the following conclusions. The Swedish air transport sector plays a significant role for spinning the cogs of the Swedish economy. Nonetheless, when pit against all the other sectors of the economy, it ranks in the lower end in terms of its contribution to output, value added, investment, employment and income. In general, it seems to more heavily depend on other sectors, than these sectors depend on it. These conclusions are also consistent with those of other international studies in the field. Moreover, the COVID-19 pandemic has a significant effect on the Swedish air transport sector, but the spin-off effects, meanwhile, are milder. A few sectors with particularly tight links to the air transport sector are heavily impacted, but the total effects on the economy, while notable, are less drastic than initially expected.

In terms of short-term implications, it can as such be concluded that out of a pure economic standpoint, using the methodologies chosen for this thesis, no conclusive evidence has been found based on which policy makers should as a rule always pour significant resources into saving their national airlines. At the same time, it is also important to consider non-economic aspects that lie on their tables, such as the benefits national airlines can bring for national security at times of dodgy global cooperation.

Lastly, to accurately estimate more long-term effects, and therefore implications, further research in this field is recommended. Upcoming studies could consider the usage of models with a time variable, which is one of the major limitations of traditional input-output analysis. This could be done either through alternative, causality-based frameworks, or the application of dynamic input-output analysis models. To further enrich the results found in this thesis, future research could also consider the usage of an input-output model with consumer demand endogenous, in order to capture additional, induced impacts on national economies.

# Appendix

## Appendix A: Air Cargo Volumes

***Table 18:*** *2020 Air Freight Volumes (tons)*

| | **2020** | **2018**[27] |
|---|---|---|
| January | 11 825 | 13 060 |
| February | 12 057 | 11 807 |
| March | 12 639 | 13 681 |
| April | 9 452 | 12 574 |
| May | 10 039 | 12 702 |
| June | 9 759 | 11 552 |
| July | 10 126 | 11 490 |
| August | 10 435 | 11 811 |
| September | 11 243 | 12 579 |
| October | 11 682 | 13 063 |
| November | 13 000 (estimate)[28] | 11 408 |
| December | 15 000 (estimate) | 7 224 |
| **Sum** | **137 257 (estimate)** | **142 951** |

Source: The Swedish Transport Agency (2018)

[27] 2019 data not available due to technical error on the side of the Swedish Transport Agency.
[28] November and December data not yet available. Estimations include a moderate lift in cargo volumes as a result of the holiday season.

# Appendix B: Swedish Output Multipliers

***Table 19:*** *Swedish Output Multipliers in Descending Order*

| Sector | Output multiplier |
|---|---|
| Manufacture of wood and of products of wood and cork, except furniture; manufacture of articles of straw and plaiting materials | 2,19710 |
| Forestry and logging | 2,08673 |
| Manufacture of paper and paper products | 1,95702 |
| Advertising and market research | 1,94359 |
| Crop and animal production, hunting and related service activities | 1,92811 |
| Water transport | 1,92359 |
| Manufacture of food products, beverages and tobacco products | 1,91141 |
| Motion picture, video and television programme production, sound recording and music publishing activities; programming and broadcasting activities | 1,86134 |
| Manufacture of other non-metallic mineral products | 1,84692 |
| Sewerage; waste collection, treatment and disposal activities; materials recovery; remediation activities and other waste management services | 1,83925 |
| Manufacture of basic metals | 1,83867 |
| Warehousing and support activities for transportation | 1,82700 |
| Printing and reproduction of recorded media | 1,81952 |
| Water collection, treatment and supply | 1,76674 |
| Construction | 1,74797 |
| Travel agency, tour operator reservation service and related activities | 1,73578 |
| Telecommunications | 1,71532 |
| Sports activities and amusement and recreation activities | 1,71481 |
| Imputed rents of owner-occupied dwellings | 1,70544 |
| Other professional, scientific and technical activities; veterinary activities | 1,69717 |
| Accommodation and food service activities | 1,69227 |
| Postal and courier activities | 1,68621 |
| Land transport and transport via pipelines | 1,67721 |
| Electricity, gas, steam and air conditioning supply | 1,63417 |
| Manufacture of electrical equipment | 1,63410 |
| Creative, arts and entertainment activities; libraries, archives, museums and other cultural activities; gambling and betting activities | 1,62866 |
| Retail trade, except of motor vehicles and motorcycles | 1,61914 |
| Repair and installation of machinery and equipment | 1,61391 |
| Manufacture of fabricated metal products, except machinery and equipment | 1,61139 |
| Real estate activities excluding imputed rents | 1,61087 |
| Manufacture of machinery and equipment n,e,c, | 1,59682 |
| Publishing activities | 1,59026 |
| Manufacture of chemicals and chemical products | 1,59023 |
| Architectural and engineering activities; technical testing and analysis | 1,58668 |
| Mining and quarrying | 1,57310 |
| Rental and leasing activities | 1,57241 |
| Wholesale trade, except of motor vehicles and motorcycles | 1,56340 |
| Manufacture of rubber and plastic products | 1,55928 |
| Manufacture of motor vehicles, trailers and semi-trailers | 1,55089 |
| Manufacture of furniture; other manufacturing | 1,54363 |
| Repair of computers and personal and household goods | 1,53817 |
| Legal and accounting activities; activities of head offices; management consultancy activities | 1,53110 |
| Computer programming, consultancy and related activities; information service activities | 1,51244 |
| Manufacture of basic pharmaceutical products and pharmaceutical preparations | 1,49275 |
| Activities of membership organisations | 1,48374 |
| Manufacture of other transport equipment | 1,47892 |
| Public administration and defence; compulsory social security | 1,46611 |
| Security and investigation activities; services to buildings and landscape activities; office administrative, office support and other business support activities | 1,45963 |
| Wholesale and retail trade and repair of motor vehicles and motorcycles | 1,43057 |
| Manufacture of textiles, wearing apparel and leather products | 1,43044 |
| Air transport | 1,42806 |
| Scientific research and development | 1,40627 |
| Employment activities | 1,39582 |
| Education | 1,38408 |
| Human health activities | 1,36238 |
| Financial service activities, except insurance and pension funding | 1,35290 |
| Other personal service activities | 1,35172 |
| Manufacture of computer, electronic and optical products | 1,35060 |
| Fishing and aquaculture | 1,33821 |
| Insurance, reinsurance and pension funding, except compulsory social security | 1,31780 |
| Activities auxiliary to financial services and insurance activities | 1,28694 |
| Social work activities | 1,27889 |
| Manufacture of coke and refined petroleum products | 1,23726 |
| Activities of households as employers; undifferentiated goods- and services-producing activities of households for own use | 1,00000 |

## Appendix C: Inoperability Analysis Tables

***Table 20:*** *Scenario 1: Inoperability Analysis: Loss in Operability (highest to lowest loss)*

| **Sector** | **Loss in Operability** |
|---|---|
| Air transport | -0,426686 |
| Warehousing and support activities for transportation | -0,003935 |
| Manufacture of other transport equipment | -0,003519 |
| Rental and leasing activities | -0,003137 |
| Repair and installation of machinery and equipment | -0,002649 |
| Land transport and transport via pipelines | -0,001498 |
| Employment activities | -0,001381 |
| Education | -0,001359 |
| Security and investigation activities; services to buildings and landscape activities; office administrative, office support and other business support activities | -0,001230 |
| Advertising and market research | -0,001149 |
| Real estate activities excluding imputed rents | -0,001110 |
| Other professional, scientific and technical activities; veterinary activities | -0,000981 |
| Printing and reproduction of recorded media | -0,000944 |
| Wholesale trade, except of motor vehicles and motorcycles | -0,000826 |
| Postal and courier activities | -0,000820 |
| Insurance, reinsurance and pension funding, except compulsory social security | -0,000739 |
| Repair of computers and personal and household goods | -0,000702 |
| Water collection, treatment and supply | -0,000686 |
| Manufacture of basic pharmaceutical products and pharmaceutical preparations | -0,000649 |
| Water transport | -0,000647 |
| Publishing activities | -0,000644 |
| Financial service activities, except insurance and pension funding | -0,000635 |
| Computer programming, consultancy and related activities; information service activities | -0,000592 |
| Motion picture, video and television programme production, sound recording and music publishing activities; programming and broadcasting activities | -0,000584 |
| Legal and accounting activities; activities of head offices; management consultancy activities | -0,000558 |
| Travel agency, tour operator reservation service and related activities | -0,000551 |
| Manufacture of food products, beverages and tobacco products | -0,000543 |
| Accommodation and food service activities | -0,000525 |
| Sewerage; waste collection, treatment and disposal activities; materials recovery; remediation activities and other waste management services | -0,000479 |
| Wholesale and retail trade and repair of motor vehicles and motorcycles | -0,000471 |
| Activities of membership organisations | -0,000464 |
| Telecommunications | -0,000462 |
| Electricity, gas, steam and air conditioning supply | -0,000445 |
| Crop and animal production, hunting and related service activities | -0,000434 |
| Architectural and engineering activities; technical testing and analysis | -0,000380 |
| Manufacture of other non-metallic mineral products | -0,000335 |
| Construction | -0,000294 |
| Manufacture of electrical equipment | -0,000288 |
| Public administration and defence; compulsory social security | -0,000277 |
| Manufacture of fabricated metal products, except machinery and equipment | -0,000275 |
| Fishing and aquaculture | -0,000256 |
| Activities auxiliary to financial services and insurance activities | -0,000246 |
| Manufacture of motor vehicles, trailers and semi-trailers | -0,000242 |
| Sports activities and amusement and recreation activities | -0,000240 |
| Retail trade, except of motor vehicles and motorcycles | -0,000236 |
| Manufacture of wood and of products of wood and cork, except furniture; manufacture of articles of straw and plaiting materials | -0,000232 |
| Manufacture of machinery and equipment n,e,c, | -0,000215 |
| Manufacture of coke and refined petroleum products | -0,000214 |
| Manufacture of computer, electronic and optical products | -0,000205 |
| Forestry and logging | -0,000203 |
| Manufacture of rubber and plastic products | -0,000200 |
| Manufacture of basic metals | -0,000199 |
| Creative, arts and entertainment activities; libraries, archives, museums and other cultural activities; gambling and betting activities | -0,000197 |
| Manufacture of chemicals and chemical products | -0,000197 |
| Manufacture of paper and paper products | -0,000196 |
| Mining and quarrying | -0,000175 |
| Other personal service activities | -0,000126 |
| Manufacture of textiles, wearing apparel and leather products | -0,000109 |
| Manufacture of furniture; other manufacturing | -0,000104 |
| Scientific research and development | -0,000079 |
| Social work activities | -0,000027 |
| Human health activities | -0,000026 |
| Imputed rents of owner-occupied dwellings | 0,000000 |
| Activities of households as employers; undifferentiated goods- and services-producing activities of households for own use | 0,000000 |

***Table 21:*** *Scenario 2: Inoperability Analysis: Change in Operability (highest to lowest)*

| Sector | Change in operability |
|---|---|
| Air transport | -0,426424 |
| Manufacture of other transport equipment | -0,003429 |
| Rental and leasing activities | -0,002851 |
| Repair and installation of machinery and equipment | -0,002269 |
| Education | -0,001335 |
| Warehousing and support activities for transportation | -0,001333 |
| Real estate activities excluding imputed rents | -0,000837 |
| Employment activities | -0,000830 |
| Security and investigation activities; services to buildings and landscape activities; office administrative, office support and other business support activities | -0,000818 |
| Advertising and market research | -0,000745 |
| Other professional, scientific and technical activities; veterinary activities | -0,000737 |
| Wholesale trade, except of motor vehicles and motorcycles | -0,000701 |
| Insurance, reinsurance and pension funding, except compulsory social security | -0,000592 |
| Financial service activities, except insurance and pension funding | -0,000498 |
| Publishing activities | -0,000453 |
| Travel agency, tour operator reservation service and related activities | -0,000449 |
| Printing and reproduction of recorded media | -0,000443 |
| Accommodation and food service activities | -0,000424 |
| Computer programming, consultancy and related activities; information service activities | -0,000401 |
| Activities of membership organisations | -0,000369 |
| Water collection, treatment and supply | -0,000339 |
| Water transport | -0,000337 |
| Electricity, gas, steam and air conditioning supply | -0,000257 |
| Sewerage; waste collection, treatment and disposal activities; materials recovery; remediation activities and other waste management services | -0,000255 |
| Motion picture, video and television programme production, sound recording and music publishing activities; programming and broadcasting activities | -0,000223 |
| Manufacture of electrical equipment | -0,000219 |
| Legal and accounting activities; activities of head offices; management consultancy activities | -0,000201 |
| Manufacture of motor vehicles, trailers and semi-trailers | -0,000197 |
| Public administration and defence; compulsory social security | -0,000197 |
| Construction | -0,000194 |
| Architectural and engineering activities; technical testing and analysis | -0,000187 |
| Manufacture of other non-metallic mineral products | -0,000178 |
| Manufacture of fabricated metal products, except machinery and equipment | -0,000175 |
| Manufacture of machinery and equipment n,e,c, | -0,000170 |
| Manufacture of computer, electronic and optical products | -0,000163 |
| Sports activities and amusement and recreation activities | -0,000157 |
| Activities auxiliary to financial services and insurance activities | -0,000155 |
| Manufacture of basic metals | -0,000151 |
| Manufacture of wood and of products of wood and cork, except furniture; manufacture of articles of straw and plaiting materials | -0,000147 |
| Creative, arts and entertainment activities; libraries, archives, museums and other cultural activities; gambling and betting activities | -0,000144 |
| Fishing and aquaculture | -0,000124 |
| Manufacture of paper and paper products | -0,000120 |
| Forestry and logging | -0,000113 |
| Mining and quarrying | -0,000100 |
| Manufacture of rubber and plastic products | -0,000093 |
| Manufacture of chemicals and chemical products | -0,000084 |
| Other personal service activities | -0,000076 |
| Manufacture of coke and refined petroleum products | -0,000074 |
| Manufacture of textiles, wearing apparel and leather products | -0,000065 |
| Wholesale and retail trade and repair of motor vehicles and motorcycles | -0,000061 |
| Scientific research and development | -0,000061 |
| Manufacture of furniture; other manufacturing | -0,000032 |
| Social work activities | -0,000021 |
| Repair of computers and personal and household goods | -0,000020 |
| Imputed rents of owner-occupied dwellings | 0,000000 |
| Activities of households as employers; undifferentiated goods- and services-producing activities of households for own use | 0,000000 |
| Crop and animal production, hunting and related service activities | 0,000011 |
| Manufacture of food products, beverages and tobacco products | 0,000347 |
| Land transport and transport via pipelines | 0,000581 |
| Human health activities | 0,000774 |
| Retail trade, except of motor vehicles and motorcycles | 0,000784 |
| Manufacture of basic pharmaceutical products | 0,002742 |
| Telecommunications | 0,004730 |
| Postal and courier activities | 0,007517 |

# Appendix D: Hypothetical Extraction Tables

***Table 22:*** *Scenario 1: Hypothetical Extraction: Normalized output loss (high to low)*

| Sector | Normalized Output Loss |
|---|---|
| Air transport | -0,669200 |
| Warehousing and support activities for transportation | -0,006171 |
| Manufacture of other transport equipment | -0,005518 |
| Rental and leasing activities | -0,004920 |
| Repair and installation of machinery and equipment | -0,004154 |
| Land transport and transport via pipelines | -0,002349 |
| Employment activities | -0,002166 |
| Education | -0,002131 |
| Security and investigation activities | -0,001929 |
| Advertising and market research | -0,001802 |
| Real estate activities excluding imputed rents | -0,001741 |
| Other professional, scientific and technical activities; veterinary activities | -0,001539 |
| Printing and reproduction of recorded media | -0,001480 |
| Wholesale trade, except of motor vehicles and motorcycles | -0,001296 |
| Postal and courier activities | -0,001286 |
| Insurance, reinsurance and pension funding, except compulsory social security | -0,001159 |
| Repair of computers and personal and household goods | -0,001101 |
| Water collection, treatment and supply | -0,001076 |
| Manufacture of basic pharmaceutical products and pharmaceutical preparations | -0,001018 |
| Water transport | -0,001015 |
| Publishing activities | -0,001010 |
| Financial service activities, except insurance and pension funding | -0,000995 |
| Computer programming, consultancy and related activities; information service activities | -0,000928 |
| Motion picture, video and television programme production, sound recording and music publishing activities; programming and broadcasting activities | -0,000916 |
| Legal and accounting activities; activities of head offices; management consultancy activities | -0,000875 |
| Travel agency, tour operator reservation service and related activities | -0,000864 |
| Manufacture of food products, beverages and tobacco products | -0,000852 |
| Accommodation and food service activities | -0,000824 |
| Sewerage; waste collection, treatment and disposal activities; materials recovery; remediation activities and other waste management services | -0,000751 |
| Wholesale and retail trade and repair of motor vehicles and motorcycles | -0,000738 |
| Activities of membership organisations | -0,000727 |
| Telecommunications | -0,000724 |
| Electricity, gas, steam and air conditioning supply | -0,000698 |
| Crop and animal production, hunting and related service activities | -0,000681 |
| Architectural and engineering activities; technical testing and analysis | -0,000596 |
| Manufacture of other non-metallic mineral products | -0,000525 |
| Construction | -0,000461 |
| Manufacture of electrical equipment | -0,000452 |
| Public administration and defence; compulsory social security | -0,000435 |
| Manufacture of fabricated metal products, except machinery and equipment | -0,000431 |
| Fishing and aquaculture | -0,000401 |
| Activities auxiliary to financial services and insurance activities | -0,000386 |
| Manufacture of motor vehicles, trailers and semi-trailers | -0,000380 |
| Sports activities and amusement and recreation activities | -0,000377 |
| Retail trade, except of motor vehicles and motorcycles | -0,000370 |
| Manufacture of wood and of products of wood and cork, except furniture; manufacture of articles of straw and plaiting materials | -0,000365 |
| Manufacture of machinery and equipment n,e,c, | -0,000337 |
| Manufacture of coke and refined petroleum products | -0,000335 |
| Manufacture of computer, electronic and optical products | -0,000321 |
| Forestry and logging | -0,000318 |
| Manufacture of rubber and plastic products | -0,000314 |
| Manufacture of basic metals | -0,000312 |
| Creative, arts and entertainment activities; libraries, archives, museums and other cultural activities; gambling and betting activities | -0,000309 |
| Manufacture of chemicals and chemical products | -0,000308 |
| Manufacture of paper and paper products | -0,000308 |
| Mining and quarrying | -0,000274 |
| Other personal service activities | -0,000198 |
| Manufacture of textiles, wearing apparel and leather products | -0,000170 |
| Manufacture of furniture; other manufacturing | -0,000162 |
| Scientific research and development | -0,000124 |
| Social work activities | -0,000042 |
| Human health activities | -0,000040 |
| Imputed rents of owner-occupied dwellings | 0,000000 |
| Activities of households as employers | 0,000000 |

***Table 23:*** *Scenario 2: Hypothetical Extraction: Normalized Output Loss (high to low)*

| **Sector** | **Normalized Output Change** |
|---|---|
| Air transport | -0,669003 |
| Manufacture of other transport equipment | -0,005429 |
| Rental and leasing activities | -0,004635 |
| Repair and installation of machinery and equipment | -0,003775 |
| Warehousing and support activities for transportation | -0,003570 |
| Education | -0,002107 |
| Employment activities | -0,001615 |
| Security and investigation activities | -0,001517 |
| Real estate activities excluding imputed rents | -0,001468 |
| Advertising and market research | -0,001398 |
| Other professional, scientific and technical activities; veterinary activities | -0,001295 |
| Wholesale trade, except of motor vehicles and motorcycles | -0,001171 |
| Insurance, reinsurance and pension funding, except compulsory social security | -0,001012 |
| Printing and reproduction of recorded media | -0,000979 |
| Financial service activities, except insurance and pension funding | -0,000859 |
| Publishing activities | -0,000819 |
| Travel agency, tour operator reservation service and related activities | -0,000762 |
| Computer programming, consultancy and related activities; information service activities | -0,000737 |
| Water collection, treatment and supply | -0,000729 |
| Accommodation and food service activities | -0,000723 |
| Water transport | -0,000705 |
| Activities of membership organisations | -0,000632 |
| Motion picture, video and television programme production, sound recording and music publishing activities; programming and broadcasting activities | -0,000555 |
| Sewerage; waste collection, treatment and disposal activities; materials recovery; remediation activities and other waste management services | -0,000527 |
| Legal and accounting activities; activities of head offices; management consultancy activities | -0,000518 |
| Electricity, gas, steam and air conditioning supply | -0,000510 |
| Repair of computers and personal and household goods | -0,000419 |
| Architectural and engineering activities; technical testing and analysis | -0,000403 |
| Manufacture of electrical equipment | -0,000383 |
| Manufacture of other non-metallic mineral products | -0,000368 |
| Construction | -0,000361 |
| Public administration and defence; compulsory social security | -0,000355 |
| Manufacture of motor vehicles, trailers and semi-trailers | -0,000335 |
| Manufacture of fabricated metal products, except machinery and equipment | -0,000331 |
| Wholesale and retail trade and repair of motor vehicles and motorcycles | -0,000328 |
| Activities auxiliary to financial services and insurance activities | -0,000295 |
| Sports activities and amusement and recreation activities | -0,000294 |
| Manufacture of machinery and equipment n,e,c, | -0,000293 |
| Manufacture of computer, electronic and optical products | -0,000279 |
| Manufacture of wood and of products of wood and cork, except furniture; manufacture of articles of straw and plaiting materials | -0,000279 |
| Land transport and transport via pipelines | -0,000270 |
| Fishing and aquaculture | -0,000270 |
| Manufacture of basic metals | -0,000265 |
| Creative, arts and entertainment activities; libraries, archives, museums and other cultural activities; gambling and betting activities | -0,000256 |
| Crop and animal production, hunting and related service activities | -0,000236 |
| Manufacture of paper and paper products | -0,000232 |
| Forestry and logging | -0,000229 |
| Manufacture of rubber and plastic products | -0,000207 |
| Mining and quarrying | -0,000200 |
| Manufacture of chemicals and chemical products | -0,000196 |
| Manufacture of coke and refined petroleum products | -0,000196 |
| Other personal service activities | -0,000148 |
| Manufacture of textiles, wearing apparel and leather products | -0,000127 |
| Scientific research and development | -0,000106 |
| Manufacture of furniture; other manufacturing | -0,000091 |
| Social work activities | -0,000036 |
| Imputed rents of owner-occupied dwellings | 0,000000 |
| Activities of households as employers | 0,000000 |
| Manufacture of food products, beverages and tobacco products | 0,000038 |
| Retail trade, except of motor vehicles and motorcycles | 0,000650 |
| Human health activities | 0,000759 |
| Manufacture of basic pharmaceutical products and pharmaceutical preparations | 0,002373 |
| Telecommunications | 0,004467 |
| Postal and courier activities | 0,007051 |

## Appendix E: Sector Abbreviations

***Table 24:*** *Sector Abbreviations (recap)*

| Abbreviation | Full description |
|---|---|
| ACM | Accommodation and food service activities |
| AMR | Advertising and market research |
| AT | Air transport |
| CP | Computer programming, consultancy and related activities; information service activities |
| CST | Construction |
| CRO | Crop and animal production, hunting and related service activities |
| ED | Education |
| EMP | Employment activities |
| HH | Human health activities |
| LTP | Land transport and transport via pipelines |
| MP | Manufacture of basic pharmaceutical products and pharmaceutical preparations |
| MF | Manufacture of food products, beverages and tobacco products |
| MMA | Manufacture of machinery and equipment n,e,c, |
| MM | Manufacture of motor vehicles, trailers and semi-trailers |
| MOT | Manufacture of other transport equipment |
| POS | Postal and courier activities |
| RAI | Repair and installation of machinery and equipment |
| RE | Real estate activities excluding imputed rents |
| RL | Rental and leasing activities |
| RT | Retail trade, except of motor vehicles and motorcycles |
| SEC | Security/services to buildings/, landscape, office administration and support |
| WHS | Warehousing and support activities for transportation |
| TEL | Telecommunications |
| WHT | Wholesale trade, except of motor vehicles and motorcycles |
| WRT | Wholesale and retail trade and repair of motor vehicles and motorcycles |